\documentclass[11pt]{article}
\pdfoutput=1
\usepackage{jheppub}
\usepackage{upgreek}
\usepackage{float, extarrows, tikz-cd}
\usepackage{graphicx}
\usepackage{slashed}
\usepackage{tabularx,ragged2e}
\usepackage{amssymb}
\usepackage{subcaption}
\usepackage{amsmath,amssymb}
\usepackage{slashed}
\usepackage{caption}
\usepackage{xcolor}
\usepackage{dsfont}
\usepackage{verbatim}
\usepackage{mathtools, xcolor,ytableau, amsfonts,tikz}
\usepackage{graphicx}
\usepackage{physics}
\usepackage{simpler-wick}
\usepackage{microtype}
\usepackage[nobottomtitles*]{titlesec}
\newcolumntype{C}{>{\Centering\arraybackslash}X}

\numberwithin{equation}{section}

\newcommand{\mL}{\mathcal{L}}
\newcommand{\mO}{\mathcal{O}}

\newcommand{\pd}{\partial}

\newcommand{\st}{s}

\newcommand{\bz}{\bar{z}}

\def\le{\left}
\def\ri{\right}
\newcommand\ov{\over}

\newcommand{\es}[2] {\begin{equation} \label{#1} \begin{split} #2 \end{split} \end{equation}}

\def\<{\langle}
\def\>{\rangle}
\newcommand\al{{\alpha}}
\newcommand\ep{\varepsilon}

\newcommand\lam{\lambda}

\newcommand\ga{{\ensuremath{{\gamma}}}}

\newcommand\de{{\ensuremath{{\delta}}}}

\author[a,b]{Gabriel Cuomo,}   
\author[a,b]{Zohar Komargodski,}        
\author[a]{M\'ark Mezei}                         
\author[a]{and Avia Raviv-Moshe}   
              \affiliation[a]{Simons Center for Geometry and Physics, SUNY, Stony Brook, NY 11794, USA}      
              \affiliation[b]{C. N. Yang Institute for Theoretical Physics, Stony Brook University, Stony Brook, NY 11794, USA}                                                              
                                            
\emailAdd{gcuomo@scgp.stonybrook.edu}					 \emailAdd{zkomargodski@scgp.stonybrook.edu}  
     \emailAdd{mmezei@scgp.stonybrook.edu}   
          \emailAdd{araviv-moshe@scgp.stonybrook.edu}

\begin{document}

\title{Spin Impurities,  Wilson Lines  and Semiclassics}

\abstract{We consider line defects with large quantum numbers in conformal field theories.  First, we consider spin impurities, both for a free scalar triplet and in the Wilson-Fisher $O(3)$ model.  For the free scalar triplet, we find a rich phase diagram that includes a perturbative fixed point,  a new nonperturbative fixed point, and runaway regimes. To obtain these results, we develop a new semiclassical approach. For the Wilson-Fisher model, we propose an alternative description, which becomes weakly coupled in the large spin limit. This allows us to chart the phase diagram and obtain numerous rigorous predictions for large spin impurities in $2+1$ dimensional magnets. Finally,  we also study $1/2$-BPS Wilson lines in large representations of the gauge group in rank-1 $\mathcal{N}=2$ superconformal field theories. We contrast the results with the qualitative behavior of large spin impurities in magnets.}

\maketitle

\section{Introduction and summary}

The study of line defects (i.e. one-dimensional defects) in critical conformal bulk theories is of fundamental importance to the study of Quantum Field Theory (QFT).  Line defects have a variety of applications ranging from condensed matter and statistical physics, such as models of magnetic impurities in metals and magnets \cite{sachdev1999quantum,vojta2000quantum}, to high energy physics, such as Wilson and ’t Hooft lines in four-dimensional gauge theories \cite{Wilson:1974sk,tHooft:1977nqb}. Studies of the Kondo problem, which emerged from models of impurities in two-dimensional systems, led to remarkable progress in the study of the renormalization group, as well as to developments in integrability; see \cite{Affleck:1995ge} for a review.  

Even if the bulk is tuned to a critical point, i.e. a conformal field theory (CFT), it is well known that line operators can undergo a nontrival defect RG flow,  which generically leads to a critical line at long distances.
 In two dimensions, Affleck and Ludwig conjectured that a renormalization group flow on a line defect leads to $g_\text{UV}\geq g_\text{IR}$, where $g_\text{UV}$ ($g_\text{IR}$) refers to the universal part of the defect free energy in the UV (IR) \cite{Affleck:1991tk}. This was subsequently proven in \cite{Friedan:2003yc,Casini:2016fgb}. A generalization of this statement to line defects in bulk CFTs of arbitrary number of spacetime dimensions $d$ was recently obtained in \cite{Cuomo:2021rkm}. This was done by identifying the following scheme-independent quantity\footnote{In two dimensions, a quantity equivalent to the defect entropy and its monotonicity properties were originally identified in the context of string field theory \cite{Witten:1992qy,Witten:1992cr,Shatashvili:1993kk,Shatashvili:1993ps,Kutasov:2000qp}. }
\begin{equation} \label{eq:DefectEntropy}
s_D= \left(1- R\frac{\partial}{\partial R} \right)\log{g(MR)},
\end{equation}
where $R$ is the radius of the circular line defect, and $M$ is a mass scale associated with the RG flow on the defect. The defect $g$-function $g(MR)$ is formally defined as the partition function of the full theory normalized by the partition function of the bulk theory without the defect. 
The above quantity $s_D$ in eq.~\eqref{eq:DefectEntropy}, which is referred to as the {\it defect entropy}, monotonically decreases under a defect RG flow  and hence must obey the  inequality $g_\text{UV} \geq g_\text{IR}$ for line defects in any number of spacetime dimensions $d$.

In quantum critical models, point-like impurities in space at zero temperature can be thought of as one-dimensional defects in spacetime. In this way the study of line defects in CFTs makes contact with the study of the phases of impurities in condensed matter. 

\begin{figure}[t]
   \centering
		{\includegraphics[width=0.6\textwidth]{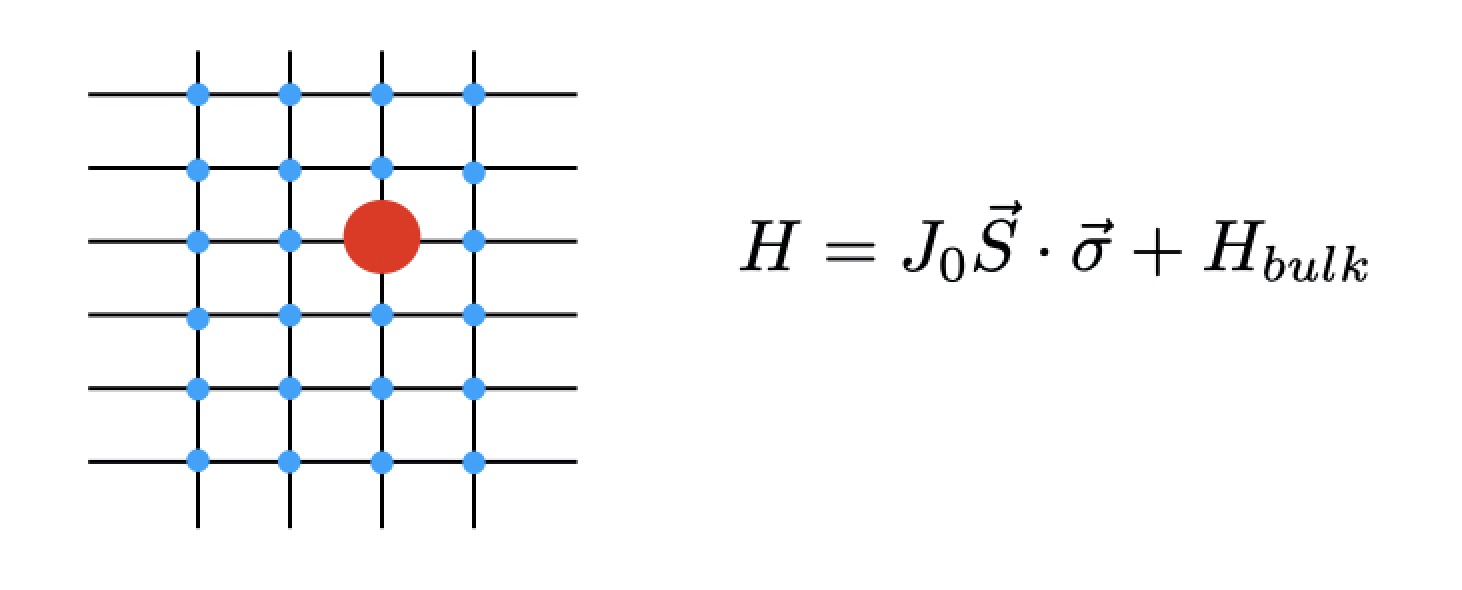}}
        \caption{ An impurity of spin $s$ under $SO(3)$ interacts with the $SO(3)$ symmetric bulk. The operators $\vec S$ are the spin $s$ representation of $so(3)$ while the operators $\vec \sigma$ are the bulk spins (typically in the spin $1/2$ representation) of the nearest neighbors and the bulk Hamiltonian $H_{bulk}$ is tuned to a critical point.}
\label{fig:DiagramLattice}
\end{figure}

The class of models of interest to us here are bulk models with global symmetry $SO(3)$ where an impurity in the spin $s$ representation is present and interacts with the bulk in an $SO(3)$ invariant fashion, see figure~\ref{fig:DiagramLattice}. Models in this family are particularly interesting due to their relation with magnets in three spacetime dimensions. Indeed, lattice realizations of $SO(3)$ bulk critical points are known and the insertion of a spin $s$ impurity is rather straightforward to implement. Such spin $s$ impurities are sometimes referred to as magnetic impurities but we will refer to them as spin impurities throughout this manuscript.

An interesting question concerning spin impurities is about their infrared behavior. Since the effective coupling on the impurity grows towards the infrared this is a very difficult problem in three spacetime dimensions. The main focus of this paper is to solve this problem in the large spin limit $s\gg1$. This limit can be taken, of course, in any number of dimensions $3\leq d\leq 4$, but it goes without saying that the most interesting case for the experimental setting is $d=3$. 

In a different context,
several works focused on the study of conformal gauge theories in the presence of line operators, especially in supersymmetric theories (see e.g. \cite{Andrei:2018die,Agmon:2020pde,Penati:2021tfj} and references therein).  Building on similarities with the description of large spin impurities, in this work we will address the large representation limit of supersymmetric Wilson lines in $\mathcal{N}=2$ superconformal gauge theories,  that is $1/2$-BPS Wilson loops in which the size $s$ of the labeling representation becomes large.

It has been recently become clear that the bulk physics of CFTs simplifies when various quantum numbers are taken to be large.\footnote{Examples include CFTs in the regimes of large scaling dimensions \cite{Lashkari:2016vgj,Belin:2020hea,Delacretaz:2020nit}, large spin \cite{Alday:2007mf,Komargodski:2012ek,Fitzpatrick:2012yx}
and large global charges \cite{Hellerman:2015nra,Monin:2016jmo,Jafferis:2017zna,Komargodski:2021zzy}.} 
A natural question which arises in this context is whether any simplification occurs for line defects with large quantum numbers and in particular for spin impurities in the large $s$ limit. 

 We will show that indeed vast simplifications occur for impurities with large spin and, furthermore, we will see that similar simplifications occur in the context of supersymmetric Wilson lines in the large representation limit. 
The sections about Wilson loops and the spin impurities can be read independently of each other.

Let us now state briefly our main results before describing the setup in more detail. 
\begin{itemize} 
\item For the large $s$ limit of spin impurities in the $O(3)$ Wilson Fisher model, we find a description which becomes increasing more useful as $s\to\infty$. The description consists of two sectors, which are weakly interacting with each other: a quantum mechanical sector  with $S^2$ target space and a first-order kinetic term, and another sector with no free parameters, which describes a previously studied conformal defect called the ``pinning field defect'' or ``localized magnetic field defect" \cite{2017PhRvB..95a4401P,Cuomo:2021kfm}.  While the pinning field defect is a strongly coupled conformal defect, some properties of it are known exactly and many others are known approximately.  
An example of an unexpected prediction that stems from this analysis is that there exists a primary defect operator in the vector representation of $SO(3)$ which is nearly marginal $\Delta_\text{vector}=1+O({1\over s})$. Another prediction is that the dimension of the lightest $SO(3)$ singlet should be approximately $\Delta_\text{singlet}\simeq1.55$. Finally, we predict that the spin operator $S$ on the defect, which acts on the  defect Hilbert space, has dimension $\Delta_S\sim 1/s^2$.

\item The second subject of this paper is 1/2-BPS Wilson lines in four-dimensional rank-1 ${\mathcal N}=2$ SCFTs in a large representation of the gauge group. (This setup also makes sense for non-Lagrangian theories, as we later explain.) Here we argue using results from localization that the large $s$ limit for protected observables leads to physics on the Coulomb branch. $1/s$ corrections are captured by higher derivative terms on the Coulomb branch. 
This allows us to make some universal predictions. For instance, for the $g$ function of such line operators we find
\begin{equation}\label{gintro}
\log g=\frac{g_\text{CB}^2 s^2}{4}+4\Delta a\log s+O\left(s^0\right)\,.
\end{equation}
where $\Delta a= a_\text{UV}-a_\text{IR}$ is the difference of the $a$-anomalies between the ultraviolet and the Coulomb branch\footnote{We work in units such that an Abelian free vector multiplet contributes with $a_\text{VM}=5/24$ and a free Hypermultiplet with $a_\text{HM}=1/24$. } and $g_\text{CB}$ is a parameter in the effective theory that is model-dependent. 
Interestingly, the $g$ function of the 1/2-BPS Wilson loops grows exponentially fast as $s\to \infty$~\eqref{gintro} while for the spin impurities the $g$ function grows only linearly as $s\to\infty$.
We propose that~\eqref{gintro} is valid also in non-Lagrangian theories.

\end{itemize}

We now delve into a more detailed summary of the content of the paper.

\subsection{Spin Impurities} 
\label{IntroMagneticImpGen}

We will consider two different scenarios of bulk theories with $O(3)$ global symmetry: a free field theory with global $O(3)$ symmetry, and the interacting $O(3)$ Wilson-Fisher model~\cite{Pelissetto:2000ek,Poland:2018epd,Henriksson:2022rnm}.

In both cases, we will consider the theory in the presence of the following line defect operator:
\begin{equation}\label{eq_ImpIntro}
D_s=\text{Tr}_{2s+1}\left[P\exp\left( \gamma_0\int d\tau\,  \phi\right)\right]\,,
\end{equation}
where $\phi_a$ is the bulk scalar field ($a=\{1,2,3\}$), $\phi=\phi_aT^a$, and the matrices $\{T^a\}$ form a $2s+1$ dimensional irreducible representation of the $su(2)$ algebra.  Such a setting describes a spin $s$ impurity inserted into a lattice site in the critical bulk and interacting with the bulk in an $SO(3)$-invariant fashion, as in figure~\ref{fig:DiagramLattice}.
The parameter $\gamma_0$ is a coupling constant and it is relevant for $d<4$. We will see that for $d=4$ it is marginally irrelevant for all $s$.

Even in the case of a free bulk theory,  we cannot at present solve the model with the defect~\eqref{eq_ImpIntro} for arbitrary number of dimensions and arbitrary $s$. The complication lies in the path-ordering in eq.~\eqref{eq_ImpIntro} that makes the diagrammatic expansion rather intricate due to the appearance of an increasing number of commutators between $su(2)$ matrices at each order in perturbation theory.

The limit we will focus on in this paper is the $s\gg 1$ limit.  Understanding the large $s$ behavior of the defect QFT in both the free bulk case and the interacting $O(3)$ Wilson-Fisher bulk case will be our main goal throughout sections~\ref{sec_magnetic_defects_free_bulk} and~\ref{sec_magnetic_defects_interacting_bulk}.  
Roughly speaking, the impurity backreacts on the bulk substantially and a new saddle-point emerges at large $s$. Then a new effective scale for quantum fluctuations emerges $s^{-1}\sim\hbar$. We will see that this intuition is partially true, indeed.

Several previous works already studied spin impurities in free theories and the $O(3)$ WF model \cite{PhysRevB.61.4041,sachdev1999quantum,vojta2000quantum,Sachdev:2001ky,Sachdev:2003yk,PhysRevLett.98.087203,PhysRevLett.99.027205,Liu:2021nck}.\footnote{See also \cite{PhysRevLett.96.036601,PhysRevB.74.165114,PhysRevB.75.224420,Biswas:2007vh,PhysRevB.77.054411} and references therein for other field-theoretical studies of impurities in different models.}
 Of particular relevance for us are \cite{PhysRevB.61.4041,sachdev1999quantum,vojta2000quantum}, that initiated the study of impurities from the field-theoretical viewpoint within the $\varepsilon$ expansion.  We also mention the Quantum Monte Carlo analysis of \cite{PhysRevLett.98.087203} for $s=1/2$ impurities. No prior work addressed the large spin limit to our knowledge.

While this paper is focused on the spin impurities, there are various other interesting defects in the $O(N)$ model. For instance, the effect of a magnetic field localized in space,  a setup which is particularly relevant for Monte Carlo simulations \cite{Assaad:2013xua},  was considered in \cite{2014arXiv1412.3449A,2017PhRvB..95a4401P,Allais:2014fqa,Cuomo:2021kfm}. The line defect that describes a localized magnetic field will be referred to as the ``pinning field defect QFT" (DQFT). The infrared conformal defect, when it exists, is referred to as the ``pinning field DCFT." Perhaps surprisingly, the results of these works, in particular of \cite{Cuomo:2021kfm}, will play an important role in our analysis later on. We will briefly review it in due course. Symmetry (twist) defects (which are not genuine line defects, since they are attached to a nontrivial topological surface) were considered in \cite{Billo:2013jda,Gaiotto:2013nva,Soderberg:2017oaa,Bianchi:2021snj,Giombi:2021uae,Gimenez-Grau:2021wiv}. Finally, let us mention that the multi-channel Kondo problem has a rich set of various large $N$ and large representation limits~\cite{tsvelick1985exact,PhysRevB.46.10812,Affleck:1995ge,PhysRevB.58.3794}.

\paragraph{Free bulk}   In sec.~\ref{Sec_FreeTheory} we discuss the defect~\eqref{eq_ImpIntro} for a free scalar triplet.
For any given fixed $s$, the model can be studied in $d=4-\varepsilon$ spacetime dimensions with $\varepsilon\ll 1$.  In the limit where $\varepsilon$ is the smallest parameter, the model admits an IR stable perturbative fixed point,  that was studied in \cite{PhysRevB.61.4041,vojta2000quantum}.

As we will explain in detail in section~\ref{sec_free_diag},  the perturbative expansion breaks down for sufficiently large spin, when $s\gtrsim 1/\varepsilon$.
In section~\ref{sec_free_semiclassics}, we find that the model can be solved in a semiclassical expansion in powers of $1/s$ for arbitrary values of the spacetime dimensions $d$. Using this approach, we are able to chart the phases of the line defect~\eqref{eq_ImpIntro}. Let us now summarize our main findings:
\begin{itemize}
\item For $\varepsilon=4-d\ll 1$,  the theory can be studied perturbatively in the double-scaling limit $s\gg 1$ with $\gamma_0^2 s=$fixed.  This regime includes the perturbative fixed point mentioned above, which occurs at any fixed $s$ for sufficiently small $\varepsilon$.
However we find a richer phase diagram. For $\varepsilon s<1/\pi$ we find two fixed points, one of which is novel and non-perturbative in the standard approach. For $\varepsilon s> 1/\pi$ we argue instead that no infrared fixed point exists and the defect $g$-function approaches zero in the IR, similarly to the free theory example discussed in \cite{Cuomo:2021kfm}. The approach of $g$ to zero means that the flow does not terminate in a healthy conformal defect in the infrared and instead one finds a certain runaway behavior.\footnote{ As we derive in subsection \ref{subsec_phi2_and_beta}, this  implies that one-point functions of local operators decay with a slower rate than in a DCFT as a function of the distance from the defect for $d>3$, while in $d=3$ they grow logarithmically with the distance, see eqs.~\eqref{eq_phi2_LO} and \eqref{eq_phi2_LO3}.} This is presumably only possible because the theory of a free triplet of scalars has a moduli space of vacua.

\item For $\varepsilon=4-d$ fixed, the model can also be studied in a $1/s$ expansion, which is similar in spirit to the usual large $N$ expansion for the $O(N)$ models \cite{Moshe:2003xn} in the sense that $s$ becomes effectively $1/\hbar$.  We find that there is no fixed point in the IR in this limit for any finite $\varepsilon>0$, and thus the flow never terminates in a DCFT. This result also applies to the large $s$ limit of the theory in $d=3$ spacetime dimensions.

\end{itemize}

\begin{figure}[t!]
\centering
\includegraphics[scale=0.45]{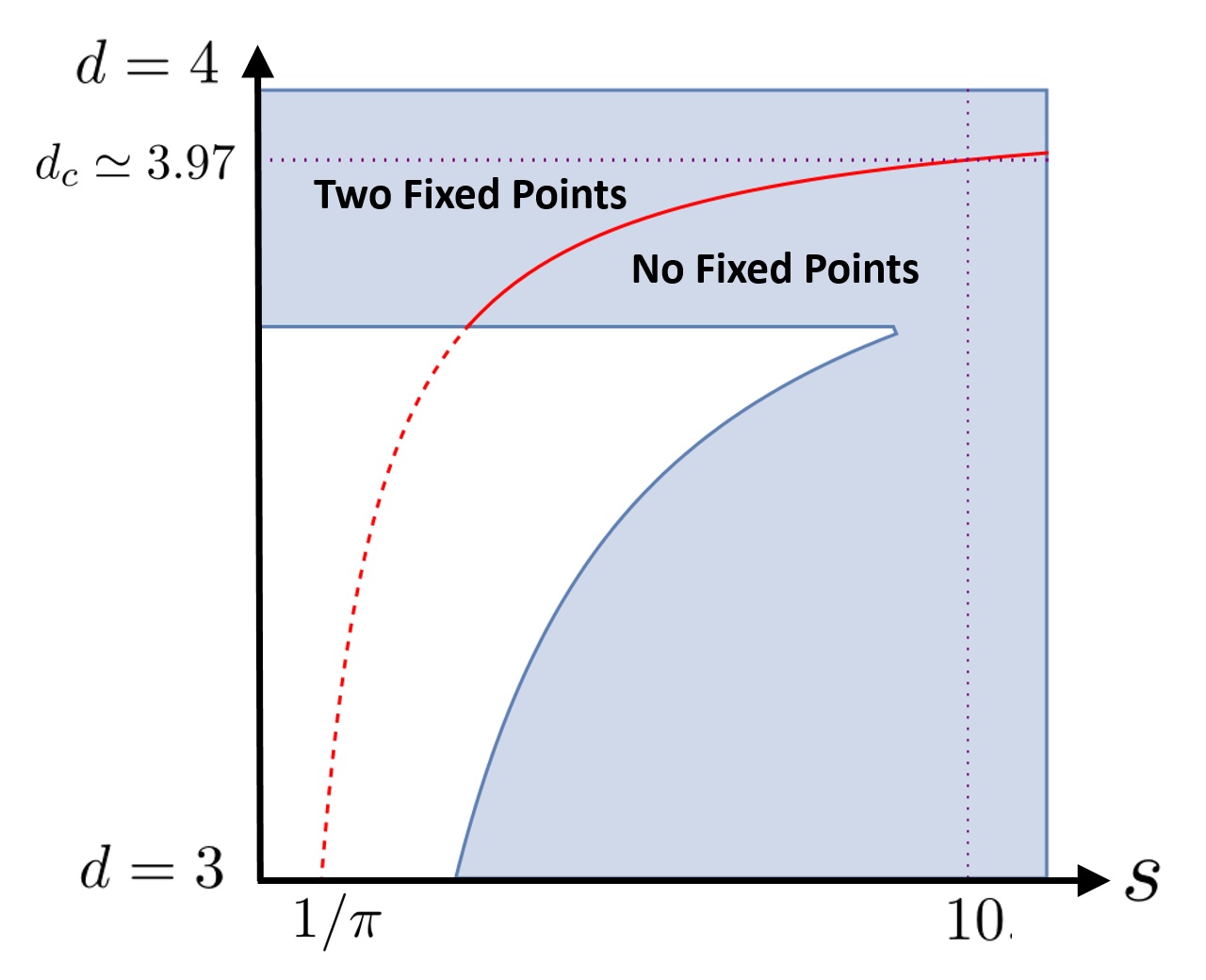}
\caption{Phase diagram of the impurity~\eqref{eq_ImpIntro} in a free bulk theory.  The blue shaded region schematically represents the one that we could reliably study with our methods. The plot is obtained from the union of the double-scaling regime,  applicable for small $4-d=\varepsilon$ but arbitrary $s$, and the fixed $\varepsilon$ large $s$ region, which requires $\varepsilon s$ to be sufficiently large (see sec.~\ref{sec_free_semiclassics} for details). The red solid line separates the region in which the theory admits two fixed points from the one in which the RG flow never terminates in a DCFT. The red dashed line is its \emph{naive} extrapolation in the region that we do not control with present techniques. Notice that the extrapolation suggests that all physical impurities ($s\geq 1/2$) in $d=3$ have no stable fixed points. Finally, the purple dotted lines refer to the numerical example for $s=10$ in the main text.}
\label{fig:FreePD}
\end{figure}

The phase diagram of the theory is summarized in figure~\ref{fig:FreePD}, where we colored blue the region of the $(d,s)$ plane that we could analyze with our methods.

To further clarify the phase diagram we propose, let us give a ``numerical'' example: Consider for instance the case with spin $s=10$ (corresponding to the purple dotted line in fig.~\ref{fig:FreePD}). For $3\leq d<d_c\simeq 3.97$ we expect no infrared DCFT to exist, and instead the flow from the trivial fixed point should never terminate and the defect entropy $s_D$ would tend to $-\infty$ in the infrared. At some critical $d_c\simeq 3.97$ a nontrivial infrared fixed point would emerge.\footnote{Such a merger of a stable and unstable fixed points and disappearance to the complex plane is the standard situation, where Miransky scaling arises \cite{Miransky:1984ef,Kaplan:2009kr,Gorbenko:2018ncu}.} This infrared fixed point has a smaller defect entropy $s_D$ than the trivial fixed point. It has an operator which is marginally relevant if added with one sign and marginally irrelevant if added with the other sign. For $d_c<d<4$ there exist two fixed points, where one of them is continuously connected to the perturbative fixed point and is stable in the infrared for $SO(3)$ symmetric perturbations. The other fixed point  has an $SO(3)$ invariant relevant operator, and has an increasingly large defect coupling as $d\to 4$, which is why it is non-perturbative. At $d=4$ the latter fixed point drifts to infinite coupling while the former fixed point merges with the trivial fixed point.

As we said, for any fixed $3\leq d<4$, at large enough $s$, the flow never terminates in a healthy conformal infrared defect. We find that this runaway behavior is analogous to the one which is obtained considering  the pinning field defect $\delta S\propto\int d \tau\phi_1$ in the free theory, see \cite{Cuomo:2021kfm}.  Also in that case, the defect renormalization group (DRG) flow never terminates, and the defect entropy $s_D$ tends to $-\infty$ in the infrared. In fact, we will argue that to leading order in $1/s$, correlation functions of $SO(3)$ invariant operators in the presence of the defect~\eqref{eq_ImpIntro} coincide with the ones in the presence of the pinning field defect.  This relation between the large $s$ limit of the spin impurity and the pinning field defect (which is a theory with no free parameters) will be especially useful in the interacting $O(3)$ model. 

It is tempting to conjecture that in 3d the DRG flow never terminates in a healthy DCFT in the IR also for $s=O(1)$. At present, we can only prove this in a $1/s$ expansion. 

The recent general results of \cite{Lauria:2020emq,Nishioka:2021uef} guarantee that free scalar theories in $d=3$ do not admit any nontrivial DCFT. This implies that the perturbative fixed points observed in the epsilon expansion cannot be extrapolated to $\varepsilon=1$ also for small values of $s$. This is consistent with the DRG runaway behavior we find at large $s$.\footnote{ In principle, it could be that the DRG for $s=O(1)$ in $d=3$ terminates in a decoupled line defect, such as one with $s_\text{IR}<s$. This is why the case of  $s=O(1)$ in $d=3$ is not yet entirely settled, however, given the results about large $s$ and fixed $d$ and the results about the double scaling limit, it is reasonable to expect that the DRG flow for $s=O(1)$ in $d=3$ indeed never terminates.} 
 Let us reiterate that we expect the runaway DRG behavior to be related to the existence of a moduli space of vacua in the bulk.

\paragraph{Interacting bulk} In sec.~\ref{sec_magnetic_defects_interacting_bulk} we consider the impurity~\eqref{eq_ImpIntro} with an interacting $O(3)$ Wilson-Fisher bulk theory with potential $\lambda(\phi_a^2)^2$.
For $d<4$, both the bulk and the defect couplings are relevant, so that the physical three-dimensional model is strongly coupled in the IR.

The simplest approach is to perform a perturbative analysis for small $\varepsilon=4-d$ for finite values of $s=O(1)$ (see e.g. \cite{sachdev1999quantum,vojta2000quantum,Sachdev:2001ky,Sachdev:2003yk}). In this limit it was found that, tuning the bulk to the critical point, the defect coupling admits a nontrivial IR stable fixed point for which $\gamma^2_*\sim\lambda_*\sim\varepsilon$. This fixed point is analogous to the one mentioned at the beginning of sec.~\ref{IntroMagneticImpGen} for the free theory with $O(3)$ symmetry.

We are interested in the phases of this impurity for arbitrary $s$, including $s\gg 1$.  As in the previous section, we should not trust the small $\varepsilon$ expansion and some resummation is required in order to understand the phase diagram.

The central questions we would like to address are whether the theory 
admits new fixed points beyond the one seen in perturbation theory and whether the large $s$ limit of the impurity in three spacetime dimensions can be understood.  A particularly important point is that, unlike the free theory, the bulk interacting theory does not have a moduli space of vacua due to the potential $(\phi_a^2)^2$. Therefore, one should not expect an instability and consequently we do expect a healthy DCFT in any $3\leq d<4$ for any $s$.\footnote{One can hope that there exist rigorous lower bounds on $s_D$ in $d>2$ theories with no moduli space of vacua. See~\cite{Friedan:2012jk,Collier:2021ngi} for results in $d=2$.}

Our main results are the following:
\begin{itemize}
\item The model can be studied for all $s$ as long as we have $d=4-\varepsilon$ with $\varepsilon\ll 1$. For fixed small $s$ this can be accomplished using a standard perturbative analysis, while for $s 
\gtrsim \varepsilon^{-2}$ a resummation is required. We are able to achieve this resummation and obtain results that are trustworthy for all $s$ using a new semiclassical limit, which allows to reorganize the perturbative series and to make non-perturbative statements at large $s$. (In particular, in this semiclassical limit, various terms in the beta function are obtained from the solution of a \emph{classical} differential equation.) 

\item There is a unique nontrivial zero of the beta function for all values of $s$,  describing an IR stable fixed point.  A major simplification occurs for $s\to \infty$ for all $d$, including both $d\to 4$ and also for $d=3$ which is the most interesting case experimentally. 
The prediction is that for $s\to \infty$ the theory breaks up into a weakly-decoupled sector of fluctuations with target space $S^2$ and a special DCFT that was studied before~\cite{Allais:2014fqa,2014arXiv1412.3449A,2017PhRvB..95a4401P,Cuomo:2021kfm} called hereafter the pinning field DCFT. 
 
\item We are able to verify this prediction for the large $s$ limit of the spin impurity within the $\varepsilon$ expansion. Additionally, we present the consequences of this prediction for the physically interesting case $d=3$, including a determination of the scaling dimensions of certain operators, as well as some other observables. These predictions for $d=3$ should be in principle testable. 

\item Finally, the nearly-decoupled sector of fluctuations with target space $S^2$ and the pinning field DCFT do couple to each other at finite large $s$, leading to some $1/s$ corrections to observables. We determine the leading coupling and use it to compute the anomalous dimension of the spin operator on the defect.

 \end{itemize}

\begin{figure}[t]
\centering
\includegraphics[scale=0.45]{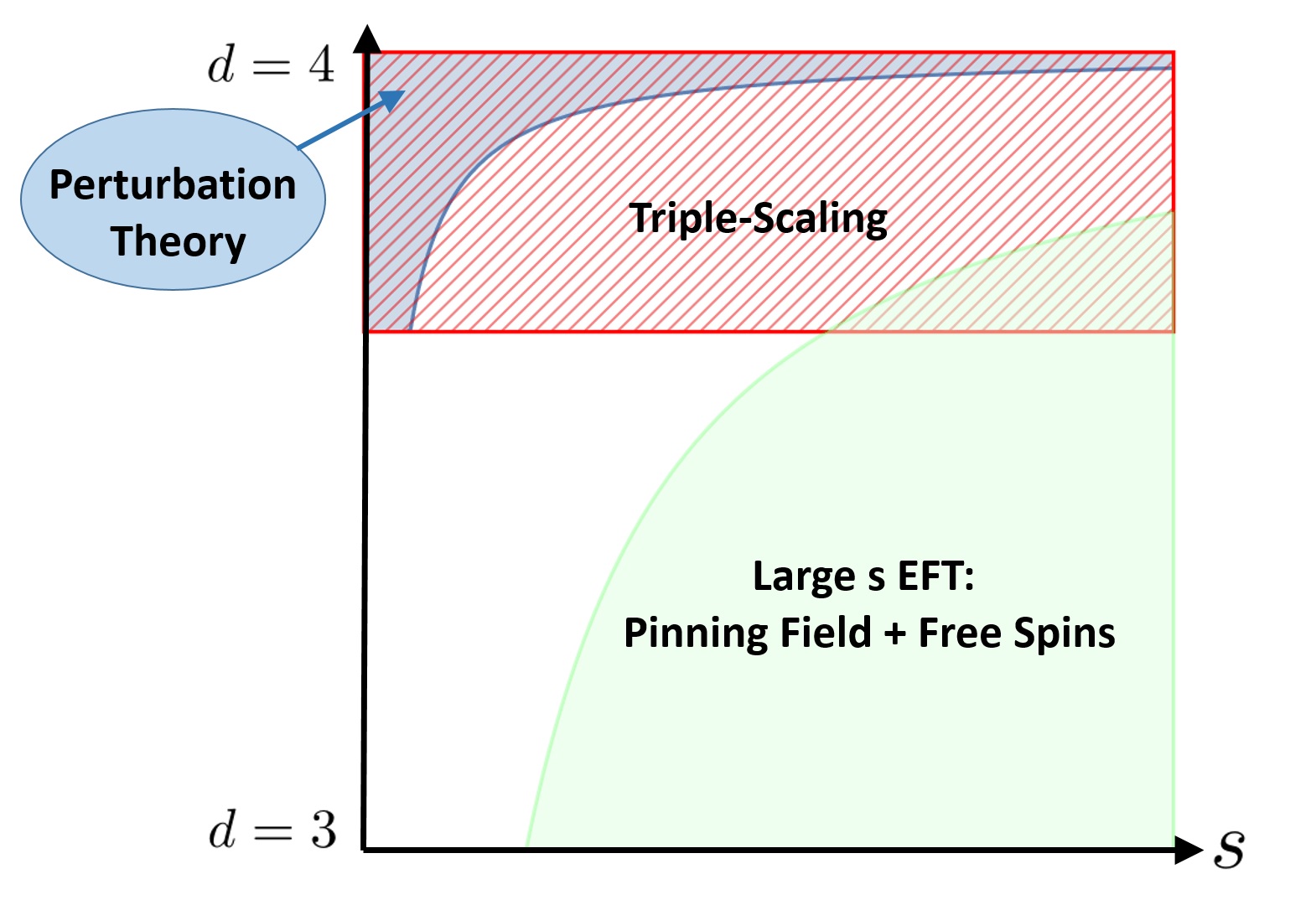}
\caption{Regimes of applicability of different methods to capture the nontrivial DCFT fixed point of the spin impurity~\eqref{eq_ImpIntro} in the $O(3)$ WF model. We expect that this DCFT exists for all $d$ and $s$.  The blue region corresponds to the standard $\ep$ expansion, the red hatched region to the new semiclassical resummation method, and the green region to an effective field theory (EFT) involving the pinning field DCFT weakly coupled to a first-order $S^2$ sigma model.   As we will explain in sec.~\ref{subsec_int_fix}, the resummed perturbation approach theory is applicable for arbitrary $s$ at small $\varepsilon$, while the pinning field effective description holds for sufficiently large $\varepsilon s^2$ (including $\varepsilon\sim O(1)$, see sec.~\ref{subsec_int_anyD} for details).}
\label{fig:IntPD}
\end{figure}

In fig.~\ref{fig:IntPD} we summarize the validity regimes of the various approaches, namely standard perturbation theory, the resummed $\varepsilon$ expansion that we introduce, and the \emph{effective} description that we propose in terms of the pinning field defect and a weakly coupled sector. As fig.~\ref{fig:IntPD} clearly shows, there are overlapping regimes between the different approaches. As a nontrivial benchmark of our ideas, we will verify explicitly the agreement between the different approaches in these regions.

\subsection{Wilson Lines in Large Representations} 

The line defect~\eqref{eq_ImpIntro} representing an impurity is remarkably similar to the familiar presentation of Wilson lines in gauge theories.  It is therefore natural to wonder if ideas analogous to those discussed in the previous section can be applied to Wilson lines in large representations of the gauge group.  In this paper we analyze in detail the case of  some protected observables of $1/2$-BPS Wilson lines in $\mathcal{N}=2$ superconformal field theories (SCFTs) in four dimensions. For concreteness, we focus on rank-1 theories.\footnote{For a complete classification of rank-1 $\mathcal{N}=2$ SCFTs see \cite{Argyres:2015ffa,Argyres:2015gha,Argyres:2016xmc,Argyres:2016xua}.}

$1/2$-BPS Wilson loops in $\mathcal{N}\geq 2$ superconformal field theories (SCFTs) are among the most studied examples of DCFTs in the literature \cite{Maldacena:1998im,Rey:1998ik,Erickson:2000af, Drukker:2000rr,Zarembo:2002an,Gomis:2006sb,Pestun:2007rz,Gaiotto:2010be,Correa:2012at,Cordova:2013bza,Lewkowycz:2013laa,Fucito:2015ofa,Fiol:2015spa,Cordova:2016uwk,Giombi:2018qox,Bianchi:2018zpb,Bianchi:2019dlw,Galvagno:2021qyq} (for a general approach to supersymmetric line defects in diverse dimensions see~\cite{Agmon:2020pde,Giombi:2021zfb,Penati:2021tfj}).  Notice that, in general, the large representation limit is of interest also for the study of non-Lagrangian theories, in which case superconformal defects are roughly labeled by the electric and magnetic charges of their IR representative in the Coulomb branch of the theory \cite{Gaiotto:2010be,Cordova:2013bza,Cordova:2016uwk}.  Importantly for us, localization techniques allow for exact predictions for certain supersymmetric observables \cite{Pestun:2007rz,Hosomichi:2016flq}.

Analogously to the large $s$ limit of the impurity in the free $O(3)$ model, the analysis of Wilson lines in large representations is carried out by identifying a new saddle-point.  As a reminder, in Lagrangian theories, the 1/2-BPS loops includes both the gauge and scalar components of the vector multiplet: 
\begin{equation}\label{intro_WL_susyprime}
D^\text{BPS}_s=\text{Tr}_{2s+1}\left[P\exp\left( \int_{\mathcal{C}}dt( i\dot x^\mu  A_\mu +|\dot x|\Phi)\right)\right]\,.
\end{equation}
Therefore, for $s\gg 1$, we expect that the classical trajectory dominating the path integral is characterized by a large scalar profile $\Phi\sim s$ and by a large Coulomb potential $A\sim s$. 
Localization arguments show that 
protected DCFT observables can be computed in a $1/s$ expansion through the effective field theory on the Coulomb branch. We consider in particular the $g$-function of the theory and the coefficient $h_D$ of the one-point function of the stress-tensor.  The result depends on a Wilson coefficient $g_\text{CB}$, identified with the IR gauge coupling, and on the difference in the ``$a$''-conformal anomaly between the CFT and the Coulomb branch contribution as quoted in~\eqref{gintro}. We additionally find a general relation between $h_D$ and $\log g$.

\subsection{Structure of the paper}

The rest of this paper is organized as follows. In section~\ref{Sec_FreeTheory} we study the spin impurity theory in the free bulk case in the large $s$ regime. In section~\ref{sec_magnetic_defects_interacting_bulk} we study the impurity theory in the interacting $O(3)$ Wilson-Fisher bulk case in the large $s$ regime.  In section~\ref{Sec_WL} we study the large representation limit of $1/2$-BPS Wilson lines in rank-1, $\mathcal{N}=2$ superconformal field theories. Section~\ref{Sec_WL} can be read without reading sections~\ref{Sec_FreeTheory}  and~\ref{sec_magnetic_defects_interacting_bulk}. In appendix~\ref{App_Details_Diagrammatics}, technical details associated with the diagrammatic calculations of section~\ref{Sec_FreeTheory} are given. Appendix~\ref{App_Details_Semiclassics} contains technical details related to the semiclassical calculations of section~\ref{Sec_FreeTheory}. In appendix~\ref{App_ClassicalBeta} we obtain the four-dimensional beta function for the defect coupling in the interacting bulk theory studied in section~\ref{sec_magnetic_defects_interacting_bulk} from the classical saddle-point equations.

\paragraph{Note added:} while we were completing this work, we were informed of the upcoming papers~\cite{Beccaria:2022bcr} and \cite{Nahum:2022fqw}, whose results overlap with part of our section~\ref{Sec_FreeTheory}. In particular,~\cite{Beccaria:2022bcr} and \cite{Nahum:2022fqw} also analyze a model equivalent to the spin $s$ impurity in free theory in the double-scaling limit $\gamma_0\rightarrow 0,\,s\rightarrow\infty$ with $\gamma_0^2 s=\text{fixed}$.  We are grateful to the authors of both works for sharing with us a preliminary version of their draft. 

\section{Spin defects at large \texorpdfstring{$s$}{s} in free theory}\label{Sec_FreeTheory}

\label{sec_magnetic_defects_free_bulk}
\subsection{Setup}
\label{subsec_setup_free_bulk}  

In this section we consider a free, massless $O(3)$-symmetric scalar field theory in $d$ spacetime dimensions. The bulk action is given by the free, massless action in flat space:
\begin{equation}\label{eq_free_bulk}
S_{\text{bulk}} = \frac{1}{2}\int d^dx \left(\partial\phi_a \right)^2,
\end{equation}
where $a=\{1,2,3\}$, $\left(\partial\phi_a\right)^2 = \partial_{\mu}\phi_a \partial^\mu \phi_a$, and $\mu$ stands for spacetime indices, $\mu = 1,\cdots, d$. As was explained in the introduction, we couple the free bulk theory to a line defect, physically representing an impurity in the spin $s$ representation of the bulk global $SO(3)$ symmetry. This is achieved by adding the following line operator to the partition function:
\begin{equation}\label{eq_free_Defect}
D_s=\text{Tr}_{2s+1}\left[P\exp\left( \gamma_0\int d\tau\,  \phi\right)\right]\,,
\end{equation}
where $\gamma_0$ is the bare coupling, and, as was explained in sec.~\ref{IntroMagneticImpGen}, $\phi=\phi_aT^a$ and $\{T^a\}$ are the $2s+1$ dimensional representation of the $su(2)$ algebra. The defect worldline is parametrized by the embedding $x^\mu=x^\mu(\tau)$, where $\tau$ is a normalized affine parameter such that $|dx/d\tau|=1$.

We will be interested both in circular and linear defects throughout this section.  An equivalent representation of the defect in eq.~\eqref{eq_free_Defect} can be given in terms of a bosonic $su(2)$ spinor $z=\{z_1,z_2\}$ on the line, subject to the constraint $\bar{z} z=2s$. In this formulation, the total action (bulk and defect) of the defect quantum field theory (DQFT) reads (see e.g. \cite{sachdev_2011})\footnote{Similar actions to~\eqref{eq_free_DCFT0} have been studied in several different contexts, see e.g. \cite{Lieb:1973vg,Rabinovici:1984mj,Clark_1997}. In particular, the constrained spinor $z$ is equivalent to a charged particle moving on a sphere with a charge $s$ monopole at the center in the lowest Landau level  \cite{WU1976365,Dunne:1989hv,Hasebe:2010vp}. } 
\begin{equation}\label{eq_free_DCFT0}
S=\frac{1}{2}\int d^dx(\pd\phi_a)^2+\int_D d\tau\left[\bz\dot z-\gamma_0\bz\frac{\sigma^a}{2}z\,\phi_a\right]\,,\qquad
\bz z=2s\, .
\end{equation}
 The action~\eqref{eq_free_DCFT0} is invariant under $U(1)$ gauge transformations $z\rightarrow e^{i\alpha(\tau)}z$, which makes the target space into the two sphere.  The canonical commutation relations imply that the operators
  $\{S^a=\bz \frac{\sigma^a}{2}z\}$ satisfy the $su(2)$ algebra $[S^a,S^b]=i\varepsilon^{abc} S^c$, while the constraint implies $S^a S^a=s(s+1)$,  so that the worldline Hilbert space indeed corresponds to that of a spin $s$ representation of $su(2)$. Since the kinetic term of $z$ is first order in derivative, eq.~\eqref{eq_free_Defect} is just the evolution operator of the worldline variable. This proves the equivalence between the action~\eqref{eq_free_DCFT0} and the expectation value of the operator~\eqref{eq_free_Defect}.

For future purposes, it is important to comment on the ordering in the definition of $S^a$.  The variables $\bar{z}$ and $z$ form a canonical pair, and therefore do not commute.  It turns out that the correct definition of the composite spin operator is obtained via the following point-splitting procedure (see e.g.  \cite{negele2018quantum,Clark_1997,sachdev_2011} for similar discussions):
\begin{equation}\label{eq_def_z_ordering}
S^a(\tau)=\left(\bz\frac{\sigma^a}{2}z\right)(\tau)\equiv\lim_{\eta\rightarrow 0^+}
\bz(\tau+\eta)\frac{\sigma^a}{2}z(\tau)\,.
\end{equation}
In terms of the path integral, such a definition prevents any issues with singularities at coincident points. The ordering in eq.~\eqref{eq_def_z_ordering} ensures that $\langle S^a(\tau) S^a(\tau')\rangle=s(s+1)$ for $\gamma_0=0$, as required.\footnote{To verify this assertion, it is important to use the constraint in the form $(\bz z)(\tau)=\lim_{\eta\rightarrow 0^+}\bz(\tau+\eta)z(\tau)=2s$.}

As was mentioned in the introduction, the coupling $\gamma_0$ is relevant for $d<4$,  and we will see that it is marginally irrelevant for $d=4$.  We are implicitly fine tuning to zero a defect cosmological constant term  $\sim M\int_D d\tau$ in the action~\eqref{eq_free_DCFT0}. Note also that there are no wavefunctions renormalizations for the fields in the action~\eqref{eq_free_DCFT0}. For $\phi$ this is obvious since it is a free field. For $z$ this is because a nontrivial wavefunction renormalization factor would contradict the $U(1)$ gauge invariance. To see this it is convenient to promote the sliding scale to a spurionic function of the defect coordinate, $M=M(\tau)$, which transforms trivially under the action of the gauge group. Since the kinetic term for $z$ is only invariant up to a total derivative under $U(1)$ gauge transformations, the coefficient of $\bar{z}\dot{z}$ cannot depend on the sliding scale $M(\tau)$, and it is therefore not renormalized at the quantum level.\footnote{This result is similar to the nonrenormalization of the Chern-Simons term \cite{Coleman:1985zi}.}

Our main findings which arise from the analysis presented in this section, including the phase diagram of the model, were already summarized in the introduction part of this paper (see sec.~\ref{IntroMagneticImpGen}).
In addition, we would like to comment that the results of this section will also prove useful at a technical level, as a warm-up for the analysis of a large spin impurity in the interacting $O(3)$ Wilson-Fisher fixed point. 

This section is organized as follows. In section~\ref{sec_free_diag} we review the diagrammatic analysis of the theory in $d=4-\varepsilon$. 
The results will be useful later on when matching between results obtained by the semiclassical analysis with those obtained using standard perturbative techniques in their overlapping regime of validity at weak coupling. In section~\ref{sec_free_semiclassics} we study the model in the large $s$ limit.  We will calculate various quantities, including the defect $g$-function and, for $\varepsilon\ll 1$, the $\beta$ function associated with the defect coupling.

\subsection{Diagrammatic results}\label{sec_free_diag}

\subsubsection{Perturbation theory and the beta function}

In this section we review the standard diagrammatic approach by considering the calculation of the one-point function of the operator $\phi_a^2$ in the presence of a straight line defect.  We focus on the regime of small $\varepsilon=4-d$ where the coupling is only weakly relevant and the full flow can be studied perturbatively. This will also allow us to extract the beta function of the renormalized coupling $\gamma$.\footnote{As usual, the coupling is renormalized according to:
\begin{equation}\label{eq_ren_gamma}
\gamma_0^2=M^{\varepsilon}\left[\gamma^2+\frac{\delta\gamma^2}{\varepsilon}+\frac{\delta_2\gamma^2}{\varepsilon^2}+\ldots\right]\,,
\end{equation}
where $M$ is the sliding scale and $\gamma$ is the renormalized coupling constant. We work in the minimal subtraction scheme (MS) to one-loop order, for which only $\delta\gamma^2$ is non-vanishing, and the beta function can be extracted 
 by requiring that $\gamma_0^2$ is independent of the sliding scale.  This yields \cite{Weinberg:1996kr}:
\begin{equation}\label{eq_general_beta}
\beta_{\gamma^2}=-\varepsilon\gamma^2+\gamma^2\frac{d\delta\gamma^2}{d\gamma^2}-\delta\gamma^2  \, .
\end{equation}
\label{footnote_convention}}

\begin{figure}[t]
   \centering
		{\includegraphics[width=0.3\textwidth]{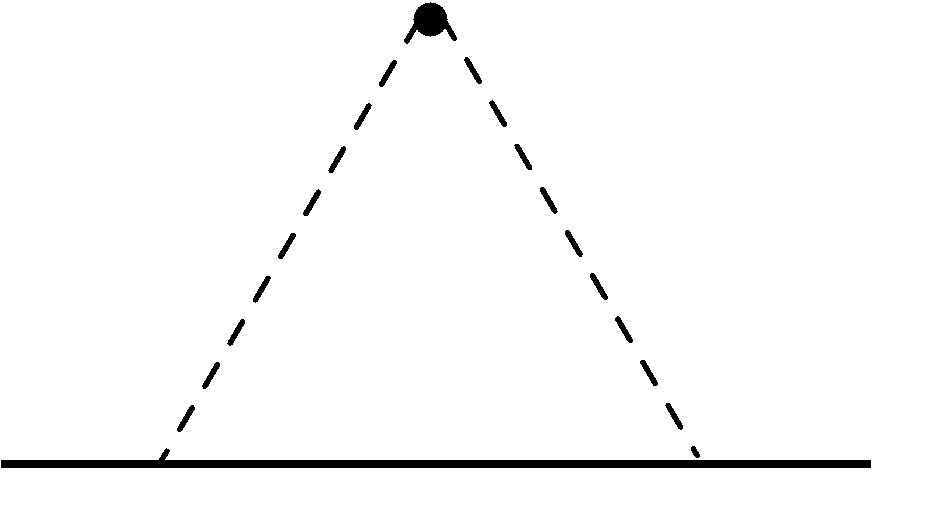}}
        \caption{ Diagram that contributes the leading order term in the one-point function $\langle\phi_a^2(x)\rangle$. Dashed lines represent bulk scalar field propagators.  The solid line represents the defect.  }
\label{fig:Diagram1point}
\end{figure}
To calculate the one-point function of $\phi_a^2$, it is convenient to work in coordinates $x^\mu=(\mathbf{x},\tau)$ with the defect located at $x^i=0$, where $i=1,\cdots,d-1$.  The propagator of the free scalar is given by 
\begin{equation}\label{eq_free_propagator}
\langle\phi_a\left(x\right)\phi_b\left(0\right)\rangle_{\gamma_0=0}=\delta_{ab} G(x)\,,\qquad
G(x)\equiv\frac{1}{(d-2)\Omega_{d-1}}\,\frac{1}{\left(x^2\right)^{\frac{d-2}{2}}}\,,
\end{equation}
where $\Omega_{d-1}=\frac{2\pi^{d/2}}{\Gamma(d/2)}$ is the volume of the $d-1$-dimensional sphere.  Working with the representation~\eqref{eq_free_Defect} of the defect, the leading contribution to the one-point correlation function $\langle\phi_a^2(x)\rangle$ arises at order $\gamma_0^2$ from the diagram in fig.~\ref{fig:Diagram1point}. It reads:\footnote{Here and in the following all correlation functions are normalized by the expectation value of the unit operator in the presence of a straight line defect. The expectation value of the unit operator  is $2s+1$ up to order $\gamma_0^4$ ( this is because the $O(\gamma_0^2)$ correction vanishes in dimensional regularization), which is all we shall use for the results in the main text.}
\begin{equation}\label{phi2_diag}
\begin{split}
\langle\phi_a^2(\mathbf{x},0) \rangle &\simeq\gamma_0^2\frac{\text{Tr}[T^a T^a]}{2s+1}
\left[\int d\tau \,G\left(x-x(\tau)\right)\right]^2
=\frac{\gamma_0^2 s(s+1)}{16\pi^{d-1} |\mathbf{x}|^{2 d-6}} \Gamma \left(\frac{d-3}{2}\right)^2
\\
& \overset{d=4}{=}\frac{\gamma_0^2s(s+1)}{16\pi^2 \mathbf{x}^2}\,,
\end{split}
\end{equation}
where we used $T^aT^a=s(s+1)$.

\begin{figure}[t]
   \centering
		\subcaptionbox{  \label{fig:NLO1}}
		{\includegraphics[width=0.22\textwidth]{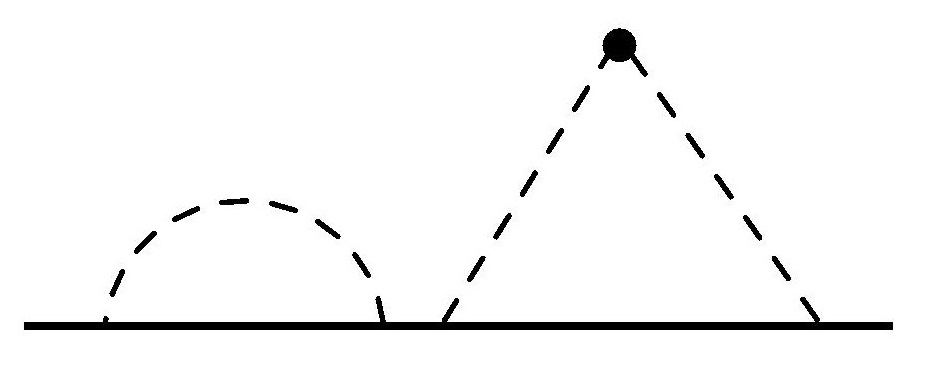}}
	\qquad \qquad 
		\subcaptionbox{ \label{fig:NLO2}}
		{\includegraphics[width=0.22\textwidth]{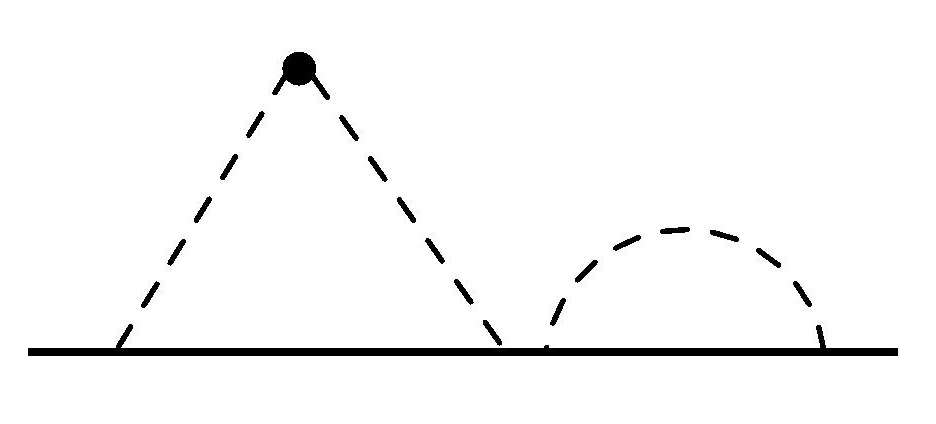}} \qquad \qquad
				\subcaptionbox{ \label{fig:NLO3}}
		{\includegraphics[width=0.22\textwidth]{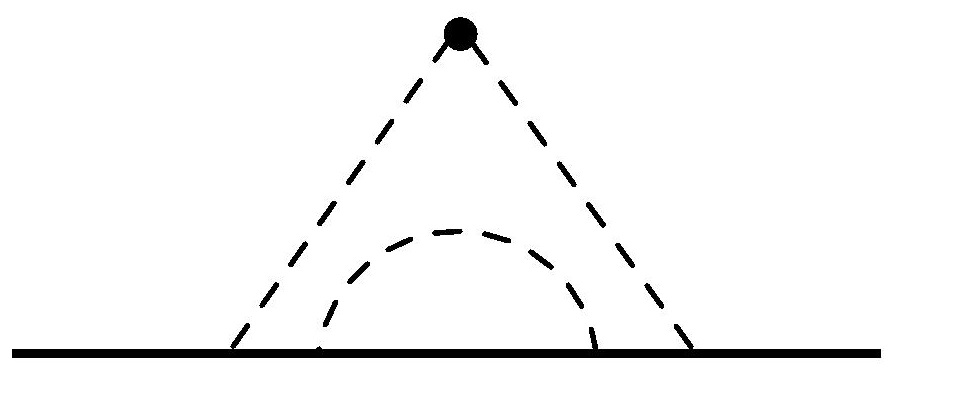}} \\
						\subcaptionbox{ \label{fig:NLO4}}
		{\includegraphics[width=0.22\textwidth]{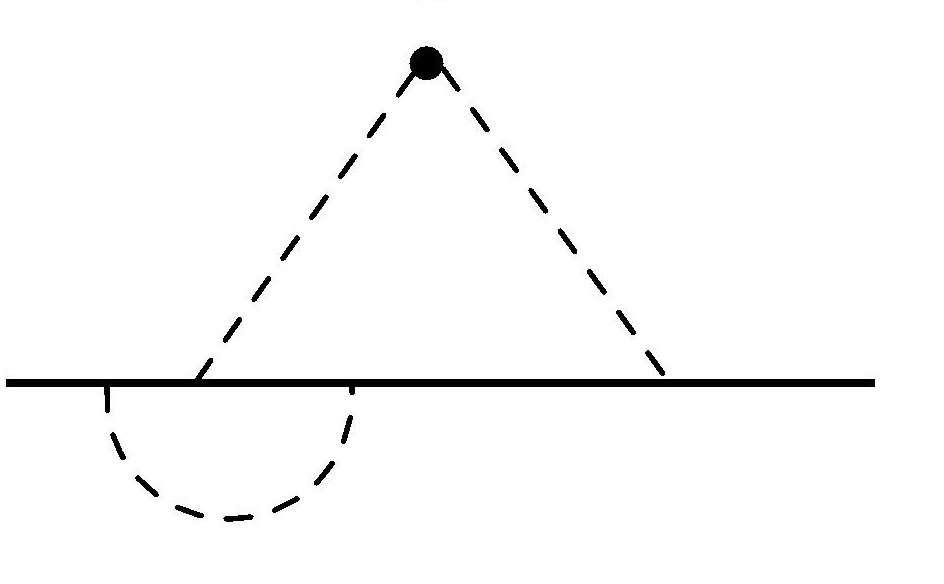}} \qquad \qquad 
						\subcaptionbox{ \label{fig:NLO5}}
		{\includegraphics[width=0.22\textwidth]{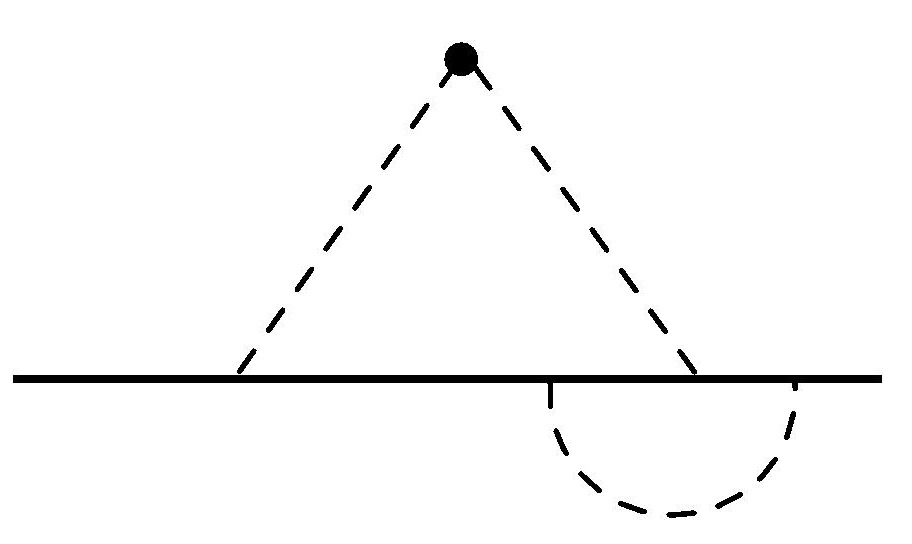}} \qquad \qquad 
						\subcaptionbox{ \label{fig:NLO6}}
		{\includegraphics[width=0.22\textwidth]{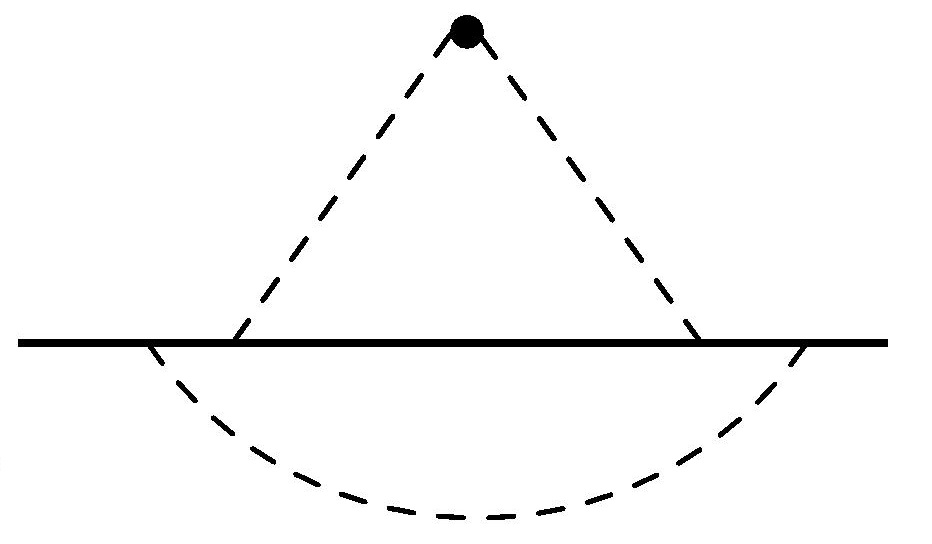}}
        \caption{Diagrams contributing the next to leading order terms in the one-point function $\langle\phi_a^2(x)\rangle$.   }
\label{fig:DiagramsNLO}
\end{figure}

The first correction to the tree-level result~\eqref{phi2_diag} in four dimensions is given by the diagrams displayed in figure~\ref{fig:DiagramsNLO}. After some matrix algebra only the diagrams in figs.~\ref{fig:NLO4} and~\ref{fig:NLO5} remain, giving:
\begin{equation}\label{phi2_diag2}
\begin{split}
(\text{1-loop})&=-4\gamma_0^4 s(s+1) \int\limits_{\tau_1>\tau_2>\tau_2>\tau_4} d^4[\tau] \,
G\left(x-x(\tau_1)\right) G\left(x(\tau_2)-x(\tau_4)\right)G\left(x-x(\tau_3)\right)
\\
&=-\frac{1}{\varepsilon}\frac{\gamma_0^4 s (s+1)}{32 \pi ^4 \mathbf{x}^2}
+O\left(\varepsilon^0\right)
\,.
\end{split}
\end{equation}
The evaluation of the integral in eq.~\eqref{phi2_diag2} is detailed in appendix~\ref{app:EasyInt}. 

Requiring that the one-point function $\langle\phi_a^2(\mathbf{x},0) \rangle$  is finite for $\varepsilon\rightarrow 0$ we obtain the counterterm $\delta\gamma^2=\gamma^4/(2\pi^2)$ and the one-loop beta function for the physical coupling $\gamma^2$ (see footnote~\ref{footnote_convention} for our conventions)
\begin{equation}\label{eq_Sachdev_ct}
\beta_{\gamma^2}=-\varepsilon\gamma^2+\frac{\gamma^4}{2\pi^2}+O\left(\gamma^6\right)\,.
\end{equation}
Eq.~\eqref{eq_Sachdev_ct} is in agreement with previous studies in the literature \cite{PhysRevB.61.4041,Sachdev:2001ky}, and it implies the existence of a perturbative IR stable fixed point at
\begin{equation} \label{eq:FixedPointPerFree}
\gamma^2_*=2\pi^2\varepsilon+O\left(\varepsilon^2\right)\,.
\end{equation}
Note that the coupling $\gamma$ is marginally irrelevant in four dimensions so the free DCFT with decoupled $2s+1$ states is attractive in $d=4$. For $0< \varepsilon\ll 1$ sufficiently small, for any fixed $s$, we get a nontrivial infrared DCFT.

We also report the result for the one-point function to one-loop order:
\begin{equation}\label{eq_phi2_diag3pre}
\begin{split}
\langle\phi_a^2(\mathbf{x},0)\rangle =&\frac{\mathcal{N}_d}{\mathbf{x}^{d-2}}
\frac{\gamma^2 s(s+1) }{4\sqrt{6}}\left\{1+\varepsilon  \left[\log (4 M |\mathbf{x}|)+\frac{1}{2} (\gamma_E +\log \pi )\right]\right. \\[0.7em]
&\left. -\gamma^2\frac{2 \log (M |\mathbf{x}|)+\gamma _E+2+\log 4 \pi }{4 \pi ^2}
+O\left(\gamma^4,\gamma^2\varepsilon,\varepsilon^2\right)\right\}\,,
\end{split}
\end{equation}
where $\gamma_E$ is the Euler constant and $\mathcal{N}_d$ is the normalization of the bulk two point-function without the defect:
\begin{equation}\label{eq_Nd}
\langle  \phi_a^2(x)\phi_a^2(0)\rangle=\frac{\mathcal{N}_d^{\,2}}{x^{2d-4}}\,,\qquad\mathcal{N}_d=\frac{\sqrt{6}}{(d-2)\Omega_{d-1}}\,.
\end{equation}
At the infrared fixed point this implies the result:
\begin{equation}\label{eq_phi2_diag3}
\langle\phi_a^2(\mathbf{x},0)\rangle=\frac{\mathcal{N}_d}{\mathbf{x}^{d-2}}
\frac{\pi^2 s(s+1) \varepsilon}{2\sqrt{6}}\left[1-\varepsilon\left(1-\log 2\right)+O\left(\varepsilon^2\right)\right]\,.
\end{equation}

We will later see that the various higher order corrections that we have neglected in~\eqref{eq_phi2_diag3pre} and~\eqref{eq_phi2_diag3} are enhanced by powers of $s$ for $s\gg 1$ and can become important if $\varepsilon$ is not the smallest parameter in the problem.

Further perturbative results involving this defect, including the two-loop beta function and various thermal susceptibilities, can be found in \cite{PhysRevB.61.4041,vojta2000quantum}. In particular in \cite{vojta2000quantum} it was argued diagrammatically that the defect spin operator $S^a$ has exactly scaling dimension $\Delta_S=\varepsilon/2$, without further corrections.  Let us briefly comment on an alternative proof of this fact which relies on the representation~\eqref{eq_free_DCFT0} of the defect. To this aim, notice that the bulk theory, besides the $su(2)$ currents, has three dimension $d/2$ conserved currents $J^{\mu\,a}_\text{shift}=-\pd^\mu\phi_a$ associated with the invariance under shifts of the scalars. This symmetry is explicitly broken at the defect. Indeed from eq.~\eqref{eq_free_DCFT0} we see that the bulk Ward identity is modified to
\begin{equation}\label{eq_tilt_operator}
\pd_\mu J^{\mu\,a}_\text{shift}=\gamma_0 S^a\delta^{d-1}_D\,,
\end{equation}
where $\delta^{d-1}_D$ is a delta function localized at the defect.  
Since the bulk current has protected dimension, the  Ward identity~\eqref{eq_tilt_operator} implies that at the fixed point the scaling dimension of $S^a$ is $\Delta_S=\varepsilon/2$.\footnote{Defect operators which appear in Ward identities for internal symmetries broken by the defect are sometimes called \emph{tilt} operators. The arguments about tilt operators having protected dimension date back to \cite{Bray_1977}, a modern treatment is given in \cite{Cuomo:2021cnb,Padayasi:2021sik}  (see also \cite{Herzog:2017xha}).}

\subsubsection{The \texorpdfstring{$g$}{g}-function and the breakdown of perturbation theory at large \texorpdfstring{$s$}{s}}\label{subsec_free_g}

We denote by $g_\gamma$ the defect g-function, defined according to the conventions in \cite{Cuomo:2021rkm}, as the partition function in the presence of the defect on a circle of radius $R$ normalized by the partition function without it:
\begin{equation}\label{eq_defect_gfunction}
\log g_\gamma \equiv \log Z^{\text{bulk}+{\text{defect}}}-\log Z^{\text{bulk}}, 
\end{equation}
where $ Z^{\text{bulk}+{\text{defect}}}$ refers to the  partition function of the full theory including the defect~\eqref{eq_free_Defect}, and $Z^{\text{bulk}}$ refers to the partition function of the bulk theory alone.  
Note that for $\gamma=0$ the defect is completely decoupled and $g_0=2s+1$ regardless of the radius of the circle (in a scheme where a cosmological constant term is absent).

In this subsection we compute the defect $g$-function for a circular defect of radius $R$ diagrammatically by expanding eq.~\eqref{eq_free_Defect} in terms of the bare coupling constant $\gamma_0$.  We will use this calculation to illustrate the structure of the diagrammatic expansion at large $s$.  The discussion in this section will be largely analogous to the one in \cite{Badel:2019oxl}, where similar properties were observed in the study of the multi-legged amplitudes  associated with large charge operators in the $O(2)$ Wilson Fisher point in $4-\varepsilon$ dimensions.\footnote{The breakdown of perturbation theory for multilegged amplitudes, and its relation to semiclassics, was first analyzed in the context of multi-particle production - see e.g. \cite{Rubakov:1995hq,Son:1995wz}.}
We will also use the result to provide an explicit check of the $g$-theorem recently proven in \cite{Cuomo:2021rkm}.

\begin{figure}[t]
   \centering
		\subcaptionbox{  \label{fig:diagram1}}
		{\includegraphics[width=0.22\textwidth]{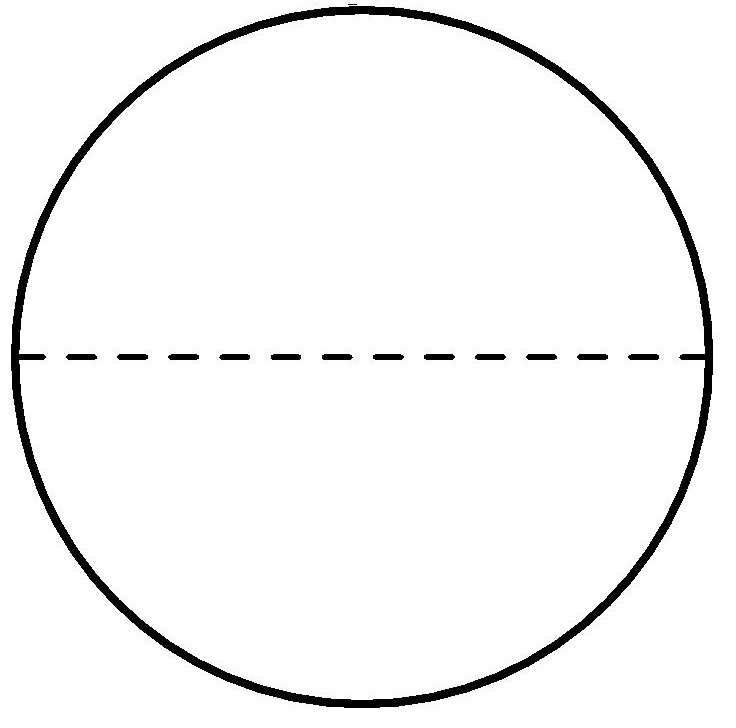}}
	\qquad \qquad \qquad \qquad 
		\subcaptionbox{ \label{fig:diagram2}}
		{\includegraphics[width=0.22\textwidth]{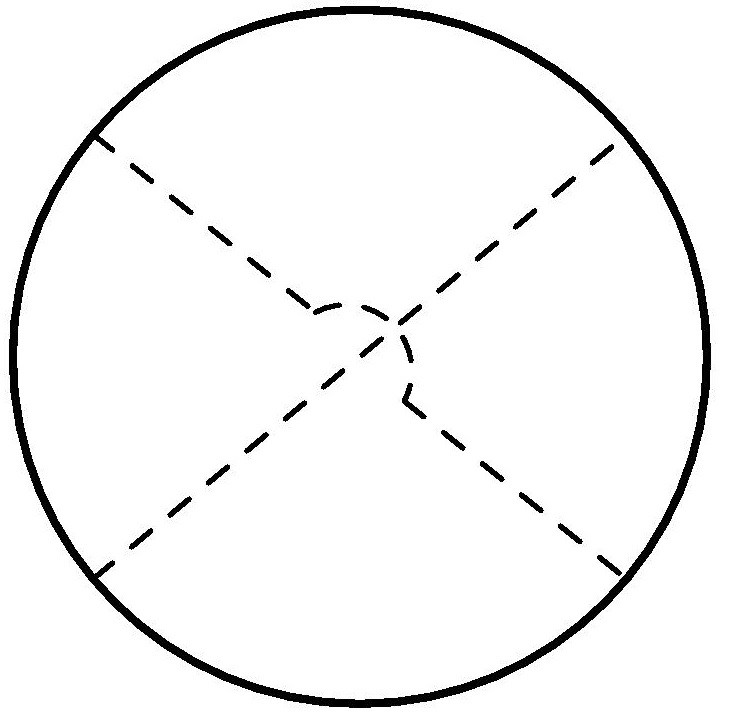}}
        \caption{Sample diagrams contributing to the defect $g$-function.   }
\label{fig:DiagramsGFuncLO}
\end{figure}

We can compute the defect partition function expanding the exponential in eq.~\eqref{eq_free_Defect}:
\begin{equation}\label{eq_g_expansion}
\begin{split}
g_{\gamma}/g_{0}&=1+\frac{\gamma_0^2}{2}\frac{\text{Tr}\left[T^a T^b\right]}{2s+1}\int_{\mathcal{C}} d^2[\tau] 
\langle P\left[\phi_a\left(x(\tau_1)\right)\phi_b\left(x(\tau_2)\right)\right]\rangle_{\gamma_0=0} \\
&
+\frac{\gamma_0^4}{4!}\frac{\text{Tr}\left[T^a T^b T^c T^d\right]}{2s+1}\int_{\mathcal{C}} d^4[\tau]  
\langle P\left[\phi_a\left(x(\tau_1)\right)\phi_b\left(x(\tau_2)\right)
\phi_c\left(x(\tau_3)\right)\phi_d\left(x(\tau_4)\right)\right]
\rangle_{\gamma_0=0}
\\
&+\ldots\,,
\end{split}
\end{equation}
where $P$ denotes the path-ordering and all the integrals are over the circle $\mathcal{C}$ parametrized by $\tau\in [0,2\pi R)$ via the embedding $x^\mu(\tau)=\{R \cos\left(\tau/R\right),
\,R\sin\left(\tau/R\right),0,\ldots\}$.\footnote{The bare coupling is related to the renormalized one as in eq.~\eqref{eq_ren_gamma}.} At order $\gamma_0^2$ we have a unique contraction, represented in the diagram in fig.~\ref{fig:diagram1}, while at order $\gamma_0^4$ the path-ordering allows for several inequivalent contractions, see the diagram in fig.~\ref{fig:diagram2} for an example.

We focus now on the regime $s\gg 1$.  We can evaluate the traces in the expansion~\eqref{eq_g_expansion} using $T^a T^a=s(s+1)$ and the commutator $[T^a,T^b]=i\varepsilon^{abc} T^c$. At each loop order $\ell$ we find contributions that range from $\gamma_0^{2\ell} s^{2\ell}$ down to $\gamma_0^{2\ell} s$. Every time we commute two matrices a suppression factor $1/s$ is brought about, as follows from the schematic scaling $T^a \sim s$. One might therefore conclude that perturbation theory breaks down when $s\gtrsim 1/\gamma_0$. This would be too quick however. A more careful analysis indeed shows that a remarkable exponentiation takes place.\footnote{For instance, it is easy to verify that the sum over diagrams of order $\gamma_0^{2\ell} s^{2\ell}$, which are simply obtained by neglecting the path-ordering (i.e. dropping all the commutators) exponentiates the first $\gamma_0^2 s^2$ contribution in fig~\ref{fig:diagram1}. A similar exponentiation property of the one-loop contribution to the $g$-function was observed in the study of supersymmetric Wilson loops in $\mathcal{N}=4$ SYM in \cite{Correa:2015kfa,Correa:2015wma}.} As a result, the logarithm of the $g$-function admits the following expansion in perturbation theory:
\begin{equation}\label{eq_g_diag_exp}
\log g_{\gamma}/g_0=s\sum_{\ell=1}\gamma_0^{2\ell} P_{\ell}(s)\,,
\end{equation}
where the $P_{\ell}(s)$ are polynomials of order $\ell$.  We have checked eq.~\eqref{eq_g_diag_exp} diagrammatically only up to two-loops, but in the next section we shall give a simple general argument which bypasses the intricate diagrammatic analysis. 

The structure of eq.~\eqref{eq_g_diag_exp} suggests the existence of a different expansion, directly in powers of $1/s$. Indeed by formally collecting all the leading order terms of the polynomials $P_{\ell}(s)$ in a new function $\tilde{f}_{-1}(\gamma_0^2 s)$, and similarly for the subleading powers, we recast the partition function in a double expansion as
\begin{equation}\label{eq_g_diag_exp2}
\log g_{\gamma}/g_0=\sum_{k=-1}s^{-k} \tilde{f}_{k}(\gamma_0^2 s)=s \tilde{f}_{-1}(\gamma_0^2 s)+\tilde{f}_0(\gamma_0^2 s)+\ldots\,.
\end{equation}
As eq.~\eqref{eq_g_diag_exp2} already suggests, we will show in the next section that the rewriting~\eqref{eq_g_diag_exp2} is associated with a different loop expansion, valid for $s\rightarrow \infty$ with $\gamma_0^2 s=\text{fixed}$. This will be obtained by expanding the path integral around a new non-trivial classical trajectory.

We now present the explicit diagrammatic computation of the $g$-function to order $O(\gamma_0^4)$: 
\begin{equation}\label{eq_gFreeDiagBare}
\log g_\gamma/g_0=
\frac{\pi\gamma_0^2 s(s+1)}{(d-2)\Omega_{d-1}R^{d-4}}I_1^{(d)}-
\frac{\gamma_0^4 s(s+1)}{\left[(d-2)\Omega_{d-1}R^{d-4}\right]^2}I_2^{(d)}+O\left(\gamma_0^6\right)\,,
\end{equation}
where we defined the following integrals 
\begin{align}\label{eq_diagram_int1}
I_1^{(d)}&=\int_0^{2\pi} d\phi\frac{1}{\left[4\sin^2\frac{\phi}{2}\right]^{\frac{d-2}{2}}}=-\frac{\pi  \sec \left(\frac{\pi  d}{2}\right) \Gamma \left(\frac{d}{2}-1\right)}{\Gamma (d-2) \Gamma \left(-\frac{d}{2}+2\right)}
=-\frac{\pi}{2}\varepsilon+O\left(\varepsilon^2\right)\,,\\ \label{eq_diagram_int2}
I_2^{(d)}&=
\int_0^{2\pi} d\phi_1\int^{\phi_1}_0 d\phi_2 
\int^{\phi_2}_0 d\phi_3 \int^{\phi_3}_0 d\phi_4 
\frac{1}{\left(16\sin^2\frac{\phi_{13}}{2} \sin^2\frac{\phi_{24}}{2}\right)^{\frac{d-2}{2}}}=-\frac{3\pi^2}{2}+O\left(\varepsilon\right)\,.
\end{align}
To obtain these expressions we used that the propagator of the scalar fields on the circle is given by:
\begin{equation}\label{eq_circle_prop}
\langle\phi_a\left(x(\tau)\right)\phi_b\left(x(0)\right)\rangle_{\gamma=0}=\frac{\delta_{ab}}{(d-2)\Omega_{d-1}}\,\frac{1}{\left[4R^2\sin^2\frac{\tau}{2 R}\right]^{\frac{d-2}{2}}}\,.
\end{equation}
The integral in eq.~\eqref{eq_diagram_int2} is computed in appendix~\ref{app:NotSoHardInt}.  Notice that the $O(\gamma_0^2)$ contribution in eq.~\eqref{eq_gFreeDiagBare} vanishes for $\varepsilon=0$ due to the vanishing of the integral~\eqref{eq_diagram_int1} in four dimensions. 
This is because in four dimensions $\gamma_0$ is a marginal parameter at the classical level, and the $g$-function cannot depend on marginal defect couplings \cite{Cuomo:2021rkm}.

Rewriting the answer in terms of the physical coupling $\gamma$ we obtain:\footnote{ Notice that the renormalizability of the defect action ensures that all terms proportional to inverse powers of $\varepsilon$ are canceled by the coupling counterterm.}
\begin{equation}\label{eq_gFree_diag}
\log g_\gamma/g_0=-\frac{\varepsilon s(s+1)}{8}\gamma^2+\frac{\gamma^4 s(s+1)}{32\pi^2}+O\left(\gamma^6,\varepsilon\gamma^4,\varepsilon^2\gamma^2\right)\,.
\end{equation}
The result~\eqref{eq_gFree_diag} depends on the radius $R$ through the beta function~\eqref{eq_Sachdev_ct} of the coupling $\gamma=\gamma\left( M R\right)$ (where $M$ is the sliding scale); in particular, evaluating the coupling at the scale $M=1/R$ resums the leading logarithimic corrections from higher orders in eq.~\eqref{eq_gFree_diag}. Specializing to the fixed point 
\eqref{eq:FixedPointPerFree} we find the $g$-function of the DCFT:
\begin{equation}\label{eq_gFree_fix}
\log (g_{\gamma_*}/g_0)=-\frac{\pi^2}{8} s (s+1) \varepsilon ^2+O\left(\varepsilon^3\right)\,.
\end{equation}
We will use the results~\eqref{eq_gFree_diag} and~\eqref{eq_gFree_fix} to verify the validity of the semiclassical approach that we present in the next section.

We end this section by using our results to test the $g$-theorem recently proven in \cite{Cuomo:2021rkm} (see also \cite{Affleck:1991tk,Friedan:2003yc,Beccaria:2017rbe,Kobayashi:2018lil}). To this aim, we consider the defect entropy $s_D$ defined as:\footnote{The definition of $s_D$ in eq.~\eqref{eq_sD_def} differs by a constant amount with respect to the definition~\eqref{eq:DefectEntropy} in the introduction due to the normalization factor $g_0$.}
\begin{equation}\label{eq_sD_def}
s_D=\left(1-R\frac{\pd}{\pd R}\right)\log g_{\gamma}/g_0\,.
\end{equation}
The differential operator cancels the contribution from a possible cosmological constant counterterm on the defect (which we have ignored thus far)  and ensures that $s_D$ is a scheme-independent observable.
The defect entropy is an important observable of the theory,  since it decreases monotonically under the defect renormalization group flow.
Using the Callan-Symanzik equation $(R\pd/\pd R+\beta_{\gamma^2}\pd/\pd \gamma^2)\log(g_\gamma/g_0)=0$, we see that $\log(g_\gamma/g_0)$ and $s_D$ in general coincide up to order $O(\gamma^6)$ corrections, and they are equal at the fixed points.\footnote{This is true only in mass independent schemes,  such as the one we are using, where no cosmological constant counterterm is generated.}
This implies:
\begin{equation}
g_{\gamma_*}\leq g_0\,,
\end{equation}
in agreement with eq.~\eqref{eq_gFree_fix}. Additionally, the defect entropy obeys the following gradient equation \cite{Cuomo:2021rkm}:
\begin{equation}\label{eq_gradient_formula}
M\frac{\pd s_D}{\pd M}=-\int_0^{2\pi R} d\tau_1 \int_0^{2\pi R} d\tau_2\,\langle T_D(\tau_1) T_D(\tau_2)\rangle\left[1-\cos\left(
\frac{\tau_1-\tau_2}{R}\right)\right]\,,
\end{equation}
where $T_D$ is the defect stress tensor.  We may verify this equation in perturbation theory using that $T_D=\beta_{\gamma} T^a\phi_a$, where $2\gamma\beta_{\gamma}=\beta_{\gamma^2}$. Evaluating the derivative on the left hand side of eq.~\eqref{eq_gradient_formula} using the Callan-Symanzik equation, the formula
\eqref{eq_gradient_formula} to the leading non-vanishing order is equivalent to the following equality
\begin{equation}\label{eq_gradient_check_diag}
\frac{\pd \log g_{\gamma}}{\pd\gamma^2}=\frac{\beta_{\gamma^2}}{4\gamma^2}\,\frac{\text{Tr}\left[T^a T^b\right]}{2s+1}\int_0^{2\pi R} d\tau_1 \int_0^{2\pi R} d\tau_2\,\langle\phi_a\left(x(\tau_1)\right) \phi_b\left(x(\tau_2)\right)\rangle \left[1-\cos\left(
\frac{\tau_1-\tau_2}{R}\right)\right]\,.
\end{equation}
This is easily verified using eq.~\eqref{eq_circle_prop} and the beta function~\eqref{eq_Sachdev_ct}.

\subsection{Semiclassics and the double-scaling limit}\label{sec_free_semiclassics}

\subsubsection{General considerations}\label{subsec_free_semiclassics_general}

In sec.~\ref{subsec_free_g} we showed that as the impurity spin $s$ becomes large, standard diagrammatic perturbation theory breaks down. An alternative framework should be used in order to address the physics at large $s$. For instance, in~\eqref{eq_gFree_diag} there could well be terms of order $\gamma^6s^3$ which would render our analysis invalid for $\varepsilon s\sim O\left(1\right)$.
Similarly, the analysis of the fixed point in~\eqref{eq_Sachdev_ct} would have to be revisited for $\varepsilon s\sim O\left(1\right)$ due to terms such as $\gamma^6 s$ which we have not yet computed. Physically this is associated with the fact that a large spin has a strong backreaction on the bulk, and thus the expansion around the trivial bulk background becomes inadequate.

We now introduce a different \emph{semiclassical} approach to study the theory in the large $s$ regime. 
That  a quasi-classical approach should exist is intuitively obvious, since   an impurity with large spin \emph{classically} sources a large response in the bulk order paramater $\phi_a\sim \gamma s$.  The proper classical profile therefore resums all the $s$-enhanced contributions,  allowing for a perturbative study of the theory. 

Concretely, consider the one-point function of the operator $\phi_a^2$. Rescaling the bulk fields in eq.~\eqref{eq_free_DCFT0} as $\phi_a\rightarrow \sqrt{s} \phi_a$ and $z\rightarrow\sqrt{s} \,z$,  one writes the corresponding path integral as\footnote{$\langle \phi_a^2(\mathbf{x},0)\rangle$ stands for the expectation value of the unscaled bulk field; we only do the field redefinition under the path integral.}$^,$\footnote{In the following,  to account for the trace in eq.~\eqref{eq_free_Defect}, periodic boundary conditions on $z$ are understood in the path integral: $z(\tau_i)=z(\tau_f)$.}
\begin{equation}
\langle \phi_a^2(\mathbf{x},0)\rangle=
\frac{\displaystyle \st\,\int\mathcal{D}\phi_a\mathcal{D}z\,
 \phi_a^2(\mathbf{x},0)
\exp\left[-
\st\, S_\text{rescaled}\right]
}{\displaystyle\int\mathcal{D}\phi_a\mathcal{D}z \exp\left[-\st\, S_\text{rescaled}\right]} \,,
\end{equation}
where we defined a rescaled action which depends only on $\gamma_0\sqrt{s}$:
\begin{equation}\label{eq_free_DCFT_after_rescaling}
S=\st\left[\frac{1}{2}\int d^dx(\pd\phi_a)^2+\int_D d\tau\left(\bz\dot z-
\gamma_0\sqrt{\st}\,
\bz\frac{\sigma^a}{2}z\phi_a\right) \right]\equiv \st\, S_\text{rescaled}\,, \qquad
\bar{z}z=2\,.
\end{equation}
It is clear from the above action that the model can be analyzed in a saddle-point expansion in the limit $s\rightarrow\infty$ by treating $\gamma_0 \sqrt{s}$ as a fixed parameter.  Therefore the correlator admits the following double expansion
\begin{equation}\label{eq_free_phi_expansion1}
\langle \phi_a^2(\mathbf{x},0)\rangle
=\frac{\mathcal{N}_d}{|\mathbf{x}|^{d-2}} \left[s \tilde h_{-1}(\gamma_0^2 s,|\mathbf{x}|,d)+\tilde h_0(\gamma_0^2 s,|\mathbf{x}|,d)+\ldots \right]\,,
\end{equation}
where for convenience we isolated the factor $\mathcal{N}_d$, defined in eq.~\eqref{eq_Nd}, in front of the right-hand-side.  From eq.~\eqref{eq_free_phi_expansion1} it is clear that $s^{-1}$ plays the role of the loop counting parameter similarly to $\hbar$. $\gamma^2_0 s$ instead is a fixed coupling, which near $4d$ is analogous to a 't Hooft coupling.  

The interpretation of~\eqref{eq_free_phi_expansion1} is slightly different between the case of $\varepsilon=4-d\ll 1$ and the case of finite $\varepsilon$. For $\varepsilon=4-d\ll 1$ there are logarithmic corrections (e.g. terms such as $\gamma_0^2s\log  |\mathbf{x}|$) that can be nicely accounted for using the power of the renormalization group. Therefore we will  switch to the physical coupling $\gamma$ and consider the double-scaling limit\footnote{This double-scaling limit is analogous to similar ones considered in the context of the large charge expansion \cite{Alvarez-Gaume:2019biu,Badel:2019oxl,Badel:2019khk,Antipin:2020abu,Jack:2021ypd}.}
\begin{equation}\label{eq_double_scaling_limit}
\gamma\rightarrow 0\,,\qquad s\to \infty, \qquad 
\gamma^2 s=\text{fixed}\,.
\end{equation} 
For $\varepsilon$ that is $O(1)$ (and in particular in $d=3$) one might similarly worry that terms such as $\gamma_0^2s|\mathbf{x}|^{4-d}$ become increasingly large in the infrared and would destroy the utility of the expansion~\eqref{eq_free_phi_expansion1}. We will see that this does not happen and no large enhancement occurs in the infrared. The large $s$ limit is fully analogous to the usual large $N$ limit in the $O(N)$ model. All large IR effects are consistently resummed by the saddle-point, and the full renormalization group flow can be studied perturbatively in a $1/s$ expansion as in~\eqref{eq_free_phi_expansion1}. Similar comments apply to other observables.

As promised, for $\varepsilon=4-d\ll 1$, we rewrite equation~\eqref{eq_free_phi_expansion1} using the physical coupling $\gamma$:
\begin{equation}\label{eq_free_phi_expansion}
\begin{split}
\langle \phi_a^2(\mathbf{x},0)\rangle
&=\frac{\mathcal{N}_d}{|\mathbf{x}|^{d-2}} \left[s h_{-1}(\gamma^2 s,|\mathbf{x}|M,\varepsilon)+h_0(\gamma^2 s,|\mathbf{x}|M,\varepsilon)+\ldots \right]\,,
\end{split}
\end{equation}
where $\gamma$ is defined so that 
$h_{-1}$ and $h_0$ are finite for $\varepsilon\rightarrow 0$. 

The beta function in the double-scaling limit takes the general form:
\begin{equation}\label{eq_beta_structure_double_scaling}
\beta_{\gamma^2}=\gamma^2\left[-\varepsilon+\beta_0^{(4d)}(\gamma^2 s)+\frac{1}{s} \beta_1^{(4d)}(\gamma^2 s)+O\left(\frac{1}{s^2}\right)\right]\,.
\end{equation}
In section~\ref{subsec_phi2_and_beta} we will find that 
\begin{equation}
\beta_0^{(4d)}(\gamma^2 s)=0\,.
\end{equation}
Furthermore, we will compute $\beta_1^{(4d)}(\gamma^2 s)$ (see eq.~\eqref{eq_BETA}). This will enable us to obtain the phase diagram of the theory summarized in sec.~\ref{IntroMagneticImpGen}.  At the fixed points the dependence on $|\mathbf{x}|$ in eq.~\eqref{eq_free_phi_expansion} of course drops out.

Analogous considerations about the existence of a double scaling limit apply to other observables of the theory.\footnote{It would also be interesting to analyze in a semiclassical expansion the fusion of two line defects \cite{Gadde:2016fbj,Isachenkov:2018pef}, maybe along the lines of earlier studies of OPE coefficients of large charge operators \cite{Cuomo:2021ygt}.  See \cite{Soderberg:2021kne,Rodriguez-Gomez:2022gbz} and references therein for some results on the fusion of defects in similar models.} 
We will consider in particular the $g$-function. Rescaling the fields as in~\eqref{eq_free_DCFT_after_rescaling}, we see that the $g$-function admits the double scaling expansion 
\begin{equation}\label{eq_free_g_expansion1}
\log g_{\gamma}/g_0 =
s \tilde{f}_{-1}(\gamma_0^2s,R,d)+
\tilde{f}_0(\gamma_0^2 s,R,d)+s^{-1}\tilde{f}_1(\gamma_0^2 s,R,d)+\ldots 
\end{equation}
where $R$ is the circle radius. For $\varepsilon\ll 1$ eq.~\eqref{eq_free_g_expansion1} is conveniently rewritten in terms of the physical coupling as:
\begin{equation}\label{eq_free_g_expansion}
\begin{split}
\log g_{\gamma}/g_0 
&=
s f_{-1}(\gamma^2s,RM,\varepsilon)+
f_0(\gamma^2 s,RM,\varepsilon)+s^{-1}f_{1}(\gamma^2 s,RM,\varepsilon)+
\ldots \,.
\end{split}
\end{equation}
When specialized to a fixed point, there is no dependence on the size of the defect $R$ and one finds the conformal defect entropy.

For $d<4$ with fixed $\varepsilon=O(1)$, $s_D\rightarrow -\infty$ for $R\rightarrow\infty$ for large enough $s$. This indicates the lack of an infrared DCFT, as we anticipated  in sec.~\ref{IntroMagneticImpGen}. Instead, for $\varepsilon\ll 1$ there is a rich phase diagram.

As a final comment, we notice that the above expansions in the double-scaling limit~\eqref{eq_double_scaling_limit} should match the result of the diagrammatic calculations discussed in sec.~\ref{sec_free_diag} for small $\gamma^2 s$.   This is because in this regime the term proprotional to $\gamma\sqrt{s}$ in eq. \eqref{eq_free_DCFT_after_rescaling} only represents a small perturbation of the free action, and consequently the saddle-point profile is close to the trivial one (around which the usual loop expansion is performed). As previously noticed, this exponentiation is a very nontrivial fact from the diagrammatic viewpoint. We will therefore use the diagrammatic results as a benchmark of our semiclassical approach in the overlapping regime, thus providing a strong consistency check of our methodology.  

The rest of this section is organized as follows.  In subsec.~\ref{subsec_NonLocalQM} we consider a nonlocal one-dimensional theory on the defect that we obtain upon integrating out explicitly the bulk scalar field. This will set the stage for all the other calculations that we perform in this section. In subsec.~\ref{subsec_phi2_and_beta} we study the one-point function of the operator $\phi_a^2$ and derive the phase diagram of the theory in the large $s$ limit as a function of $d$. Finally in subsec.~\ref{subsec_g_free_double} we compute the partition function of a circular defect and use our results to test the $g$-theorem. 

\subsubsection{The nonlocal theory on the line and the saddle-point}\label{subsec_NonLocalQM}

Let us consider the DQFT~\eqref{eq_free_DCFT0} for an arbitrary line geometry $x(\tau)$. Due to the simplicity of the bulk theory, we can integrate out explicitly $\phi_a$ on its equations of motion. Rescaling $z\rightarrow \sqrt{s} z$, this gives:
\begin{equation} \label{eq_bulk_field_sol}
\phi_a(x)=-\frac{\gamma_0 s}{(d-2)\Omega_{d-1}} \int_D d\tau\frac{\left(\bz\frac{\sigma^a}{2}z\right)(\tau)}{\left|x-x(\tau)\right|^{d-2}}+\delta\phi_a(x)\,,
\end{equation}
where $\delta\phi_a(x)$ is a free field fluctuation that completely decouples from the line.  The defect action then reduces to the following nonlocal quantum mechanical model:
\begin{equation}\label{eq_non-local1}
S=s\left[\int_D d\tau\bz\dot{z}-\frac{\alpha_0}{2 }\int_D d\tau \int_D d\tau'
\frac{\bz\frac{\sigma^a}{2}z\,\bz'\frac{\sigma^a}{2}z'}{\left|x(\tau)-x(\tau')\right|^{d-2}}\right]\,,
\end{equation}
where $\bz z=2$, $z'$ stands for $z(\tau')$ and we defined 
\begin{equation}\label{eq_Alpha}
\alpha_0\equiv\frac{\gamma_0^2 s}{(d-2)\Omega_{d-1}}\,.
\end{equation}

To proceed with the large $s$ limit, we need to expand the action~\eqref{eq_non-local1} around its saddle-point solution. By regulating the short distance divergence for $\tau\rightarrow \tau'$ in dimensional regularization, the saddle-point is simply given by:
\begin{equation}\label{eq_saddle}
z=z_0=\text{const}\,.
\end{equation}
There is a $S^2$ manifold of saddle-points: this is accounted for by the integration over the zero modes which rotate the solution as $z_0\rightarrow U z_0$, where $U$ is an arbitrary element of $U(2)$ modulo the $U(1)$ gauge transformations. The integration over the zero modes enforces the symmetry; for instance it implies that only $SO(3)$ singlet bulk operators can acquire a non-trivial one-point function.

It is useful to write $z$ in terms of polar and azimuthal angles $\theta$ and $\phi$, in the so called Bloch sphere parametrization:
\begin{equation}\label{eq_Bloch_par}
z=\sqrt{2}\left(\begin{array}{c}
\cos\frac{\theta}{2}\\
\sin\frac{\theta}{2}e^{i\phi}
\end{array}\right)\,.
\end{equation}
We choose to expand around $\theta=\frac{\pi}{2}$ and $\phi=0$.\footnote{Notice that the parametrization~\eqref{eq_Bloch_par} is singular at the south and north poles $\theta=0$ and $\theta=\pi$.} It is further convenient to recast the fluctuations $\delta\theta$ and $\delta\phi$ in terms of a complex variable $\chi$ as
\begin{equation}\label{eq_fluct}
\chi=\sqrt{\frac{s}{2}}\left(\delta\theta+i \delta\phi\right)\,.
\end{equation}
The action then reads:
\begin{equation}\label{eq_free_defect_action_nonlocal}
\begin{aligned}
S=&-s\frac{\alpha_0}{2}\int_Dd\tau\int_D d\tau'
\frac{1}{\left|x(\tau)-x(\tau')\right|^{d-2}}\\[0.3em]
&+\int d\tau\,\bar{\chi}\dot{\chi}-\frac{\alpha_0}{2}
\int d\tau \int d\tau'
\frac{\left(\bar{\chi}\chi'+\bar{\chi}'\chi-\bar{\chi}\chi-\bar{\chi}'\chi'\right)}{\left|x(\tau)-x(\tau')\right|^{d-2}} \\[0.3em]
&+O\left(\frac{1}{s}\right)
\end{aligned}
\end{equation}
The first line of eq.~\eqref{eq_free_defect_action_nonlocal} will be important when we  study the defect partition function, even though it does not depend on the fluctuation $\chi$. The second line is the quadratic action for the fluctuations around the saddle-point, and it allows studying $1/s$ corrections to observables. The previous remark above eq.~\eqref{eq_def_z_ordering} implies that the equal time product $\bar{\chi}\chi$ is a formal notation for $\lim_{\eta\rightarrow 0^+}\bar{\chi}(\tau+\eta)\chi(\tau)$, and similarly for $\bar{\chi}'\chi'$.  Finally $1/s$ suppressed quartic vertices arise both from the expansion of the kinetic term and the nonlocal interaction.
The theory nonetheless remains under perturbative control at all scales for large $s$, since in the semiclassical approach the $\alpha_0$ term behaves like a mass term for the fluctuations, so that the $1/s$ suppressed relevant couplings remain always parametrically small.\footnote{This is analogous to the case of a three-dimensional theory with potential $V(\phi)=m^2\phi^2+\lambda \phi^4$, which is perturbative at all scales for $\lambda/m\ll 4\pi$. } In this sense, the large $s$ limit of the theory~\eqref{eq_non-local1} resembles the large $N$ limit of the three-dimensional $O(N)$ model, since in both cases the saddle-point allows resumming the leading effects of the relevant interaction term \cite{Moshe:2003xn}.

\subsubsection{The one-point function of \texorpdfstring{$\phi_a^2$}{phi2} and the phase diagram}\label{subsec_phi2_and_beta}

Here we illustrate our ideas by performing the semiclassical calculation of $\langle\phi_a^2(x)\rangle$, for a straight defect located at $x^i=0$, to the first subleading order in the $1/s$ expansion. We will use this calculation to extract the beta function~\eqref{eq_beta_structure_double_scaling} in the double-scaling limit~\eqref{eq_double_scaling_limit}, and thus extract the phase diagram of the theory as a function of $d$ in the large $s$ limit. We will also comment on other observables.

To perform the calculation, we use~\eqref{eq_bulk_field_sol} to express the one-point function of $\phi_a^2$ as:
\begin{equation} \label{eq_one_point0}
\langle\phi_a^2(\mathbf{x},0)\rangle =
\frac{s \alpha_0 }{(d-2)\Omega_{d-1}}
\int d\tau\int d\tau'
\frac{\langle\bz\frac{\sigma^a}{2}z\,\bz'\frac{\sigma^a}{2}z'\rangle}{\left(\mathbf{x}^2+\tau^2\right)^{\frac{d-2}{2}}\left(\mathbf{x}^2+\tau'^{\,2}\right)^{\frac{d-2}{2}}} \,.
\end{equation}
To obtain the leading order result for the one-point correlation function, we simply plug the saddle-point solution~\eqref{eq_saddle} in eq.~\eqref{eq_one_point0}. This gives:
\begin{equation}\label{eq_phi2_LO}
\begin{split}
\langle\phi_a^2(\mathbf{x},0)\rangle
&=\frac{s \alpha_0 }{(d-2)\Omega_{d-1}}\left(
\int d\tau\frac{1}{\left(\mathbf{x}^2+\tau^2\right)^{\frac{d-2}{2}}}\right)^2+O\left(s^0\right)
\\&
=s\frac{\gamma_0^2 s}{16\pi^{d-1} |\mathbf{x}|^{2 d-6}} \Gamma \left(\frac{d-3}{2}\right)^2+O\left(s^0\right)\,.
\end{split}
\end{equation} 
In terms of the expansion~\eqref{eq_free_phi_expansion1} this implies:
\begin{equation}\label{eq_htminus1_free}
\tilde{h}_{-1}(\gamma^2_0 s,|\mathbf{x}|,\varepsilon)=
\mathcal{N}_d^{-1}\frac{\gamma_0^2 s}{16\pi^{d-1} |\mathbf{x}|^{4-d}} \Gamma \left(\frac{d-3}{2}\right)^2\,.
\end{equation}
Eq.~\eqref{eq_htminus1_free} exactly agrees with the leading order diagrammatic result~\eqref{phi2_diag}, which is therefore \emph{exact} in the double-scaling limit. We shall see in a moment that the first correction $\tilde{h}_0$ takes a more intricate (and interesting) form.

Before focusing on the next to leading order correction, a few comments are in order. The one-point function~\eqref{eq_phi2_LO} in $d=4$ reads:
\begin{equation}\label{eq_phi2_LO4}
\langle\phi_a^2(\mathbf{x},0)\rangle\stackrel{d=4}{=}\frac{\gamma_0^2s^2}{16\pi^2 \mathbf{x}^2}+O\left(s^0\right)\,.
\end{equation}
The one-point function is conformally invariant in $d=4$, in agreement with the marginal nature of the coupling at tree level. We will soon show that quantum corrections provide logarithmic corrections in four dimensions, leading to a rich phase diagram for $\varepsilon=4-d\ll 1$. For $\varepsilon=O(1)$, the result instead deviates from the conformal scaling $\langle\phi_a^2(\mathbf{x},0)\rangle\propto 1/|\mathbf{x}|^{d-2}$, due to the relevant nature of the coupling. \footnote{Note that in general, in $d=3$, a term of the form $(\phi^a)^2$ on the line must be taken into account as well, as it is classically marginal.   However, as was shown in \cite{Cuomo:2021kfm}, such a term turns out to be marginally irrelevant.} As anticipated in the discussion below eq.~\eqref{eq_free_defect_action_nonlocal},  we shall see that $1/s$ corrections do not change this qualitative behavior at long distances, and thus the theory never reaches an infrared fixed point.  Finally we notice that the result~\eqref{eq_phi2_LO} has a double pole in $d=3$, associated with an infrared logarithmic divergence of the integral in three dimensions. To regulate the result, we introduce a cutoff length $L\gg |\mathbf{x}|$ on the extension of the line, so that to leading logarithmic accuracy the result reads:\footnote{Alternatively, we could regulate the infrared divergence by considering a circular defect.}
\begin{equation}\label{eq_phi2_LO3}
\langle\phi_a^2(\mathbf{x},0)\rangle
\stackrel{d=3}{\simeq} \frac{ \gamma_0^2 s^2 }{(4\pi)^2}\log^2\left(\mathbf{x}^2/L^2\right)+O\left(s^0\right)\,.
\end{equation}

One conceptual point that will be crucial later is that the leading in $s$ behavior is analogous to the one which is obtained by considering a symmetry breaking source localized on a line in the free theory, $\delta S\propto\int d \tau\phi_1$,  i.e. the external field (or pinning field) defect,  see \cite{Cuomo:2021kfm}. In fact,  as eq.~\eqref{eq_bulk_field_sol} shows, the impurity behaves precisely as a localized source up to the zero-mode integration.  This is also reflected in the result for the $g$-function, that we discuss in the next subsection. From this point of view, the lack of a fixed point for large $s$ at any fixed $d<4$ is therefore due to the same physics as the lack of a fixed point in the external field defect, which was explained in~\cite{Cuomo:2021kfm}. In other words, it is due to the moduli space of vacua in the bulk.

Let us now focus on the $O\left(s^0\right)$ correction to the result~\eqref{eq_phi2_LO}. To this aim,  we write explicitly the quadratic action for the fluctuations in eq.~\eqref{eq_free_defect_action_nonlocal} for a straight line:
\begin{equation}
\begin{split}
S^{(2)}&\simeq\int d\tau\,\bar{\chi}\dot{\chi}-\frac{\alpha_0}{2}
\int d\tau \int d\tau'
\frac{\left(\bar{\chi}\chi'+\bar{\chi}'\chi-\bar{\chi}\chi-\bar{\chi}'\chi'\right)}{\left|\tau-\tau'\right|^{d-2}}\\
&=\int \frac{d\omega}{2\pi}\bar{\chi}(\omega)G_\chi^{-1}(\omega)\chi(\omega)\,,
\end{split}
\end{equation}
where  $G_{\chi}(\omega)$ is the propagator associated with the fluctuations:
\begin{equation}\label{eq_psi_prop}
G_{\chi}(\omega)=\frac{1}{-i\omega-\alpha_0 \bar{c}^{(d)}(\omega)  }\,.
\end{equation}
The function $\bar{c}^{(d)}(\omega)$ is defined by
\begin{equation}\label{eq_psi_c}
\bar{c}^{(d)}(\omega)=\int d\tau \frac{e^{-i\omega\tau}-1}{|\tau|^{d-2}}=
-2|\omega|^{d-3}\Gamma(3-d)\sin\left(\frac{d\pi}{2}\right) \qquad\text{for } d>3\,.
\end{equation}
The case of $d=3$ is special due to the infrared logarithmic divergence that we encountered before and we will discuss it separately at the end.
Expanding eq.~\eqref{eq_one_point0} in terms of the fluctuations around the saddle-point solution, we may now use the propagator~\eqref{eq_psi_prop} to write the next-to-leading order contribution to the one-point function as
\begin{equation}\label{phi2_intermediate}
\begin{split}
\delta\langle\phi_a^2(\mathbf{x},0)\rangle &=\frac{\alpha_0}{(d-2)\Omega_{d-1}}\int d\tau\int d\tau'\frac{\langle\bar{\chi}'\chi+\bar{\chi} \chi'-\bar{\chi}\chi-\bar{\chi}'\chi'\rangle}{\left(\mathbf{x}^2+\tau^2\right)^{\frac{d-2}{2}}\left(\mathbf{x}^2+\tau'^{\,2}\right)^{\frac{d-2}{2}}} \\
&=\lim_{\eta\rightarrow 0^+}
\frac{2\alpha_0}{(d-2)\Omega_{d-1}}\int \frac{d\omega}{2\pi}G_\chi(\omega)\left[\left|h_{\mathbf{x}}(\omega)\right|^2
-e^{i\omega\eta}\left|h_{\mathbf{x}}(0)\right|^2\right]\,,
\end{split}
\end{equation}
where the factor $e^{i\omega \eta}$ arise from the point-splitting regularization mentioned around eq.~\eqref{eq_def_z_ordering} and we have defined
\begin{equation}\label{eq_hxOmega}
\begin{split}
h_{\mathbf{x}}(\omega) &=\int d\tau\frac{e^{-i\omega\tau}}{\left(\mathbf{x}^2+\tau^2\right)^{\frac{d-2}{2}}}=
\frac{\sqrt{\pi } 2^{\frac{5-d}{2}} |\mathbf{x}|^{\frac{3-d}{2}} | \omega | ^{\frac{d-3}{2}} K_{\frac{d-3}{2}}(|\mathbf{x}| |\omega|  )}{\Gamma \left(\frac{d}{2}-1\right)}
\,.
\end{split}
\end{equation}
In the above $K_{\nu}(x)$ is the modified Bessel functions of the second kind.   Eq.~\eqref{eq_hxOmega} simplifies in $d=4$:
\begin{equation}
h_{\mathbf{x}}(\omega) \stackrel{d=4}{=}\frac{\pi}{|\mathbf{x}|}e^{-|\omega||\mathbf{x}|}\,.
\end{equation}

The expression~\eqref{phi2_intermediate} holds for any value of $d>3$. 
 Nonetheless it is technically hard to obtain an explicit general result. Therefore, to proceed, it is technically convenient to discuss separately the case of small $\varepsilon=4-d$ and that of $4-d=O(1)$.

We consider first the case of small $\varepsilon$. Noticing that $G_{\chi}(\omega)\sim 1/|\omega|$ for $\omega\rightarrow\infty$ in four dimensions, we see that the integral~\eqref{phi2_intermediate} would lead to a logarithmic divergence in the $\eta\rightarrow 0^+$ limit  due to the integration over the $G_{\chi}(\omega)\left|h_{\mathbf{x}}(0)\right|^2$ term. The divergence needs to be renormalized by the coupling counterterm, and therefore leads to a nontrivial RG flow.

Explicitly, studying the integral in $4-\varepsilon$ dimensions, we find that the next-to-leading order correction to the one-point function~\eqref{eq_phi2_LO} is:
\begin{equation}\label{eq_free_bulk_NLO}
\begin{split}
\delta\langle\phi_a^2(\mathbf{x},0)\rangle&=
-\frac{1}{\varepsilon}\frac{\alpha_0 \arctan(\pi  \alpha_0)}{2 \pi   \mathbf{x}^2}
+\frac{\alpha_0}{4\pi \mathbf{x}^2}
-\frac{2 \alpha^2_0 (\log |\mathbf{x}|+1+\log 2)}{4(1+\pi ^2 \alpha_0^2) \mathbf{x}^2}\\[0.5em]
&-\frac{\alpha_0\arctan(\pi  \alpha_0) (4 \log |\mathbf{x}|+\gamma_E +\log 16+\log \pi )}{4\pi \mathbf{x}^2 } +
O\left(\varepsilon\right)
\,.
\end{split}
\end{equation}
We detail the computation  in appendix~\ref{app:1Ovs_phi2}. Here we only remark that the second term on the right hand side of eq.~\eqref{eq_free_bulk_NLO}, which is linear in $\alpha_0$,  arises because of the point-splitting in eq.~\eqref{eq_def_z_ordering}. When added to the leading order~\eqref{eq_phi2_LO4},  it modifies the prefactor $\gamma_0^2s^2$ to $\gamma_0^2s(s+1)$, as expected.

Comparing eq.~\eqref{eq_free_bulk_NLO} with the leading order expression in four dimensions~\eqref{eq_phi2_LO4}, we see that one obtains a finite result for the one-point correlation function of $\langle\phi_a^2(x)\rangle$ upon renormalizing the coupling $\alpha_0$ according to:
\begin{equation}\label{eq_alpha_ct}
\alpha_0=\frac{4\pi^2M^{\varepsilon}}{(2-\varepsilon)\Omega_{3-\varepsilon}}\left(\alpha+
\frac{\delta\alpha}{\varepsilon}
\right)\,,\qquad
\delta\alpha=
\frac{2\alpha  \arctan(\pi  \alpha)}{\pi s}\,,
\end{equation}
where the prefactor $\frac{4\pi^2}{(2-\varepsilon)\Omega_{3-\varepsilon}}=1+O\left(\varepsilon\right)$ is there to compensate the $\varepsilon$-dependence in the definition~\eqref{eq_Alpha}. This ensures that eq.~\eqref{eq_alpha_ct} corresponds to the same renormalization scheme used in sec.~\ref{sec_free_diag}, allowing for a comparison with the results in that section also away from the fixed points.
Using the above result, we find the following beta function:
\begin{equation}\label{eq_BETA}
\beta_{\alpha}=-\varepsilon\alpha+\frac{1}{s}\frac{2\alpha^2}{1+\pi^2\alpha^2}\,.
\end{equation}
Notably, this result for the beta function agrees with the diagrammatic one~\eqref{eq_Sachdev_ct} in the small $\alpha$ limit, as can be seen using $\alpha=\displaystyle\frac{\gamma^2 s}{4\pi^2}$. In eq.~\eqref{eq_beta_structure_double_scaling} it implies:
\begin{equation}
\beta^{(4d)}_0=0\,,\qquad
\beta^{(4d)}_1=2\alpha/(1+\pi^2\alpha^2)\,.
\end{equation}

\begin{figure}[t]
\centering
\includegraphics[scale=0.5]{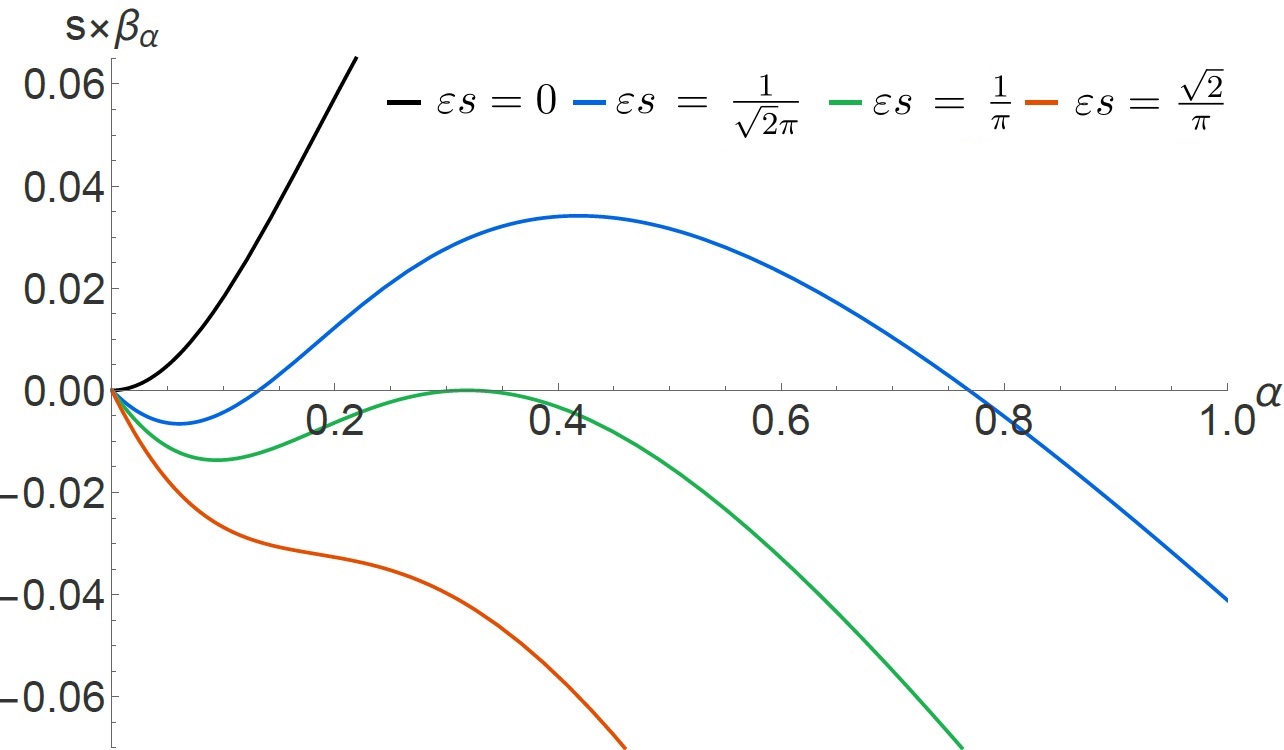}
\caption{Plot of the beta function (multiplied by a factor of $s$) as a function of $\alpha$ for various values of fixed $\varepsilon s$: strictly in four dimensions with $\varepsilon s =0$ (black), in the range $0<\varepsilon s <\frac{1}{\pi}$ with $\varepsilon s = \frac{1}{\sqrt{2}\pi}$ (blue), $\varepsilon s = \frac{1}{\pi}$ (green) and in the range $\varepsilon s >\frac{1}{\pi}$ with $\varepsilon s = \frac{\sqrt{2}}{\pi}$ (orange).}\label{plot}
\end{figure}

Eq.~\eqref{eq_BETA} admits nontrivial zeros for:
\begin{equation}\label{eq_beta_function_zero_free_case}
\varepsilon s=\frac{2\alpha}{1+\pi^2 \alpha^2}\,.
\end{equation}
The solutions depend on the value of the double-scaling parameter $\varepsilon s$ and are summarized in figure~\ref{plot}.
Strictly in $d=4$, there is only one fixed point, which is stable and trivial (at $\alpha=0$) -- see the black curve of figure~\ref{plot}.
Interestingly, there are two solutions in the regime where $0<\varepsilon s<1/\pi$, as is demonstrated by the blue curve of figure~\ref{plot}.
These are given by:
\begin{equation}\label{fixed_points}
\alpha=\begin{cases}\displaystyle
\frac{1-\sqrt{1-\pi ^2 (s\varepsilon)^2}}{\pi ^2 s\varepsilon}\equiv\alpha_1\, = \frac{\varepsilon s}{2}+O\left((\varepsilon s)^3\right),\\[1em]
\displaystyle\frac{1+\sqrt{1-\pi ^2 (s\varepsilon)^2}}{\pi ^2 (s\varepsilon)}\equiv\alpha_2= \frac{2}{\pi^2 \varepsilon s}+O(\varepsilon s) \,.
\end{cases}
\end{equation}
The fixed point with the smaller $\alpha$, $\alpha=\alpha_1$, reduces to the weak-coupling perturbative fixed point studied in sec.~\ref{sec_free_diag} and it is the only stable fixed point. However, the semiclassical approach used in this subsection reveals the existence of a new fixed point, which appears for $\alpha =\alpha_2$; this is a nonperturbative fixed point in the standard perturbative approach valid only for $\varepsilon s\ll 1$. This new fixed point is unstable towards the first fixed point for $\alpha<\alpha_2$, and it flows to the strongly coupled regime for $\alpha>\alpha_2$; we expect that this latter flow never reaches an endpoint (with a behaviour analogous to the one discussed above eq. \eqref{eq_phi2_LO3} about the large $s$ limit for $3<d<4$).

The two solutions  coincide at $\varepsilon s=1/\pi $ (as can be seen from the green curve of figure~\ref{plot}).  No solutions exist for $\varepsilon s>1/\pi$ (see the orange curve of figure~\ref{plot}).\footnote{Interestingly, a technically analogous double-scaling limit unveils a similar fixed point annihilation phenomenon for the $SU(2)_k$ Wess-Zumino-Witten model in $2+\varepsilon$ dimensions \cite{Nahum:2019fjw}.} We will indeed see  that for finite $\varepsilon$ and large $s$ the infrared limit of the defect is not described by a DCFT, rather, the flow never terminates and tends towards $s_D\to-\infty$.

We also comment on the defect operator spectrum at the fixed points. 
The comment below eq.~\eqref{eq_tilt_operator} implies that the impurity spin operator $S^a$ has dimension $\Delta_S=\varepsilon/2$ for both fixed points in eq.~\eqref{eq_beta_function_zero_free_case}.  The anomalous dimension of the operator $S^a\phi_a$ is extracted from the beta function and to the first nontrivial order it reads
\begin{equation}\label{eq_free_anomalous}
\gamma_{S\cdot\phi}=\Delta_{S\cdot\phi}-1=\left.\frac{\pd\beta_\alpha}{\pd\alpha}\right\vert_{\beta_{\alpha}=0}\simeq
\begin{cases}
\displaystyle
+\varepsilon  \sqrt{1-\pi^2  s^2 \varepsilon ^2} 
&\text{for }\alpha=\alpha_1\,,\\
\displaystyle-\varepsilon  \sqrt{1-\pi ^2 s^2 \varepsilon ^2} 
&\text{for }\alpha=\alpha_2\,.
\end{cases}
\end{equation}
As expected, the anomalous dimension is positive at the stable fixed point and negative at the unstable one.  The operator is marginal when the two fixed points collide.

Finally, we provide the result for the one-point correlation function to the next-to-leading order. In terms of the expansion~\eqref{eq_free_phi_expansion}, we find:
\begin{align}\nonumber
h_{-1}(\gamma^2 s,|\mathbf{x}|M,\varepsilon)&=\frac{\gamma^2s}{4\sqrt{6}}\left\{1+
\varepsilon  \left[\log (4 M |\mathbf{x}|)+\frac{1}{2} (\gamma_E +\log \pi )\right]+O\left(\varepsilon^2\right)
\right\}\,,
\\[0.5em] 
h_0(\gamma^2 s,|\mathbf{x}|M,\varepsilon)&=\frac{\gamma^2s}{4\sqrt{6}}-\frac{\gamma^4s^2}{\sqrt{6}\left(16 \pi ^2+\gamma^4 s^2\right)}
\left[2 \log (M |\mathbf{x}|)+\gamma _E+2+\log 4 \pi \right]
+O\left(\varepsilon\right)\,.
\label{eq_phi2_result1}
\end{align}
The expansion of eqs.~\eqref{eq_phi2_result1} for $\gamma^2 s\ll (4\pi)^2$ is in perfect agreement with the diagrammatic results~\eqref{eq_phi2_diag3pre}. From eq.~\eqref{eq_phi2_result1} we also obtain the correlator at the fixed points $\alpha_1$ and $\alpha_2$ in eq.~\eqref{fixed_points}:
\begin{equation}\label{eq_phi2_result2}
\langle\phi_a^2(\mathbf{x},0)\rangle=
\frac{\mathcal{N}_d}{|\mathbf{x}|^{2-\varepsilon}}
\frac{s\left(1\mp\sqrt{1-\pi^2\varepsilon^2 s^2}\right)}{\sqrt{6}\varepsilon s}
\left\{1+\frac{1}{s}\left[
1+\varepsilon s\left(\log 2-1\right)\right]
+O\left(\frac{1}{s^2}\right)
\right\}\,,
\end{equation}
where the $-$ and $+$ sign refer, respectively, to the fixed points $\alpha_1$  and $\alpha_2$.
This concludes the discussion for $\varepsilon\ll 1$.

We now wish to evaluate the $1/s$ correction in eq.~\eqref{phi2_intermediate} for generic $d$ with $\varepsilon =O(1)$. In this case no renormalization is required since there are no divergences in~\eqref{phi2_intermediate}. Thus, we will work directly in terms of the bare coupling $\alpha_0$.  

To obtain an analytic expression, we focus on the long distance limit of $\langle\phi_a^2(\mathbf{x},0)\rangle$, specified by $\alpha_0|\mathbf{x}|^{4-d}\gg 1$. In this limit, the leading result arises from the second term in square parenthesis in eq.~\eqref{phi2_intermediate}, which is proportional to a tadpole integral of the propagator.  This term indeed behaves as $|h_{\mathbf{x}}(0)|^2\propto 1/|\mathbf{x}|^{2d-6}$ like the leading order~\eqref{eq_phi2_LO}, while we shall soon see that the first contribution in the square parenthesis in eq.~\eqref{phi2_intermediate} decays faster at large distances. We find:
\begin{equation}\label{eq_free_sInt}
\lim_{\eta\rightarrow 0^+}\int \frac{d\omega}{2\pi}e^{i\omega\eta}G_\chi(\omega)=\frac{1}{2(4-d)}-\frac{1}{2}\,.
\end{equation}
While the propagator~\eqref{eq_psi_prop} depends on $\alpha_0$ the result is independent of it, as expected from dimensional analysis.\footnote{Technically, this can be seen rescaling $\omega\rightarrow\omega \alpha_0^{\frac{1}{4-d}}$ in the integral~\eqref{eq_free_sInt}.} The $-1/2$ arises from the point-splitting prescription~\eqref{eq_def_z_ordering}.  We may also evaluate the leading long distance contribution from the first term in the square parenthesis of eq.~\eqref{phi2_intermediate} by expanding the propagator~\eqref{eq_psi_prop} for small $\omega$:\footnote{This can be seen explicitly rescaling $\omega\rightarrow\omega/|\mathbf{x}|$ in the integral~\eqref{eq_free_sInt2}.}
\begin{equation}\label{eq_free_sInt2}
\begin{split}
\int \frac{d\omega}{2\pi}G_{\chi}(\omega)|h_{\mathbf{x}}(\omega)|^2 &=\frac{-1}{\alpha_0 \bar{c}^{(d)}(1)}
\int\frac{d\omega}{2\pi}\frac{|h_{\mathbf{x}}(\omega)|^2}{|\omega|^{d-3}}\left[1+O\left(\frac{1}{\alpha^2_0|\mathbf{x}|^{2(4-d)}}\right)\right] \\
&=\frac{ \pi \Gamma \left(\frac{d-3}{2}\right)^2}{\Gamma \left(\frac{d-2}{2}\right)^2}\frac{\sigma(d)}{\alpha_0|\mathbf{x}|^{d-2}}\left[1+O\left(\frac{1}{\alpha^2_0|\mathbf{x}|^{2(4-d)}}\right)\right]\,,
\end{split}
\end{equation}
where  for convenience  we defined a positive coefficient $\sigma(d)$ as:
\begin{equation}
\sigma(d)=\frac{(3-d) \cot \left(\frac{\pi  d}{2}\right) \Gamma \left(\frac{d}{2}-1\right)^2}{  2^{5-d} \,\pi\Gamma (d-3)}>0\qquad
\text{for }3<d<4.
\end{equation}
The coefficient $\sigma(d)$ has a pole for $d\rightarrow 4$ and vanishes in $d=3$.

Using eqs.~\eqref{eq_phi2_LO},~\eqref{eq_free_sInt} and~\eqref{eq_free_sInt2} in the expression~\eqref{phi2_intermediate}, we write the final result for the one-point function~\eqref{eq_phi2_LO} in $3<d<4$ as
\begin{equation}\label{eq_free_phi2_anyD}
\begin{split}
\langle\phi_a^2(\mathbf{x},0)\rangle &= \frac{s\alpha_0 \pi \Gamma \left(\frac{d-3}{2}\right)^2 }{(d-2)\Gamma \left(\frac{d-2}{2}\right)^2\Omega_{d-1}|\mathbf{x}|^{2d-6}} \\
&\times\left\{1+\frac{1}{s}\left[
\frac{3-d}{4-d}+\frac{\sigma(d)}{\alpha_0|\mathbf{x}|^{4-d}}+O\left(\frac{1}{\alpha_0^3
|\mathbf{x}|^{3(4-d)}}\right)\right]
+O\left(\frac{1}{s^2}\right)\right\}\,.
\end{split}
\end{equation}
Notice that the expansion breaks down for $d\rightarrow 4$, which is why we had to perform renormalization in that case. Otherwise, we see that $1/s$ corrections only change the prefactor of the leading $1/|\mathbf{x}|^{2d-6}$ term at large distances. From eq.~\eqref{eq_free_phi2_anyD} we also see that the first subleading correction at long distances is independent of $\alpha_0$ (but depends on $s$) and obeys a conformal scaling law $1/|\mathbf{x}|^{d-2}$. 

A qualitatively similar behavior describes other correlation functions. For instance, an analogous calculation shows that the two-point function of the spin operator on the line takes the following form:
\begin{equation}
\langle S^a(\tau) S^b(0)\rangle= \frac{\delta^{ab}}{3}s^2\left\{1+\frac{1}{s}\left[\frac{3-d}{4-d}+
\frac{(3-d)\cot \left(\frac{\pi  d}{2}\right) }{\pi\alpha_0|\tau|^{4-d}}
+O\left(\frac{1}{\alpha_0^3
|\tau|^{3(4-d)}}\right)\right]+O\left(\frac{1}{s^2}\right)\right\}\,.
\end{equation}
We see from these examples that $1/s$ corrections for $d<4$ are indeed small and do not alter the long distance behavior. In addition, as promised, we see that the long distance behavior is not compatible with a DCFT (which would require a leading $1/\abs{\tau}^{4-d}$ dependence because of eq.~\eqref{eq_tilt_operator}), rather, the RG flow at large $s$ and fixed $d<4$ never terminates.

While our treatment so far focused on $3<d<4$, a similar discussion applies in $d=3$, provided one carefully regulates the infrared logarithmic divergences associated with the infinite extent of the line.  In particular,  $1/s$ corrections again do not lead to a well defined DCFT at long distances.
Technically, these infrared singularities arise because $\bar{c}^{(3)}(\omega)$ in~\eqref{eq_psi_prop} reads:
\begin{equation}\label{eq_psi_c3d}
\bar{c}^{(3)}(\omega)=\int d\tau\frac{e^{-i\omega\tau}-1}{|\tau|}=-2\log(|\omega|L)+\text{const}\,,
\end{equation}
where $L$ is the IR cutoff length of the defect.
Because of the ambiguities related to how precisely we perform the IR regularization, we postpone the discussion of $d=3$ to circular defects, for which no ambiguities of this sort arise.

We summarize: at fixed $3\leq d<4$, for large $s$, there is no infrared DCFT. Our large $s$-result leads to a never-ending flow with correlation functions scaling as in the presence of a localized external source (up to the zero-mode integration). 
We expect (but cannot prove) that the DQFT behaves analogously also for $s=O(1)$ in $3\leq d<4$.

\subsubsection{The g-function}\label{subsec_g_free_double}

In this subsection we compute the defect $g$-function for a circular defect of radius $R$ in the large $s$ limit. We will also use our results to check the $g$-theorem for the fixed points we have found at $\varepsilon\ll1$.

To perform the calculation, we consider the defect on a circle of radius $R$, $x^\mu(\tau)=R(\cos\tau/R,\sin\tau/R,0\ldots)$. The leading order result arises from the classical value of the action~\eqref{eq_free_defect_action_nonlocal} on the saddle-point $z=\text{const}$. In terms of the expansion~\eqref{eq_free_g_expansion1}, we find
\begin{equation}\label{eq_ftminus1}
\tilde{f}_{-1}(\gamma_0^2 s,R,d)=
\frac{\alpha_0 R^{4-d}}{2}\int d\phi \int d\phi'
\frac{1}{\left(4\sin^2\frac{\phi-\phi'}{2}\right)^{{d-2\ov 2}}}=\pi \alpha_0 R^{4-d} I_1^{(d)}\,,
\end{equation}
where  we set $\tau=R \phi$ and $I_1^{(d)}$ is defined in eq.~\eqref{eq_diagram_int1}.  As for the leading $\tilde{h}_{-1}$ contribution to $\langle\phi_a^2(\mathbf{x},0)\rangle$ before, eq.~\eqref{eq_ftminus1} exactly agrees with the leading order diagrammatic result in eq.~\eqref{eq_gFreeDiagBare}.  As expected, the result~\eqref{eq_ftminus1} also exactly coincides with that of a localized source $\int d\tau \phi_1$ on the defect discussed in \cite{Cuomo:2021kfm},  where the result was also shown to satisfy the gradient formula~\eqref{eq_gradient_formula}.

The function $I_1^{(d)}$ in eq.~\eqref{eq_diagram_int1} vanishes for $d=4$, in agreement with the classical marginality of the coupling.  Notice that, since $I_1^{(d)}= -\pi\varepsilon/2+O\left(\varepsilon^2\right)$, to obtain the value of $\log g$ for small $\varepsilon\sim 1/s$ we also need to compute  the one-loop correction $\tilde{f}_0$ in eq.~\eqref{eq_free_g_expansion1}. We will soon do that.

Before discussing subleading corrections, let us comment on the result for $d<4$ with $\varepsilon=O(1)$. We find that $I_1^{(d)}$ has a pole for $d=3$. The divergence can be renormalized by adding a cosmological constant counterterm on the line, since it is linear in $R$, and results in a $R\log R$ contribution\footnote{As commented in footnote 9 of \cite{Cuomo:2021kfm} for the case of localized symmetry breaking source, this term is associated to an anomaly in coupling space \cite{Gomis:2015yaa,Schwimmer:2018hdl,Schwimmer:2019efk}.} to $\log g$ in $d=3$:
\begin{equation}\label{eq_ftminus1_3d}
\tilde{f}_{-1}(\gamma_0^2 s,R,3)=2\pi\alpha_0 R\log (RM)+\text{const}\times MR\,,
\end{equation}
where $M$ is an arbitrary cutoff scale. To obtain a scheme-independent quantity for arbitrary $d$ we compute the defect entropy as in eq.~\eqref{eq_sD_def},
\begin{equation}\label{eq_free_sD}
s_D=-s\pi\alpha_0 R^{4-d}\rho(d)\left[1+O\left(\frac{1}{s}\right)\right]\,,
\end{equation}
where we defined a function $\rho(d)$ which is positive and regular for $2<d<4$ and vanishes in $d=4$:
\begin{equation}
\rho(d)\equiv\frac{\sqrt{\pi } 2^{4-d} \Gamma \left(\frac{5}{2}-\frac{d}{2}\right)}{\Gamma \left(2-\frac{d}{2}\right)}=\begin{cases}
0 & \text{for }d=4\,,\\
2 & \text{for }d=3\,.
\end{cases}\
\end{equation}
The result~\eqref{eq_free_sD} therefore implies that $s_D\rightarrow -\infty$ in the infrared ($R\rightarrow\infty$) for fixed  $d<4$ and large $s$. This is compatible with the previously discussed scenario of a defect renormalization group flow that does not terminate in a healthy DCFT (that would have $g>0$).  At the end of this section we will see that $1/s$ corrections do not change the IR behaviour of $s_D$.

Let us now discuss the first $1/s$ correction to the result for small $\varepsilon\sim 1/s$. In particular, to obtain the $g$-function at the previously discussed fixed points we need to compute the next to leading order correction $\tilde{h}_0$ strictly in $d=4$.  This follows from the one-loop determinant of the fluctuations $\chi$ defined in eq.~\eqref{eq_fluct}.
The details of the calculation can be found in appendix~\ref{app_details_of_calc_f0}, while here we report the main result:
\begin{equation} \label{eq_f0_final}
\begin{split}
\tilde{f}_{0}(\gamma_0^2 s,R,4-\varepsilon)&=
\frac{1}{2} \log \left(1+\pi ^2 \alpha_0^2\right)+\pi  \alpha_0 \arctan(\pi  \alpha_0)+O\left(\varepsilon\right)\,.
\end{split}
\end{equation}
We now have all the information we need in order to compute the physical $g$-function to the leading nonvanishing order in the double-scaling limit~\eqref{eq_double_scaling_limit}. In terms of the physical, renormalized, coupling, we find the following results for the coefficient $f_{-1}$ and $f_0$~\eqref{eq_free_g_expansion}:
\begin{align}\nonumber
f_{-1}(\gamma^2 s,RM,\varepsilon)&=-\frac{\varepsilon}{2}\pi^2\alpha
+O\left(\varepsilon^2\right)
\,,\\
f_0(\gamma^2 s,RM,\varepsilon)&=\frac12\log\left(1+\pi^2\alpha^2\right)+O\left(\varepsilon\right)\,.
\label{eq_g_4}
\end{align}
Using $\alpha=\displaystyle\frac{\gamma^2 s}{4\pi^2}$ and expanding for small $\gamma^2 s$, eqs.~\eqref{eq_g_4} can be seen to agree with the previous diagrammatic result in eq.~\eqref{eq_gFree_diag}.
As already commented below eq.~\eqref{eq_gFree_diag}, the result depends on $RM$ through the running of the coupling constant $\alpha=\alpha (RM)$. This implies in particular that $\log g $ and $s_D$ coincide to the first nontrivial order.

\begin{figure}[t]
    \centering
    \begin{minipage}{0.45\textwidth}
       \centering
\includegraphics[scale=0.54]{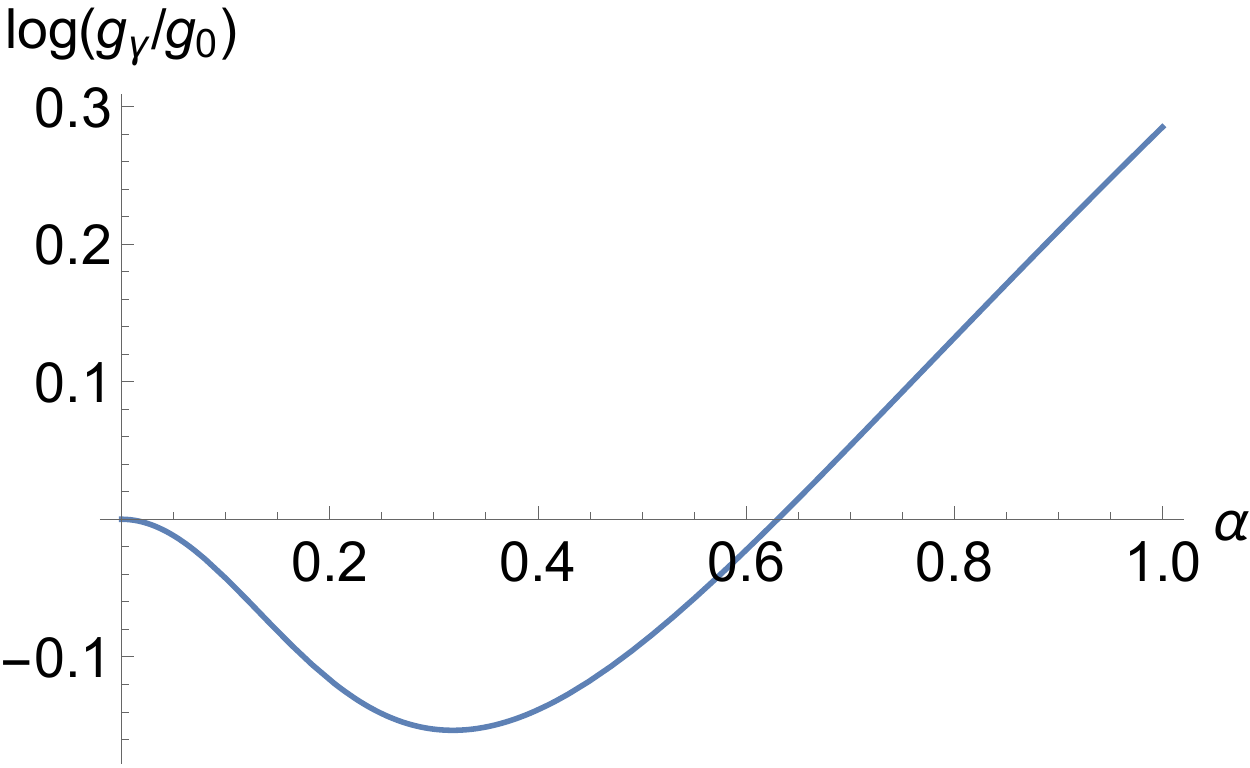}
\caption{Plot of $\log(g_{\gamma}/g_0)$ at the fixed points as a function of $\alpha$.}\label{plot2}
    \end{minipage}\hfill
    \begin{minipage}{0.45\textwidth}
        \centering
\includegraphics[scale=0.52]{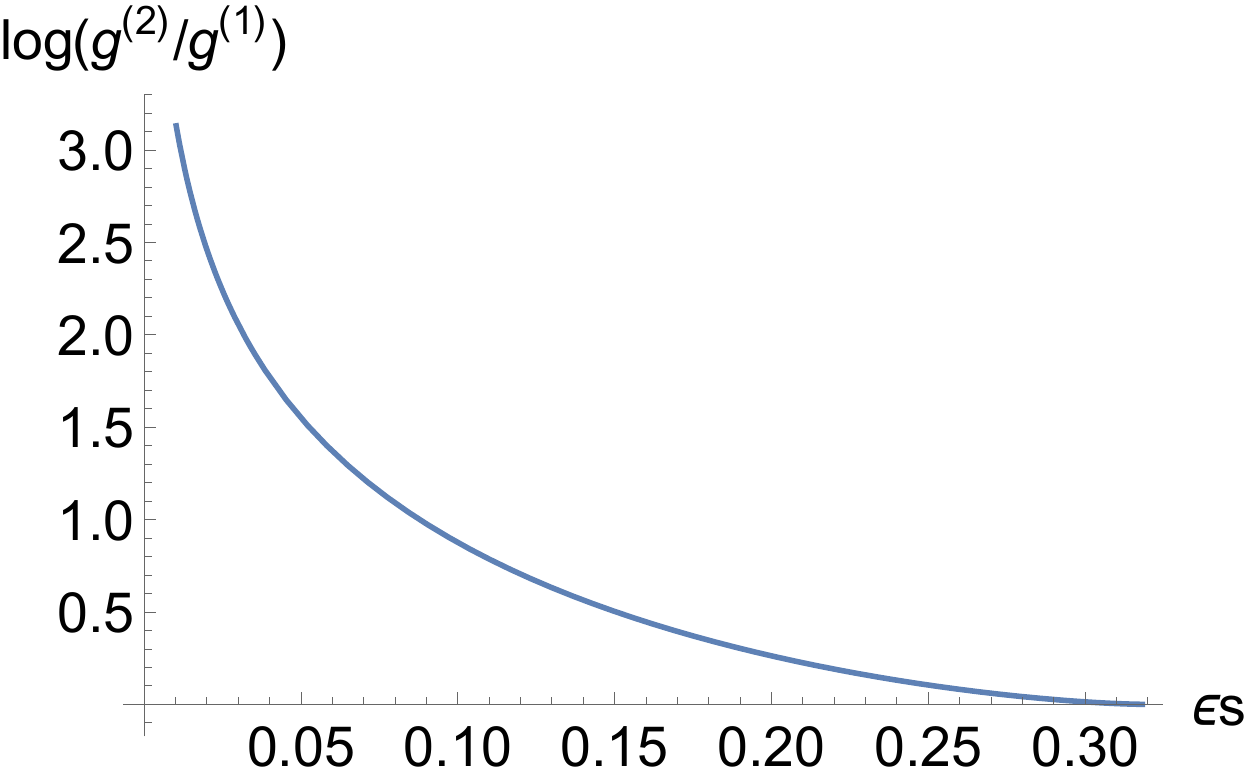}
\caption{The difference $\log g^{(2)}-\log g^{(1)}$ as a function of $\varepsilon s$.}\label{plot3}
    \end{minipage}
\end{figure}
At the fixed points that satisfy~\eqref{eq_beta_function_zero_free_case}, the $g$-function is conveniently expressed in terms of $\alpha$ as
\begin{equation}\label{eq_g_fixPT}
\log g_{\gamma}/g_0=\frac{1}{1+\pi ^2 \alpha ^2}+\frac{1}{2} \log \left(\pi ^2 \alpha ^2+1\right)-1+O\left(\frac{1}{s}\right)\,.
\end{equation}
Eq.~\eqref{eq_g_fixPT} is plotted in figure~\ref{plot2}. It
has a minimum at $\alpha=1/\pi$, where the two fixed points~\eqref{fixed_points} coincide. 

We may use eq.~\eqref{eq_g_fixPT} to further test  the $g$-theorem as in sec.~\ref{subsec_free_g}. To this aim, we denote by $g^{(i)}$ the values of the defect $g$-function at the fixed points~\eqref{fixed_points}, i.e. $g^{(i)}=g_{\gamma}/g_0\vert_{\alpha=\alpha_i}$. One can verify that the (normalized) $g$-function at the fixed points satisfies:
\begin{equation}
\log g^{(1)}\leq 0\,,\qquad
\log g^{(1)}\leq \log g^{(2)}\,,
\end{equation}
for $\varepsilon s\in [0,\pi^{-1}]$. This is in agreement with the g-theorem \cite{Cuomo:2021rkm}, which predicts that the most stable fixed point corresponds to the lowest value of $g$.  The difference $\log g^{(2)}-\log g^{(1)}$ is plotted in fig.~\ref{plot3}.

We may also use eq.~\eqref{eq_g_4} to verify explicitly the gradient formula~\eqref{eq_gradient_formula}. On the right hand side of the above,  the defect stress tensor reads $T_D=\beta_\gamma\mO$, where $\mO$ is the following defect operator: 
\begin{equation}\label{eq_operator_o}
\mO=s\,\bz\frac{\sigma^a}{2}z\phi_a=-\frac{s\alpha_0}{\gamma_0}\int d\tau'\frac{\bz\frac{\sigma^a}{2}z\,\bz'\frac{\sigma^a}{2}z'}{|x(\tau)-x(\tau')|^{d-2}}+s\,\bz\frac{\sigma^a}{2}z\,\delta\phi_a\,.
\end{equation} 
 To obtain the connected two-point function of the defect stress tensor we also have to expand the $z$ field around the saddle $z=z_0$. All field fluctuations are of order $1/\sqrt{s}$, hence to leading order we can make the replacement $\mO=s\,\bz_0\frac{\sigma^a}{2}z_0\,\delta\phi_a$, and the condition~\eqref{eq_gradient_formula} then reads:
\begin{equation}\label{eq_g_theorem_check2}
-\beta_{\alpha}\frac{\pd }{\pd \alpha}\log g_{\gamma}/g_0=-s^2\beta_\gamma^2
2\pi \int d\phi \langle\delta\phi_a(x(\phi))\delta\phi_b(x(0))\rangle
\bz_0\frac{\sigma^a}{2}z_0\,\bz_0\frac{\sigma^b}{2}z_0
\left[1-\cos(\phi)\right]\,.
\end{equation}
Proceeding similarly to the discussion around eq.~\eqref{eq_gradient_check_diag}, the above condition~\eqref{eq_g_theorem_check2} reduces to:
\begin{equation}
\frac{\pd }{\pd \alpha}\log g_{\gamma}/g_0=\frac{\pi^2 s\beta
_\alpha}{2\alpha}\,,
\end{equation}
which is easily verified using eq.~\eqref{eq_BETA} and plugging in eq.~\eqref{eq_g_4}. 

We now discuss the $1/s$ correction $\tilde{f}_0$ for fixed $d<4$.  As for the correlation function discussed in the previous section, we focus on the IR limit $\alpha_0R^{4-d}\gg 1$. The calculation is detailed in appendix~\ref{app_details_of_calc_f0}. For $d>3$ the result reads:
\begin{equation}\label{eq_f0_fixedD}
\begin{split}
\tilde{f}_{0}(\gamma_0^2 s,R,d)=\pi R^{4-d}\alpha_0I_1^{(d)}\frac{5-d}{4-d}+\text{c.c.}
+O\left(\left(R^{4-d}\alpha_0\right)^0,\left(R^{4-d}\alpha_0\right)^{1-\frac{d-3}{4-d}}\right)\,,
\end{split}
\end{equation}
where we neglected a scheme-dependent cosmological constant term, denoted by $\text{c.c.}$. Eq.~\eqref{eq_f0_fixedD} scales as $R^{4-d}\alpha_0$ in the IR, like the leading order~\eqref{eq_ftminus1}.  From eq.~\eqref{eq_f0_fixedD} we compute the $1/s$ corrections to the defect entropy~\eqref{eq_free_sD} in the $R\rightarrow\infty $ limit:
\begin{equation}
s_D\stackrel{R\rightarrow\infty}{=}
-s\pi\alpha_0 R^{4-d}\rho(d)\left[1+\frac{1}{s}\,\frac{5-d}{4-d}+O\left(\frac{1}{s^2}\right)\right]\,.
\end{equation}

The case of $d=3$ needs to be discussed separately, as the expansion in eq.~\eqref{eq_f0_fixedD} takes a more intricate form due to the logarithmic behavior of the propagator mentioned above eq.~\eqref{eq_psi_c3d}. Neglecting again a cosmological constant contribution,  the final result reads:
\begin{equation}\label{eq_f0_fixedD_3d}
\begin{split}
\tilde{f}_{0}(\gamma_0^2 s,R,3)=
-2\pi\alpha_0 R\log\left(\log(\alpha_0 R)\right)
+\text{c.c.}
+O\left(\alpha_0 R\frac{\log^2\left(\log(\alpha_0 R)\right)}{\log (\alpha_0 R)}\right)
\,.
\end{split}
\end{equation}
Eq.~\eqref{eq_f0_fixedD_3d} scales as $R\log(\log R)$ in the IR, differently than the leading order~\eqref{eq_ftminus1} which is proportional to $R\log R$. Subleading terms in the expansion are suppressed by powers of  $\frac{\log\left(\log(\alpha_0 R)\right)}{\log (\alpha_0 R)}$.  From eq.~\eqref{eq_f0_fixedD_3d}, we find that the defect entropy reads:
\begin{equation}
s_D\stackrel{d=3}{=}-s2\pi R\alpha_0 \left\{1-\frac{1}{s}\left[ \frac{1}{\log (\alpha_0 R)}+
O\left(\frac{\log^2\left(\log(\alpha_0 R)\right)}{\log^2 (\alpha_0 R)}\right)
\right]
+O\left(\frac{1}{s^2}\right)
\right\}\,,
\end{equation}
and indeed in the infrared $s_D\to-\infty$, consistently with the absence of an infrared DCFT.

\section{Spin defects at large \texorpdfstring{$s$}{s}: the interacting theory}
\label{sec_magnetic_defects_interacting_bulk}

\subsection{Setup}

In this section we consider the $O(3)$ Wilson-Fisher model:
\begin{equation}\label{eq_BulkAction}
S_\text{bulk}=\int d^dx\left[\frac{1}{2}(\pd\phi_a)^2
+\frac{\lambda_0}{4!}(\phi_a)^4\right]\,,
\end{equation}
where a mass term has been tuned to zero.  We will be interested in the model~\eqref{eq_BulkAction} in the presence of a spin $s$ impurity. As in the free theory, this is modeled by inserting in the path integral the line operator~\eqref{eq_free_Defect}. We can write the corresponding DQFT action with our constrained bosonic spinor satisfying $\bz z=2s$ similarly to eq.~\eqref{eq_free_DCFT0}:
\begin{equation}\label{secondForm}
S=\int d^dx\left[\frac{1}{2}(\pd\phi_a)^2
+\frac{\lambda_0}{4!}(\phi_a)^4\right]+\int_D d\tau\left[\bz\dot z-\gamma_0\bz\frac{\sigma^a}{2}z\phi_a\right]\,, \quad
\bar{z}z=2s\,.
\end{equation}
The same comments below eq.~\eqref{eq_free_DCFT0} apply to the action~\eqref{secondForm}. In this section we will analyze the theory~\eqref{secondForm} for $s\gg 1$.  Our main findings were already summarized in the introduction in sec. 
\ref{IntroMagneticImpGen}.

The rest of this section is organized as follows. In subsec.~\ref{subsec_int_diag} we review the diagrammatic perturbative approach to the impurity. In subsec.~\ref{subsec_int_classical} we study a certain triple scaling limit, and obtain the \emph{classical} beta-function in four dimensions.  In sec.~\ref{subsec_int_fix} we study the large spin fixed point within the $\varepsilon$ expansion. Finally in sec.~\ref{subsec_int_anyD} we analyze the large spin limit in an arbitrary number of spacetime dimensions $d<4$ and make some concrete predictions for the physical case of the $O(3)$ model in three spacetime dimensions.

\subsection{Diagrammatic results}\label{subsec_int_diag}

As well known, the theory~\eqref{eq_BulkAction} flows to a weakly coupled fixed points in $d=4-\varepsilon$ dimensions with $\varepsilon\ll 1$.\footnote{We remind that the bare coupling $\lambda_0$ in the minimal subtraction scheme is renormalized according to \cite{Kleinert:2001ax}:
\begin{equation}
\lambda_0=M^{\varepsilon}\left[\lambda+\frac{\delta\lambda}{\varepsilon}+\frac{\delta_2\lambda}{\varepsilon^2}+\ldots\right]\,,\qquad
\delta\lambda=\frac{11}{3}\frac{\lambda^2}{(4\pi)^2}
+O\left(\frac{\lambda^3}{(4\pi)^4}\right)
\,,
\end{equation}
where $\lambda$ is the renormalized coupling associated with the quartic interaction of the bulk theory, $M$ is the sliding scale and, as in the previous section, we work in the MS
scheme. We will not need higher orders or the value of $\delta_2\lambda$ for what follows. Also note that the wavefunction renormalization of the fundamental field starts at two-loop order and it will not be needed in what follows. \label{footnote_convention2}} The theory admits the following beta function:
\begin{equation}\label{eq_LambdaBeta}
\beta_{\lambda}=-\varepsilon\lambda+\frac{11}{48\pi^2}\lambda^2
+O\left(\frac{\lambda^3}{(4\pi)^4}\right)
\,,
\end{equation}
which leads to an IR stable perturbative fixed point at:
\begin{equation}\label{eq_LambdaStar}
\frac{\lambda_*}{(4\pi)^2}=\frac{3}{11}
\varepsilon
+O\left(\varepsilon^2\right)
\,.
\end{equation}
The fixed point~\eqref{eq_LambdaStar} describes the critical $O(3)$ Wilson-Fisher model. 

We now consider the theory in the presence of the defect~\eqref{secondForm}. The diagrammatic analysis in the limit where $\varepsilon$ is the smallest parameter proceeds similarly to that in sec.~\ref{sec_free_diag} and was performed first in \cite{sachdev1999quantum,vojta2000quantum}. Here we reproduce a few results that will be necessary for what follows. 

\begin{figure}[t]
   \centering
		{\includegraphics[scale=0.25]{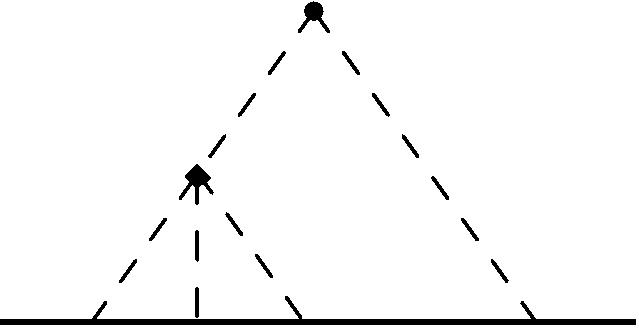}}
        \caption{Diagram for the $O\left(\lambda\gamma^4 s^4\right)$ two-loop contribution to $\langle\phi_a^2(\mathbf{x},0)\rangle$.  }
\label{fig:Diagram_interacting_bulk}
\end{figure}

By considering corrections to the one-point function $\langle\phi_a^2(\mathbf{x},0)\rangle$, one easily finds that to one-loop order the beta function of the defect coupling coincides with the free theory result~\eqref{eq_Sachdev_ct}.\footnote{There is a divergent $O(\lambda \gamma^2s^2)$ contribution to the correlation function, but this is renormalized by the $\phi_a^2$ wavefunction.} At two loops order there are several new contributions. The most important one for our purposes scales as $\lambda \gamma^4 s^4$ and arises from the two-loop diagram correction described in fig.~\ref{fig:Diagram_interacting_bulk}.  
From this diagram we extract a contribution to the counterterm, which we add to
the counterterm for the free bulk to obtain
$\delta\gamma^2=\frac{\gamma^4}{2\pi^2} +\frac{\lambda  \gamma^4 }{96 \pi ^2}s(s+1) +O\left(\lambda^3 s^0\right)$, 
from which one obtains the following beta function:
\begin{equation}\label{eq_gamma_S_beta}
\begin{split}
\beta_{\gamma^2}&=-\varepsilon \gamma^2+\frac{\gamma ^4}{2 \pi ^2}
+\frac{\lambda  \gamma^4 }{48 \pi ^2}s(s+1)
+O\left(\varepsilon^3 s^0\right)\,.
\end{split}
\end{equation}
In deriving the above result we have used the counting $\lambda\sim\gamma^2\sim\varepsilon$ and neglected $O\left(\varepsilon^4\right)$ (three-loop) contributions. Furthermore, we neglected $O\left(\varepsilon^3 s^0\right)$ contributions (of order $\gamma ^6 s^0,\,\lambda \gamma^4 s^0$ and $\lambda^2\gamma^2 s^0$), since we would like to think about $s$ which is $s\gg1$ (but not as large as to require a resummation, yet). 
Later on we will reproduce the $\lambda\gamma^4 s^2$ term from a classical calculation.
Eq.~\eqref{eq_gamma_S_beta} implies that the coupling in the IR flows to a perturbative fixed point at:
\begin{equation}\label{eq_gamma_fix_diag_int}
\gamma^2_*=2\pi^2\varepsilon\left[1-\frac{2\pi^2}{11}\varepsilon s(s+1)+O\left(\varepsilon s^0\right)\right]\,.
\end{equation}

Various quantities of interest were computed to two-loop order in the $\varepsilon$ expansion in \cite{sachdev1999quantum,vojta2000quantum,Sachdev:2001ky,Sachdev:2003yk}. For instance, differently from the free theory case discussed in the previous section, the scaling dimension of the defect spin operator is not  protected by the Ward identity~\eqref{eq_tilt_operator} anymore, and receives a correction at order $\varepsilon^2$ \cite{vojta2000quantum}:
\begin{equation}\label{Sdimen}
\Delta_S=\frac{\varepsilon}{2}\left[1-\frac{2\pi^2}{11}\varepsilon s(s+1)+O\left(\varepsilon s^0\right)\right]\,,
\end{equation}
where again we retained only the largest two-loop contribution for $s\gg 1$ and neglected three-loop corrections. Clearly, the above expressions should  only be trusted for $\varepsilon\to0$ with fixed $s$, since otherwise one may get a negative $\gamma_*^2$ and $\Delta_S$, which is of course disallowed.

Finally, we consider the $g$-function of the theory. The leading order result for the defect partition function coincides with the free theory one~\eqref{eq_gFree_diag}.  
At the next order we find a $\gamma^6 s^2$ correction, while the coupling $\lambda$ contributes at order $\lambda^2\gamma^2 s^2$ and $\lambda\gamma^4 s^4$.\footnote{One might naively expect a contribution $\lambda\gamma^2 s^2$, but this is proportional to a bulk tadpole and vanishes in dimensional regularization.} We compute this last contribution, which is dominant for $s\gg 1$. 
\begin{figure}[t]
   \centering
		{\includegraphics[width=0.25\textwidth]{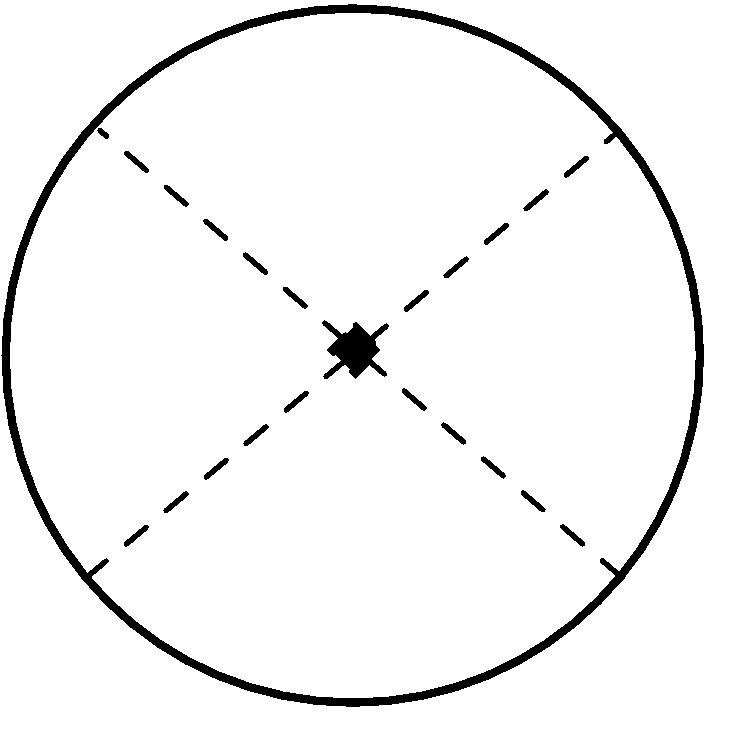}}
        \caption{ Diagram that contributes the leading order new contribution to the defect $g$-function defined in~\eqref{eq_defect_gfunction} due to the bulk self-interaction.  }
        \label{fig:diagram4}
\end{figure}
It comes from a diagram in which four defect insertions are connected through a bulk quartic vertex (see figure~\ref{fig:diagram4}). The leading $s^4$ term is obtained by neglecting all commutators and it reads: 
\es{eq_g_bulk_intPRE}{
\delta g_{\gamma}/g_0&=-\lim_{\ep\to0}\frac{\gamma_0^4\lambda_0}{4!}[s(s+1)]^2\int d^4[\tau]\int d^{2-\ep}x_\bot \,d^2x_\parallel \,\prod_{i=1}^4\frac{1}{4\pi^2\left[x_\bot+x_\parallel-x_\parallel(\tau_i)\right]^{2-\ep}}\\
&=[s(s+1)]^2 \frac{\gamma_0^4\lambda_0}{384\pi^2}\,,
}
where $x_\parallel$ and $x_\bot$ are in the plane of the defect and perpendicular to it respectively and in the last line we used the result obtained in appendix B of \cite{Cuomo:2021kfm} for the value of the integral. Adding~\eqref{eq_g_bulk_intPRE} to the leading order~\eqref{eq_gFree_diag} and writing the result in terms of the physical couplings $\gamma$ and $\lambda$ (see footnotes~\ref{footnote_convention} and~\ref{footnote_convention2} for our conventions) we obtain:
\begin{equation}\label{eq_int_g_diag}
\begin{split}
\log g_{\gamma}/g_0 &=\;\;\;-\frac{\varepsilon}{8}s(s+1)\gamma^2+\frac{\gamma^4 s(s+1)}{32 \pi ^2}+\frac{\gamma^4\lambda[s(s+1)]^2}{768\pi^2} + O\left(\varepsilon^{3}s^2\right)
\\
&\overset{\mathclap{\text{fix.pt.}}}{=}\;\;\;-
\frac{\pi^2}{8} s (s+1) \varepsilon ^2+\frac{\pi^4}{44} [s(s+1)]^2 \varepsilon ^3+O\left(\varepsilon^{3}s^2\right)\,.
\end{split}
\end{equation}

\subsection{The all-orders structure of perturbation theory}\label{subsec_int_classical}

To begin our exploration into the physics of large $s$ it is very useful to understand systematically the structure of the beta function and other physical quantities as a function of $\gamma,\lambda,s$. Our analysis in the previous subsection only allows to understand the regime of small $\gamma,\lambda$ and fixed $s$, and we clearly need to go beyond that to understand the true large $s$ limit.

By the same arguments as in the free bulk theory in section~\ref{Sec_FreeTheory}, as the spin $s$ of the impurity increases the standard perturbative approach becomes less and less accurate, and eventually breaks down. It turns out that perturbation theory nicely reorganizes as an expansion in $\gamma^2$ with arbitrary functions of $\gamma^2s, \lambda s$.  
This reorganization would be very useful to us, so let us prove it:
We implement the rescalings $\phi_a\to \phi_a/\sqrt{\lambda}$ and $z\rightarrow \sqrt{s} z$ in eq.~\eqref{secondForm}:
\begin{equation}\label{thirdForm}
S={1\ov \lambda}\int d^dx\left[\frac{1}{2}(\pd\phi_a)^2
+\frac{1}{4! }(\phi_a)^4\right]+\st\int d\tau\left[\bz\dot z-\frac{\gamma}{\sqrt{\lambda}}\bz\frac{\sigma^a}{2}z\phi_a\right]\,, \quad
\bar{z}z=2\,.
\end{equation}
From requiring that all terms in the action scale the same way, we obtain a new semiclassical limit:
\es{TripleScaling}{
&s\to\infty\,,\quad \ga\to 0\,,\quad \lam\to 0\,,\\
&\al={\ga^2 s\ov 4\pi^2}=\text{fixed}\,,\quad y\equiv{\sqrt{\lam} \,\ga s\ov 4\pi}=\text{fixed}\,.
} 
This of course reduces to the semiclassical limit~\eqref{eq_double_scaling_limit}
for $y=0$, i.e. a free bulk.\footnote{Notice that, as for the double-scaling limit \eqref{eq_double_scaling_limit}, the small $\alpha$ and $y$ limit of the semiclassical approach \eqref{TripleScaling} matches the results of the standard diagrammatic expansion.}

From this it follows that the beta function and $\Delta_S$ admit the following expansions:  
\begin{equation}\label{eq_beta_structure_double_scaling_int}
\beta_{\gamma^2}=\gamma^2\left[-\varepsilon+\beta_0^{(4d)}(\alpha,y)+\gamma^2 \beta_1^{(4d)}(\alpha,y)+\cdots\right]\,,
\end{equation}
\begin{equation}\label{DeltaS}
\Delta_S=\gamma^2\left[\Delta_0(\alpha,y)+\gamma^2\Delta_1(\alpha,y)+\cdots\right]\,.
\end{equation}
In~\eqref{DeltaS} we have extended the notion of $\Delta_S$ away from the fixed point -- to obtain the physical scaling dimension of the spin operator at the fixed point we must evaluate $\Delta_S$ for solutions of $\beta_{\gamma^2}=0$.

The perturbative result~\eqref{eq_gamma_S_beta} corresponds to
$\beta_0={1\over 3}y^2+...$ and $\beta_1={1\over 2\pi^2}+...$. Unlike in the free theory case, for the bulk interacting theory the leading term $ \beta_0^{(4d)}(\alpha,y)$ in eq.~\eqref{eq_beta_structure_double_scaling_int} is nonzero. 

Due the semiclassical nature of the expansion~\eqref{eq_beta_structure_double_scaling_int},  $\beta_0^{(4d)}$  can be completely understood from the properties of a new classical solution. Since developing this direction is somewhat tangential to our main thrust, we detail this conceptually and technically interesting analysis in appendix~\ref{App_ClassicalBeta}.

\begin{figure}[t]
\centering
\includegraphics[scale=0.8]{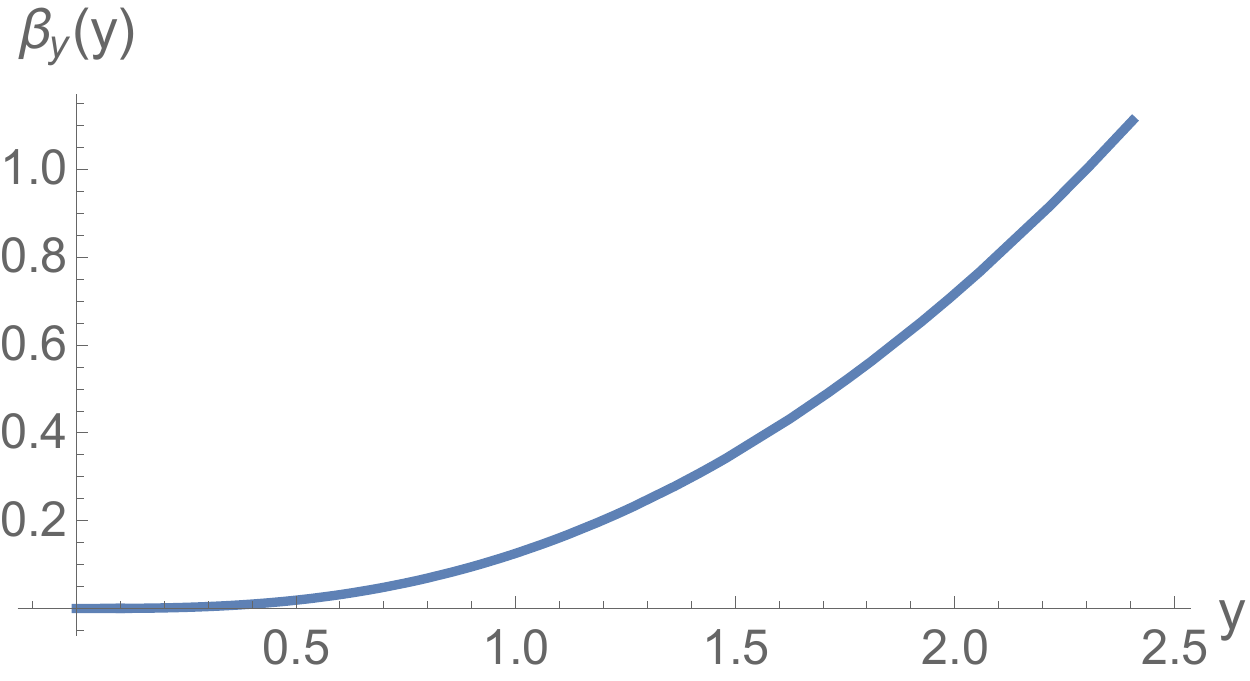}
\caption{Plot of the 4-dimensional beta function $\beta_y(y)$ as a function of $y$, extracted from the solution of the saddle-point equation (given by the ODE~\eqref{chieq}).}\label{fig:betay}
\end{figure}

For us, the most important conclusion is that $\beta_0^{(4d)}$  is only a function of $y$,  $\beta_0^{(4d)}=\beta_0^{(4d)}(y)$ and that $\beta_0^{(4d)}(y)$ is a monotonically increasing function, implying that $\beta_0^{(4d)}(y)$ has no zeroes other than at $y=0$. From this and fig.~\ref{fig:betay} we conclude that there is no interacting DCFT fixed point in 4 dimensions, which is not surprising. The only fixed point is the decoupled one with $y=0$. However, the major difference from the previous section, in that we have a nonzero $\beta_0^{(4d)}$, leads to rather different conclusions also in $4-\varepsilon$ dimensions. Since now $\beta_0^{(4d)}\neq 0$ we must solve 
$\varepsilon=\beta_0^{(4d)}(y)$ in order to find fixed points in $4-\varepsilon$ dimensions. Since $\beta_0^{(4d)}(y)$ has no zeroes other than at $y=0$ and does not tend to zero at infinity, the only solution is at infinitesimal $y$, 
\begin{equation}\label{sol}
y_*=\sqrt {3 \ep}~.
\end{equation}  

The semiclassical approach~\eqref{TripleScaling} is very useful to understand the form of the beta function, scaling dimensions etc. We will use it further below. But there are no new fixed points in this semiclassical limit.\footnote{Yet the results from the classical analysis, which allow to fix $\beta_0^{(4d)}(y)$, lead to a wealth of information about various perturbative corrections to the $\beta$ function which correspond to high-order effects in the usual diagrammatic methods. 
For instance, from the next to leading order in the expansion of $\beta_0^{(4d)}(y)$ around $y=0$ given in~\eqref{betaBehaviors} (and remembering that the running of $\lambda$ is subleading in four dimensions), we infer that in four dimensions   
\es{betagasq}{
\beta_{\ga^2}^{(4d)}&={\lam\ga^4s^2\ov 48\pi^2}-{\lam^2\ga^6s^4\ov 1536\pi^4}+\dots\,,
}
We have thus reproduced the $s^2$ piece of the third term from~\eqref{eq_gamma_S_beta} and essentially computed a new (scheme-independent) 4-loop term which should be possible, in principle, to verify also from the standard diagrammatic approach.
}

\subsection{The  phase diagram in \texorpdfstring{$4-\varepsilon$}{4-epsilon} dimensions}\label{subsec_int_fix}

With the large amount of results that we have amassed both in the standard $\varepsilon$ expansion and the semiclassical regime~\eqref{TripleScaling} we can now systematically understand the phases of the defect for small $\varepsilon$ but arbitrary $s$. We will see that a major simplification occurs in the large $s$ limit and we will argue that the same simplification happens in any number of dimensions, which would lead us, finally, to a solution of the model in $d=3$ for large $s$ as well. 

In our exploration of the phase diagram we will fix a finite small $\varepsilon$ and let $s$ vary. The bulk coupling is going to be fixed to its fixed point~\eqref{eq_LambdaStar}.  For small $s$ the standard perturbative analysis holds. We find a healthy infrared DCFT. The defect coupling scales as \begin{equation}\label{regone} 
s\sim O(1),\quad d\to4~,\qquad\gamma_*\sim \sqrt{\varepsilon}~.\end{equation} 
As we keep increasing $s$ the corrections of order $\varepsilon s^2$ in~\eqref{eq_gamma_fix_diag_int},~\eqref{Sdimen} begin to increase in importance and eventually, for $\varepsilon s^2\sim1$ we must switch to a different description, which is accurately provided by our analysis around~\eqref{sol}. Therefore for large $s$ we find 
\begin{equation}\label{regtwo}
s\to\infty~,\quad d\to4~,\qquad \gamma_*={\sqrt{11}\over s}~.\end{equation}

Throughout this whole regime, and for arbitrary $s$, the terms beyond the three terms quoted in~\eqref{eq_gamma_S_beta} always remain parametrically small, provided only that $\varepsilon$ is sufficiently small and regardless of $s$.\footnote{This can be justified by a careful analysis of the implications of the structure~\eqref{eq_beta_structure_double_scaling_int} in the perturbative regime, together with the observation that powers of $s$ are always multiplied by a greater than equal power of $\gamma$.} Therefore we can in fact find results that cover the whole range of $s$ near four dimensions: 
\begin{equation}\label{eq_int_fixptNew}
\gamma_*^2= \frac{2 \pi ^2 \varepsilon }{1+2 \pi ^2 \varepsilon s^2 /11}\left[1
-\varepsilon^{1/2}
\frac{2 \pi ^2 \sqrt{\varepsilon s^2}}{ 11+2 \pi ^2  \varepsilon s^2}
+O\left(\varepsilon\right)\right]\,.
\end{equation}
This formula clearly interpolates between the regimes~\eqref{regone} and~\eqref{regtwo}. For $\varepsilon s^2\ll 1$ the fixed point~\eqref{eq_int_fixptNew} reduces to the one in eq.~\eqref{eq_gamma_fix_diag_int} that was analyzed in \cite{sachdev1999quantum,vojta2000quantum,Sachdev:2001ky,Sachdev:2003yk}. In the opposite regime, $\varepsilon s^2\gg 1$, we get the scaling~\eqref{regtwo} and in this regime our results are new to the best of our knowledge. 

We can compute many other observables in this framework. Let us quote a few. All of them apply for arbitrary $s$ and small $\varepsilon$.

 Let us consider first the correlation function $\langle\phi_a^2(\mathbf{x},0)\rangle$.
Up to order $O(\varepsilon)$ corrections, the one-point function $\langle\phi_a^2(\mathbf{x},0)\rangle$ coincides with the tree-level free theory result~\eqref{phi2_diag}. Using eq.~\eqref{eq_int_fixptNew}, at the fixed point we obtain
\begin{align}\label{eq_phi2_1pt}
\langle\phi_a^2(\mathbf{x},0) \rangle 
\;\;&\overset{\mathclap{\text{fix.pt.}}}{\equiv}\;\;
\frac{\mathcal{N}_d}{\mathbf{x}^{d-2}}a_{\phi^2}\,,\\
\label{eq_phi2_diag_int}
a_{\phi^2}\;\;&=\;\;
\frac{\pi^2\varepsilon s^2}{2\sqrt{6}\left(1+2\pi^2\varepsilon s^2/11\right)}
\left[1+\varepsilon^{1/2}\frac{1}{\sqrt{\varepsilon s^2}\left(1+2\pi^2\varepsilon s^2/11\right)}+O\left(\varepsilon\right)
\right]\,.
\end{align}
Next we consider the spectrum of defect operators, and in particular the spin operator $S$.
Using the two-loop result in \cite{vojta2000quantum} and combining with~\eqref{eq_int_fixptNew} we can read the dimension of the spin operator $S^a$:\footnote{Incidentally,  we notice that the extrapolation of eq.~\eqref{eq_int_Delta} to $\varepsilon\rightarrow 1$ and $s=1/2$ gives $\Delta_S\approx 0.1 \divisionsymbol 0.3$, which is not too different from the result $\Delta_S=0.20(1)$ of Quantum Monte Carlo simulations \cite{PhysRevLett.98.087203}.}
\begin{equation}\label{eq_int_Delta}
\Delta_S=\frac{\gamma^2_*}{4\pi^2}\left[1 +O\left(\varepsilon\right)\right]
=\frac{\varepsilon/2}{1+2 \pi ^2 \varepsilon s^2 /11}\left[1
-\varepsilon^{1/2}
\frac{2 \pi ^2 \sqrt{\varepsilon s^2}}{ 11+2 \pi ^2  \varepsilon s^2}
+O\left(\varepsilon\right)\right]\,.
\end{equation}
From the beta-function we obtain the scaling dimension $\Delta_{S\cdot\phi}$ of the perturbation $S^a \phi_a $:
\begin{equation}\label{eq_int_anomalous}
\gamma_{S\cdot\phi}=\Delta_{S\cdot\phi}-1=\left.\frac{\pd\beta_{\gamma^2}}{\pd\gamma^2}\right\vert_{\beta_{\gamma^2}=0}=\varepsilon+O\left(\varepsilon^2\right)\,.
\end{equation}

Finally we consider the $g-$function of the theory at the fixed point. We find that this is given by the diagrammatic result~\eqref{eq_int_g_diag} up to relative $O(\varepsilon)$ corrections. At the fixed point the result reads:
\begin{equation}\label{eq_int_g_LargeS}
\log g_{\gamma}/g_0\overset{\text{fix.pt.}}{=}-\frac{\pi ^2  s^2 \varepsilon ^2 }{8\le({1+2 \pi ^2 s^2 \varepsilon/11}\ri)}
\left[1+\varepsilon^{1/2}\frac{1}{\sqrt{\varepsilon s^2}\left(1+2 \pi ^2 s^2 \varepsilon/11\right)}
+O\left(\varepsilon\right)\right]\,.
\end{equation}
The result is negative for arbitrary values of $\varepsilon s^2$ in agreement with the attractive nature of the fixed point. 
Away from the fixed points, the result \eqref{eq_int_g_diag} satisfies: 
\begin{equation}
\gamma^2\frac{\partial \log g_{\gamma}/g_0}{\partial \gamma^2} = \frac{s(s+1)}{8}\beta_{\gamma^2}\,,
\end{equation}
from which it is possible to check that the defect entropy obtained from eq.~\eqref{eq_int_g_diag} obeys the gradient formula~\eqref{eq_gradient_formula}.

\subsection{Large spin impurity as a pinning field}\label{subsec_int_anyD}

\subsubsection{An interpretation of the small \texorpdfstring{$\ep$}{epsilon} results for large \texorpdfstring{$s$}{s}}

Above we have quoted several predictions, valid for small $\varepsilon$ and large $s$. Here we would like to give an alternative description of the large $s$ limit. 
To motivative it consider~\eqref{eq_int_Delta} in the large $s$ limit: 
$\Delta_S\sim 1/s^2$. Therefore the spin operator on the defect becomes decoupled at large $s$. This suggests that the large $s$ theory might have a free sector. 

We claim that at large $s$ the defect~\eqref{eq_free_DCFT0} flows to the pinning field DCFT in which the symmetry $O(3)$ is explicitly broken to $O(2)$ but in addition there is a decoupled QM with $2s+1$ degrees of freedom which induces, at large $s$, an integral over the direction of the magnetic field inside $SO(3)$ so that the full system at large $s$ is still manifestly $SO(3)$ invariant.

Therefore, the large $s$ limit is identified with the DCFT that describes a localized (pinning) magnetic field, averaged over the two sphere. The $z$ variables do not fluctuate at large $s$, which is why the dimension $\Delta_S$ tends to zero in that limit.
This picture receives $1/s$ corrections that we will later compute. 

The advantage of this new description at very large $s$ is that a lot is known about the $O(N)$ theories with a localized magnetic field~\cite{Cuomo:2021kfm}. (We will also briefly review it in the next section.)
In particular a lot is also known about the pinning field also 
away from the $\varepsilon$ expansion, allowing us to make  rigorous predictions also in $d=3$. For now, however, we focus on matching some of the explicit calculations above at large $s$ with the previously-known results about the pinning field DCFT, finding remarkable agreement. 

The mysterious $\sqrt{11}$ factor in~\eqref{regtwo} can now be simply understood.
Since the $z$ degrees of freedom do not fluctuate at large $s$, $\gamma_*s$ is simply the magnitude of the effective localized magnetic field in the  fixed point. The magnitude of the effective localized field was found to leading order in~\cite{Cuomo:2021kfm} to be $\sqrt{N+8}$ in the $O(N)$ theory. Plugging $N=3$ we see that this nicely matches the $\sqrt{11}$ factor.
As another example, the large $s$ limit of~\eqref{eq_int_g_LargeS} gives $\log g_\gamma/g_0 \to {-11\varepsilon\over 16 }$, while for the localized magnetic field defect it was found in~\cite{Cuomo:2021kfm} that the leading order result for  $\log g$ is $ -{8+N\over 16}\varepsilon$, which again nicely matches our present result upon plugging $N=3$. 
Finally, for the one-point function~\eqref{eq_phi2_diag_int} we obtain in the large $s$ limit $a_{\phi^2}={11 \over 4\sqrt 6  }$.  The corresponding one-point function in the pinning field DCFT was computed in \cite{Cuomo:2021kfm} for small $\varepsilon$ and exactly the same result was found. 

The fact that the large $s$ limit of the spin impurity is described by the pinning field with an integration over the direction of the external magnetic field is correct not only at the infrared DCFT -- it extends to the RG flow, as the reader can verify.

\subsubsection{Equivalence with the pinning field defect at large \texorpdfstring{$s$}{s}}

On the one hand, we have the large $s$ limit of the spin impurity which, in the lattice system, would describe an external atom with large spin interacting with the bulk in an $SO(3)$ invariant fashion. On the other hand, we have the $SO(3)$ system with an external, localized in space, magnetic field, explicitly breaking $SO(3)$ to $SO(2)$. 

These two systems are essentially claimed to be equivalent at $s\to\infty$, after properly averaging over the direction of the external localized magnetic field inside $SO(3)$. We have demonstrated this equivalence close to four dimensions above. 

In fact this equivalence extends to any $d$. Furthermore, at large finite $s$ one can set up a systematic perturbation theory around the pinning field DCFT and hence we have a systematic $1/s$ expansion at hand in any $d$. The leading coupling between the two sectors is through the tilt operator of the pinning DCFT which allows us to control the $1/s$ expansion.

The purpose of this section is to formalize the above statements and show some examples of how the $1/s$ expansion around the pinning field DCFT can be performed. We will use this correspondence to write down some explicit predictions about the large $s$ limit of the spin impurity in $d=3$ as well as to discuss $1/s$ corrections in $d=3$.\footnote{As in the rest of this paper, our analysis will concern only with the DCFT at zero temperature; in the future it would be interesting to study with this method also thermal susceptibilities, for which results from Quantum Monte Carlo simulations are available \cite{PhysRevLett.98.087203,PhysRevLett.99.027205}.}

Let us formally consider the path integral for an arbitrary observable $\mO$ in the DQFT that we obtain by coupling the $O(3)$ critical model in arbitrary $d\leq 4$ to the spin $s$ impurity:
\begin{equation}\label{eq_int_formal}
\langle\mO\rangle=\mathcal{Z}^{-1}\int_{\bz z=2} \mathcal{D}\phi \mathcal{D}z\, \mathcal{O}\exp\left(-S_{O(3)}-s\int_Dd\tau\bz\dot{z}-\gamma_0 s\int_D d\tau \bz\frac{\sigma^a}{2}z\phi_a\right)\,,
\end{equation}
where as usual we work in the normalization $\bz z=2$, $\mathcal{D}\phi\, e^{-S_{O(3)}}$ is a formal notation for the measure of the (generically strongly coupled) path integral describing the $O(3)$ critical point and $\mathcal{Z}$ is the vacuum partition function. 

We have argued that in the large $s$ limit the fluctuations of the $z$ variables are suppressed and hence in the $s\gg 1$ limit we should expand around a constant profile for the quantum-mechanical spinor:
\begin{equation}
\bz\frac{\sigma^a}{2}z=\hat{n}^a=\text{const}\,.
\end{equation}
We therefore recast the path integral~\eqref{eq_int_formal} as:
\begin{equation}\label{eq_int_formal2}
\langle\mO\rangle=\mathcal{Z}^{-1}\int_{S^2} d^2\hat{n} \int\mathcal{D}\phi \mathcal{D}\chi\, \mathcal{O}\exp\left[-S_{O(3)}-\int_Dd\tau\bar{\chi}\dot{\chi}-\gamma_0 s\int_D d\tau \hat{n}^a\phi_a
+O\left(\frac{1}{\sqrt{s}}\right)\right]\,,
\end{equation}
where we separated the two-dimensional integral over the zero-mode rotating $\hat{n}$ from the path integral over the $\tau$-dependent fluctuations defined similarly to eq.~\eqref{eq_fluct}. In eq.~\eqref{eq_int_formal2} we neglected a $\sim\gamma_0\sqrt{s}$ coupling between $\phi_a$ and the fluctuation $\chi$, as this is $1/\sqrt{s}$ suppressed compared to the $\gamma_0 s\hat{n}^a\phi_a$ term. Therefore~\eqref{eq_int_formal2} is a useful description at very large $s$.

Let us now imagine performing the path integral over $\phi$ and the fluctuations $\chi$ for a fixed direction $\hat{n}$. The variable $\chi$ is decoupled from the bulk field up to $1/\sqrt{s}$ corrections. The path integral over $\phi$ coincides with the one obtained by perturbing the critical $O(3)$ model with a \emph{pinning} magnetic field $h=\gamma_0 s$ localized on a line.  This is precisely the setup which defines the pinning field DCFT studied in \cite{Cuomo:2021kfm}. Let us briefly review this construction.

Consider for a moment a general Wilson-Fisher $O(N)$ fixed point. We can consider perturbing it with a magnetic field $h$ localized on a line as:
\begin{equation}\label{eq_pinning}
\delta S=h\int_D d\tau\, \hat{n}^a\phi_a\,.
\end{equation}
The coupling $h$ breaks $O(N)\rightarrow O(N-1)$ explicitly and constitutes a relevant perturbation of the trivial line defect. The $g$-theorem implies that the RG flow results in a stable nontrivial DCFT in the IR; we will call such a DCFT the \emph{pinning} field DCFT in the following. The pinning field DCFT is obviously strongly coupled in $d=3$, but it admits a perturbative description in the $\varepsilon$ expansion or in the large $N$-limit, for which several results were obtained in \cite{Cuomo:2021kfm} (see also \cite{Allais:2014fqa}).  Additionally,  a few Monte Carlo results for $N=1$ and $N=3$ were obtained in \cite{2017PhRvB..95a4401P} (see also \cite{2014arXiv1412.3449A}).  

We conclude that, neglecting $1/s$ corrections, the $\phi$ path integral in the IR coincides with that of the $O(3)$ pinning field DCFT. We therefore obtain
\begin{equation}\label{eq_int_formal_result}
\langle\mO\rangle_\text{IR}=\mathcal{Z}_\text{IR}^{-1}\int_{S^2} d^2\hat{n} \int\mathcal{D}\phi \mathcal{D}\chi\, \mathcal{O}\exp\left[-S_\text{pinning}(\hat{n})-\int_Dd\tau\bar{\chi}\dot{\chi}+O\left(\frac{1}{\sqrt{s}}\right)\right]\,,
\end{equation}
where $\mathcal{D}\phi\, e^{-S_\text{pinning}(\hat{n})}$ is a formal notation for the path integral measure associated with the pinning field DCFT, which is obtained perturbing the $O(3)$ Wilson-Fisher critical point with a localized magnetic field in the direction $\hat{n}$. The observable $\langle\mO\rangle_\text{IR}$ is therefore computed in a product of the free theory of the fluctuations $\chi$ and the pinning field DCFT.

We will discuss the $1/\sqrt{s}$ corrections in the next subsection. Finally, the zero-mode integral in eq.~\eqref{eq_int_formal_result} restores the $SO(3)$ symmetry by averaging over the direction of the symmetry breaking field $h\hat{n}$. Notice that the partition function $\mathcal{Z}_\text{IR}$ has a similar expression, and thus for $s\gg 1$ it factorizes as follows:
\begin{equation}
\mathcal{Z}_\text{IR}=4\pi\times\mathcal{Z}_\text{pinning} \times\mathcal{Z}_{\chi}\times\left[1+O\left(\frac{1}{s}\right)\right]\,,
\end{equation}
where the factor $4\pi=\int d^2\hat{n}$ arises from the zero-mode integral, and the remaining terms are associated with the two decoupled path integrals. 

The relation~\eqref{eq_int_formal_result} between the pinning field DCFT and the $s\rightarrow\infty $ limit of the impurity is one of our main results, as it provides access to the large $s$ limit of the spin impurity and also, as we will see, allows to consider $1/s$ corrections systematically.\footnote{In fact, as eq.~\eqref{eq_int_formal2} shows, the relation holds also for the DQFTs away from the fixed points.} 
In the following we will illustrate some implications by discussing specific observables. 

The simplest observables to be discussed are bulk one-point functions. In the pinning field DCFT, one finds the following results for the normalized one-point functions of $\phi_a$ and $\phi_a^2$:\footnote{ We use the notation $\langle\mathcal{O}(\infty)\mathcal{O}(0)\rangle=\lim_{x\rightarrow\infty}x^{2\Delta_{\mathcal{O}}}\langle\mathcal{O}(x)\mathcal{O}(0)\rangle$ where $\Delta_{\mathcal{O}}$ is the scaling dimension of $\mathcal{O}$.}
\begin{equation}\label{eq_pinning_1pt}
\begin{aligned}
\frac{\langle \phi_a(\mathbf{x},0)\rangle_\text{pin}}{\sqrt{\langle \phi_1(\infty) \phi_1(0)\rangle_{O(3)}}}
&=\hat{n}_a\frac{a^\text{(pin)}_{\phi}}{|\mathbf{x}|^{\Delta_{\phi}}}\,,\\
\frac{\langle \phi_a^2(\mathbf{x},0)\rangle_\text{pin}}{\sqrt{\langle \phi_a^2(\infty) \phi_b^2(0)\rangle_{O(3)}}}
&=\frac{a^\text{(pin)}_{\phi^2}}{|\mathbf{x}|^{\Delta_{\phi^2}}}\,,
\end{aligned}
\end{equation}
where $\Delta_{\phi}$ and $\Delta_{\phi^2}$ are the corresponding bulk scaling dimensions (see e.g. \cite{Pelissetto:2000ek,Poland:2018epd,Henriksson:2022rnm}), the subscript $\langle\ldots\rangle_\text{pin}$ denotes pinning field correlators and we normalized by the two-point functions in the absence of a defect, distinguished by the subscript $\langle\ldots\rangle_{O(3)}$. The coefficients $a^\text{(pin)}_{\phi}$ and $a^\text{(pin)}_{\phi^2}$ were computed in \cite{Cuomo:2021kfm} in the $\varepsilon$ expansion (to one-loop order) and in the large $N$ limit. 

To relate~\eqref{eq_pinning_1pt} with one-point functions in the presence of a spin $s\gg 1$ impurity all we have to do is to integrate over the zero-mode.  Clearly this simply projects one-point functions onto the singlet sector and thus we find:
\begin{equation}\label{eq_1pt_from_pinning}
\begin{aligned}
\frac{\langle \phi_a(\mathbf{x},0)\rangle}{\sqrt{\langle \phi_1(\infty) \phi_1(0)\rangle_{O(3)}}}&=0\,,\\
\frac{\langle \phi_a^2(\mathbf{x},0)\rangle}{\sqrt{\langle \phi_a^2(\infty) \phi_b^2(0)\rangle_{O(3)}}}
&=\frac{a^\text{(pin)}_{\phi^2}}{|\mathbf{x}|^{\Delta_{\phi^2}}}\left[1+O\left(\frac{1}{s}\right)\right]\,.
\end{aligned}
\end{equation}
We have already checked in the previous section that this method agrees with explicit computations in the large $s$ limit of the spin impurity near four dimensions, as both approaches give  $a_{\phi^2}= {11\over  4\sqrt{6}}$. For $d=3$ there is not yet a prediction for $a^\text{(pin)}_{\phi^2}$ of the pinning field DCFT that is entirely reliable, however, even in the absence of such a prediction, our methods show that the two quantities should agree between an impurity at large spin and an localized magnetic field in the infrared.

The lowest dimensional operator (other than the unit operator) of the pinning field DCFT is the \emph{tilt} operator $\hat{t}_{\hat{a}}$, where the index $\hat{a}$ runs over the components orthogonal to $\hat{n}$. The tilt is thus an $O(2)$ vector and its dimension is protected to be $1$ nonperturbatively in the pinning field DCFT. 
This predicts that at large $s$ the spin impurity DCFT must have an $SO(3)$ vector defect operator that we denote $\hat{\phi}_a$, whose scaling dimension is therefore marginal to leading order in $s$:
\begin{equation}\label{Delta1}
\Delta_{\hat{\phi}_a}=1+O\left(\frac{1}{s}\right)\,.
\end{equation}
The existence of a marginal operator is a highly non-generic fact for the spin impurity DCFT and therefore constitutes a very nontrivial prediction of our approach! 

In fact, one can construct several marginal composites made out of $\hat{\phi}_a$ and $S$, since they are decoupled at leading order in the $1/s$ expansion and $S$ has scaling dimension $0$ at leading order. 

We will now analyze these composites and their quantum numbers.
First let us construct the $SO(3)$ vector $\hat{\phi}_a$ from the tilt operator $\hat{t}_{\hat{a}}$ of the pinning field DCFT.  The procedure is completely analogous to the one used in the pion Lagrangian to construct operators transforming in a $SU_L(N_f)\times SU_R(N_f)$ representation out of operators linearly transforming in a representation of the diagonal subgroup \cite{Weinberg:1996kr}. We discuss it below for completeness.

It is convenient to define a $SO(3)$ matrix $\tilde{\Omega}^{-1}_n$ which aligns the saddle-point in the direction ``1":
\begin{equation}
(\tilde{\Omega}^{-1}_{n})^a_{\;b}\hat{n}^b=\delta^a_1.
\end{equation}
Using this matrix we can now parametrize the space dependent fluctuations $\chi$ and $\bar{\chi}$ precisely as in eqs.~\eqref{eq_Bloch_par} and~\eqref{eq_fluct} of sec.~\ref{subsec_NonLocalQM}. For our purposes, it is convenient to do this as follows
\begin{equation}
\bz\frac{\sigma^a}{2}z=(\tilde{\Omega}_{n})^a_{\;b}(\Omega_{\chi,\bar{\chi}})^b_{\;c}\delta^c_1=(\Omega_\text{full})^a_{\;b}\hat{n}^b\,,
\end{equation}
where we defined two other matrices as
\begin{equation}\label{eq_coset}
\Omega_{\chi,\bar{\chi}}=e^{-i\delta\phi Q_3}e^{-i\delta\theta Q_2}\,,\qquad
\Omega_\text{full}=\tilde{\Omega}_{n}\Omega_{\chi,\bar{\chi}}\tilde{\Omega}_{n}^{-1}\, ;
\end{equation}
the relation between $\delta\theta,\,\delta\phi$ and $\chi,\,\bar{\chi}$ is as in eq.~\eqref{eq_fluct},  and $\left(Q_a\right)^b_{\;c}=-i\varepsilon_{abc}$ are the $SO(3)$ matrices in the fundamental representation.  

It is now straightforward to use the $SO(3)/SO(2)$ coset in eq.~\eqref{eq_coset} to construct a vector out of the pinning field tilt operator $\hat{t}_{\hat{a}}$. This is simply done as:
\begin{equation}\label{eq_hatphi_def}
\hat{\phi}=\Omega_\text{full}\cdot\hat{t}=
\tilde{\Omega}_n\cdot\Omega_{\chi,\bar{\chi}}\cdot
\left(\begin{array}{c}
0\\
\hat{t}_2\\
\hat{t}_3
\end{array}\right)=
\tilde{\Omega}_n\cdot
\left(\begin{array}{c}
\hat{t}_3\frac{   \chi +\bar{\chi}}{\sqrt{2s}}+i \hat{t}_2 \frac{\chi -\bar{\chi}}{\sqrt{2 s} }+O\left(\frac{1}{s^{3/2}}\right)\\
\hat{t}_2+O\left(\frac{1}{s}\right)\\
\hat{t}_3+O\left(\frac{1}{s}\right)
\end{array}\right)\,,
\end{equation}
where we defined in obvious notation 
\begin{equation}\label{eq_tilt_rot}
\left(\begin{array}{c}
0\\
\hat{t}_2\\
\hat{t}_3
\end{array}\right)\equiv
\tilde{\Omega}_n^{-1}\cdot \hat{t}\,.
\end{equation}
A similar analysis allows to reconstruct operators in the $2n'+1$ dimensional $SO(3)$ representation out of pinning field operators with charge $n$ under $O(2)$ as long as $n'\geq |n|$.  Notice also that the operator~\eqref{eq_hatphi_def} so defined is orthogonal to the impurity spin, $S\cdot\hat{\phi}=0$, and therefore we cannot build $O(3)$ singlets by considering composites of $S$ and $\hat{\phi}$. In the large $s$ limit there is therefore no marginal singlet of $SO(3)$ but there is a marginal vector of $SO(3)$ (and in fact there is a marginal operator of any positive integer spin), as we said.

It was argued in \cite{Cuomo:2021kfm} that the lowest dimensional neutral defect operator (besides the identity) in the pinning field DCFT is irrelevant and it is identified with the infrared version of the perturbation $\hat{\phi}_n\equiv\hat{n}^a\phi_a\vert_{\mathbf{x}=0}$ on the defect.  On the impurity side of our correspondence, this operator is naturally identified with the infrared version of the operator coupling the impurity spin with the bulk order parameter, $(S\cdot{\phi})\equiv S^a\phi_a\simeq s\hat{n}^a\phi_a+O\left(\sqrt{s}\right)$. From the results of \cite{Cuomo:2021kfm} we conclude that its scaling dimension reads:\footnote{The result for $d=3$ was obtained in \cite{Cuomo:2021kfm} interpolating between a two-loop epsilon expansion calculation and the two-dimensional exact value; it is in agreement with the large $N$ prediction and the Monte Carlo estimate of \cite{2017PhRvB..95a4401P}.}
\begin{equation}\label{eq_pinning_DeltaPhi1}
\Delta_{S\cdot\hat{\phi}}=
\Delta_{{\phi}_n}\big\vert_\text{pin}
+O\left(\frac{1}{s}\right)\,,\qquad
\Delta_{\hat{\phi}_n}\big\vert_\text{pin}
=\begin{cases}
1+\varepsilon+O\left(\varepsilon^2\right) & \varepsilon=4-d\ll 1\,,\\
\sim 1.55 & d=3\,.
\end{cases}
\end{equation}
Again, we happily find that the epsilon expansion result from the pinning field DCFT in eq.~\eqref{eq_pinning_DeltaPhi1} matches the independent calculation of the previous section, whose result in eq.~\eqref{eq_int_anomalous} does not depend on  $s$ to one-loop order. In $d=3$, eq.~\eqref{eq_pinning_DeltaPhi1} makes a concrete prediction for the scaling dimension of the leading irrelevant (symmetry preserving) perturbation of the spin impurity DCFT for $s\gg 1$.  This concludes our discussion of the defect operator spectrum to leading order. 

Finally, we comment that the equivalence~\eqref{eq_int_formal_result} implies that the $g$-function of the defect is given by the product of the pinning field $g$-function and that of a decoupled spin $s$ impurity:
\begin{equation}\label{gfactorization}
g=g_\text{pin}\, g_0\,\left[1+O\left(\frac{1}{s}\right)\right]\,,
\end{equation}
where $g_0=2s+1$. Since the pinning field $g$ function satisfies $g_\text{pin}<1$, we see that $g<2s+1$ and is hence consistent with the $g$ theorem.  Further, in the previous subsection we checked~\eqref{gfactorization} near four dimensions.\footnote{One can similarly extend the discussion to correlation functions, matching large $s$ correlators of the spin impurity and the pinning field correlation functions. Consider for concreteness the two-point function of the bulk order parameter $\phi_a$. In the pinning field DCFT it is convenient to decompose it in parallel and orthogonal components to the magnetic field $h\hat{n}$:
\begin{equation}
\phi_a\overset{\text{pinning}}{=}\hat{n}_a\phi_n+\phi^{\bot}_a\,.
\end{equation}
We then obtain for the two-point function of $\phi_a$ in the $SO(3)$ invariant impurity case:
\begin{equation}\label{eq_pinning_phi2pt}
\langle\phi_a(x)\phi_b(0)\rangle=\frac{\delta_{ab}}{3}\left[
\langle \phi_n(x)\phi_n(0)\rangle_\text{pin}+
\langle \phi^{\bot}_c(x)\phi^{\bot}_c(0)\rangle_\text{pin}
\right]+
O\left(\frac{1}{s}\right)
\,.
\end{equation}

}

\subsubsection{Subleading corrections}

The $1/s$ corrections to the factorized ansatz~\eqref{eq_int_formal_result} are quite interesting.  As we show below, their form is fixed by the requirement that the path integral is $SO(3)$ invariant. The interesting problem is to write down the couplings that restore the $SO(3)$ invariance of the fixed point using the pinning field infrared DCFT operators.
Our analysis will be analogous to and inspired by that in \cite{Metlitski:2020cqy}. In what follows, we assume for simplicity that all bulk and defect operators of the pinning field DCFT have canonically normalized two-point functions.

Working at leading order in the fields, the most general form of the interaction term involves a linear coupling between the angular fluctuations $\delta\theta,\, \delta\phi\sim O(1/\sqrt{s})$ and the pinning field tilt operator. In the notation~\eqref{eq_tilt_rot} this reads:
\begin{equation}\label{eq_int_to_be}
S_\text{int}=-\kappa\int_D d\tau( \hat{t}_2\delta\phi-\hat{t}_3 \delta\theta )+\delta S+\ldots\,,
\end{equation}
where the structure of the coupling is fixed by $SO(2)$ invariance, $\delta S$ consist of counterterms that are adjusted to ensure $SO(3)$ invariance order by order in $1/s$ and we neglected additional interactions that will not play a role in what follows, including those arising from the expansion of the kinetic term $\bz\dot{z}$ and nonlinear couplings between the tilt operator and the angular fluctuations. Because the tilt operator has dimension 1, the coupling~\eqref{eq_int_to_be} is dimensionless.

We now want to argue that the coupling $\kappa$ is fixed by $SO(3)$ invariance. To this aim, we momentarily consider the pinning field theory that we obtain by freezing $\hat{n}^a=\delta^a_1$.  In this theory we have:
\begin{equation}
\langle \phi_1(\mathbf{x},0)\rangle= \frac{a^\text{(pin)}_{\phi}}{|\mathbf{x}|^{\Delta_{\phi}}}\,,\qquad
\langle \phi_2(\mathbf{x},0)\rangle=\langle \phi_3(\mathbf{x},0)\rangle=0\,.
\end{equation}
If we rotate $\hat{n}$ by an infinitesimal angle $\alpha$ in the direction ``2",  we should find:
\begin{equation}\label{eq_rotation}
\langle \phi_2(\mathbf{x},0)\rangle\simeq \alpha\frac{a^\text{(pin)}_{\phi}}{|\mathbf{x}|^{\Delta_{\phi}}}\,.
\end{equation}
At the same time,  an infinitesimal rotation of the unit vector $\hat{n}$ can be understood as originating from an infinitesimal perturbation proportional to the tilt operator. Since for constant angular fluctuations the impurity action coincides with the pinning field one, the coefficient of this perturbation coincides with the coupling in eq.~\eqref{eq_int_to_be}. 
Therefore, setting $\delta\theta=0$ and $\delta\phi=\alpha$ in eq.~\eqref{eq_int_to_be} we conclude that we can express the one-point function of $\phi_2$ as:
\begin{equation}\label{eq_rotation_tilt}
\begin{split}
\langle \phi_2(\mathbf{x},0)\rangle &\simeq \alpha \kappa \int_D d\tau\langle \phi_2(\mathbf{x},0)\hat{t}_2(\tau)\rangle
=\alpha \kappa 
\int_D d\tau\frac{b_{\phi\hat{t}}|\mathbf{x}|^{1-\Delta_{\phi}}}{(\mathbf{x}^2+\tau^2)}\\
&=\alpha \frac{\pi \kappa \,b_{\phi\hat{t}}}{|\mathbf{x}|^{\Delta_{\phi}}}
\,.
\end{split}
\end{equation}
In eq.~\eqref{eq_rotation_tilt} we used that the scaling dimension of the tilt operator is $1$ and that the bulk-to-defect two-point function of $\phi$ and the tilt is completely fixed in terms of a single OPE coefficients $b^\text{(pin)}_{\phi\hat{t}}$ \cite{Herzog:2020bqw,Billo:2016cpy}, defined as:
\begin{equation}
\phi_{\hat{a}}(\mathbf{x},0)\sim\frac{b^\text{(pin)}_{\phi\hat{t}}\,}{|\mathbf{x}|^{\Delta_{\phi}-1}}\hat{t}_{\hat{a}}(0)+\cdots
\end{equation}
By comparing eqs.~\eqref{eq_rotation} and~\eqref{eq_rotation_tilt} we conclude:
\begin{equation}\label{obtained_coupling}
\kappa=\frac{a^\text{(pin)}_{\phi}}{\pi b^\text{(pin)}_{\phi\hat{t}}}\,.
\end{equation}
We have therefore expressed $\kappa$ in terms the pinning field DCFT data, $a^\text{(pin)}_{\phi}$ and $b^\text{(pin)}_{\phi\hat{t}}$.

We can now use the result~\eqref{obtained_coupling} to compute the leading nontrivial correction to the scaling dimension of the impurity spin operator in terms of the DCFT coefficients $a^\text{(pin)}_{\phi}$ and $b^\text{(pin)}_{\phi\hat{t}}$. To this aim, we notice that the free propagator for the fluctuations $\chi$ reads:
\begin{equation}\label{prop}
G_{\chi}(\tau)=\langle\chi(\tau)\bar{\chi}(0)\rangle_{s\rightarrow\infty}=
\frac12\text{sgn}(\tau)+\text{const}\,,
\end{equation}
where the constant term drops from all physical observables (its value depends on the boundary conditions, which are not important for what follows).  Using eq.~\eqref{prop}, we immediately find that:
\begin{align}\nonumber
\langle S^a(\tau) S^a(0)\rangle &=s^2+s
\langle \bar{\chi}(\tau)\chi(0)+\bar{\chi}(0)\chi(\tau)-
\bar{\chi}(\tau)\chi(\tau)-\bar{\chi}(0)\chi(0)\rangle_{s\rightarrow\infty}+O\left(s^0\right)\\[0.6em]
&=s(s+1)+O\left(s^0\right)\,, 
\label{S2pt_1s}
\end{align}
where we remind that, as explained below eq.~\eqref{eq_def_z_ordering}, the equal time product is regulated by a point-splitting procedure $\bar{\chi}(\tau)\chi(\tau)=\lim_{\eta\rightarrow 0^+}\bar{\chi}(\tau+\eta)\chi(\tau)$.  To obtain the anomalous dimension of the operator $S^a$ we need to compute the logarithmic $O(s^0\log|\tau|)$ corrections to eq.~\eqref{S2pt_1s}. These arise upon lowering twice the interaction term between the tilt and the angular fluctuations~\eqref{eq_int_to_be}, which in terms of $\chi$ reads:\footnote{The $1/s$ interaction terms arising from the expansion of the kinetic term exist also for a free decoupled impurity and thus cannot change the result~\eqref{S2pt_1s}; the counterterms in $\delta S$ are needed instead only to compute the $\tau$-independent term.}
\begin{equation}
S_{int}\supset-\frac{\kappa}{\sqrt{2s}}\int_D d\tau\left[\chi (\hat{t}_3+i\hat{t}_2)
+\bar{\chi}(\hat{t}_3-i\hat{t}_2)\right]\,.
\end{equation}
Using that $\langle\hat{t}_{\hat{a}}(\tau)\hat{t}_{\hat{b}}(0)\rangle=\delta_{\hat{a}\hat{b}}/\tau^2$,  the logarithmic correction arises from\footnote{A simple way to isolate the logarithmic contribution that we are interested in is to compute $\frac{\pd^2}{\pd\tau\pd\tau'} \langle S^a(\tau) S^a(\tau')\rangle$ for $\tau\neq \tau'$ and then integrate the result twice.  To order $O(s^0)$, only the terms explicitly shown in the first line of eq.~\eqref{correction} yield a nontrivial contribution to the derivative (where for arbitrary $\tau'\neq 0$ we should change $G_{\chi}(\tau_2)\rightarrow G_{\chi}(\tau_2-\tau')$ and $G_{\chi}(-\tau_1)\rightarrow G_{\chi}(\tau'-\tau_1)$). The result then follows immediately using eq.~\eqref{prop}.}
\begin{align}\nonumber
\delta\langle S^a(\tau) S^a(0)\rangle&= \kappa^2
\int_D d\tau_1 \int_Dd\tau_2
\frac{G_{\chi}(\tau-\tau_1)G_{\chi}(\tau_2)+G_{\chi}(-\tau_1)G_{\chi}(\tau_2-\tau)}{(\tau_1-\tau_2)^2}+\ldots
\\[0.6em]
&
\sim-2\kappa^2\log(|\tau|\Lambda)\,, \label{correction}
\end{align}
where $\Lambda$ is a cutoff scale and we neglected all $\tau$-independent contributions. 
Using eq.~\eqref{correction} we can write:
\begin{equation}
\langle S^a(\tau) S^a(0)\rangle=s(s+1)-2\kappa^2\log(|\tau|\Lambda)+O\left(s^0|\tau|^0\right)\approx
\frac{s(s+1)}{|\tau|^{2\kappa^2/s^2}}\,.
\end{equation}
Therefore, using~\eqref{obtained_coupling}, we finally obtain the result:
\begin{equation}\label{eq_DeltaS_anomalous}
\Delta_S\simeq\frac{\kappa^2}{s^2}=\frac{1}{s^2}\left(\frac{a^\text{(pin)}_{\phi}}{\pi b^\text{(pin)}_{\phi\hat{t}}}\right)^2\,.
\end{equation}
This prediction for the dimension of the spin field can be explicitly tested in the epsilon expansion. To this aim, we use that at leading order in the epsilon expansion the tilt and the fundamental field coincide in the pinning field DCFT, and thus $b_{\phi\hat{t}}=1$. Using $a^\text{(pin)}_{\phi}=\sqrt{11\over 4}+O(\varepsilon)$ which was derived in~\cite{Cuomo:2021kfm}, we find:
\begin{equation}\label{eq_DeltaS_anomalous_eps}
\Delta_S\approx\frac{11}{4\pi^2s^2}\,.
\end{equation}
Eq.~\eqref{eq_DeltaS_anomalous_eps} exactly agrees with the large $\varepsilon s^2$ limit of the anomalous dimension explicitly computed in eq.~\eqref{eq_int_Delta}.  This provides a very nontrivial check of our methodology.

By using eq.~\eqref{obtained_coupling} in eq.~\eqref{eq_int_to_be} we can in principle compute (or parameterize)  $1/s$ corrections to other observables as well. In practice, this is generically hard to do without more data about the pinning field DCFT, since, for instance, corrections to bulk one-  and two-point functions are proportional to integrals of three- and four-point bulk to defect correlation functions, whose form is not fully constrained by symmetry and about which we generically know little at the moment.

\section{Wilson lines in large representations}\label{Sec_WL}

In the previous section we demonstrated that impurities in the large spin limit can be studied in a semiclassical expansion in inverse powers of $s$. It is therefore natural to wonder if similar results hold for other line defects.  A natural setup is provided by Wilson and 't Hooft lines in four-dimensional conformal gauge theories, in the limit in which the size $s$ of the labelling representation becomes large.\footnote{Previously, Wilson lines in large representations were studied within holography, see e.g. \cite{DHoker:2007mci,Gomis:2008qa}. }

In this section we briefly compare our previous findings with exact localization results  for $1/2$-BPS Wilson lines at large $s$ in rank-$1$ Lagrangian SCFTs.  We show that the large $s$ expansion for protected observables coincides the derivative expansion for the theory on the Coulomb branch. This will lead us to conjecture universal formulas for non Lagrangian theories as well.

As a reminder, in Lagrangian theories, the 1/2-BPS loops take a form similar to standard Wilson loops except that we must also include the real part of the vector multiplet scalar \cite{Zarembo:2002an}: 
\begin{equation}\label{WL_susyprime}
D^\text{BPS}_s=\text{Tr}_{2s+1}\left[P\exp\left( \int_{\mathcal{C}}dt( i\dot x^\mu  A_\mu +|\dot x|\Phi)\right)\right]\,.
\end{equation}
There is a SUSY-breaking generalization of these line operators where the coefficient of the vector multiplet scalar is arbitrary but we will not discuss it here.\footnote{We note that we do not obtain a conformal line for arbitrary values of this coefficients. An interesting defect RG flow related to this coefficient in $\mathcal{N}=4$ SYM was the subject of some previous studies \cite{Polchinski:2011im} (see also \cite{Beccaria:2017rbe,Beccaria:2018ocq,Beccaria:2021rmj}).} $1/2$-BPS lines exist also for non-Lagrangian theories, in which case superconformal defects are roughly labeled by the electric and magnetic charges of their IR representation in the Coulomb branch of the theory \cite{Gaiotto:2010be,Cordova:2013bza,Cordova:2016uwk}.

\subsection{Exact results from localizations}

In this section we focus on two specific Lagrangian examples of $SU(2)$ superconformal gauge theories: $\mathcal{N}=4$ SYM and $\mathcal{N}=2$ SQCD with $N_f=4$ Hypermultiplets in the fundamental representation.  In these theories localization provides exact expressions for the $g$-function and the $h_D$ coefficient of the stress tensor one-point function in terms of a one-dimensional integral. We will evaluate these integrals in the double-scaling limit for weak coupling $g_\text{YM}^2$ and large representation $s$ with fixed $g_\text{YM}^2s$.\footnote{An analogous double-scaling limit was explored for large $R$-charge correlators within localization in \cite{Bourget:2018obm,Beccaria:2018xxl,Beccaria:2018owt,Grassi:2019txd,Beccaria:2020azj}, where it was also shown to be associated with a dual matrix model description. } We will also briefly discuss the large representation limit $s\rightarrow\infty $ with fixed arbitrary values of the coupling $g_\text{YM}^2$.

Let us study first the defect partition function on the equator of the four-sphere.  This can be written in terms of a one-dimensional integral as \cite{Pestun:2007rz,Pestun:2016zxk}:
\begin{equation}\label{eq_WL_gLoc1}
g=\frac{\int_{\mathds{R}} da (2a^2)e^{-\frac{16\pi^2}{g_\text{YM}^2} R^2 a^2}Z(a R)\sum_{q=-s}^s e^{4\pi R q a}}{\int_{\mathds{R}} da (2a^2)e^{-\frac{16\pi^2}{g_\text{YM}^2} R^2 a^2}Z(a R)}\,,
\end{equation}
where $R$ is the sphere radius and $Z(aR)$ represents the contribution from the fluctuations determinant.  The denominator is the sphere partition function with no Wilson loop insertion. In the first line, the Wilson loop is represented as a sum over all eigenvalues $q$ of the Cartan generator in the $2s+1$-dimensional representation. In $\mathcal{N}=4$ we have $Z=1$, while in $\mathcal{N}=2$ with $N_f=4$ we can write $Z$ as the product of the one-loop contribution and the instanton partition function: $Z=Z_\text{1-loop}Z_\text{inst}$. In the following we will need the explicit expression for the one-loop part:
\begin{equation}\label{eq_WL_Z1loop}
Z^{(\mathcal{N}=2)}_\text{1-loop}(aR)=\frac{H(2 i a R)H(-2i a R)}{\left|
H( i a R)H(-i a R)\right|^4}\,,\qquad
H(x)=e^{(1+\gamma_E)x^2}G(1+x)G(1-x)\,,
\end{equation}
where $\gamma_E$ is the Euler constant and $G$ is the Barnes G-function. For details about the instanton contribution see \cite{Moore:1997dj,Nekrasov:2002qd,Nekrasov:2003rj}; instantons  are exponentially suppressed in the limit of $s\to\infty$ with fixed $g_\text{YM}^2s$ 
 and therefore we will neglect them in what follows. It is further convenient to rescale  $a\rightarrow a/R$, so that the dependence on $R$ drops explicitly, and perform the sum in eq.~\eqref{eq_WL_gLoc1} in order to write the localization integral as:
\begin{equation}\label{eq_WL_gLoc}
g=
\frac{\int_{\mathds{R}} da (2a^2)
\exp\left[-\frac{16\pi^2}{g_\text{YM}^2} a^2+(2s+1)2a \pi\right]Z(a )/\sinh(2a \pi)}{\int_{\mathds{R}} da (2a^2)e^{-\frac{16\pi^2}{g_\text{YM}^2}  a^2}Z(a )}\,.
\end{equation}
Physically, the integration variable $a$ in eq.~\eqref{eq_WL_gLoc} is associated with the value of the bottom component of the vector multiplet in the theory deformed by a $Q$-exact term \cite{Pestun:2007rz}.  We will show below that the integral is peaked around large values of $a$ for $s\gg 1$. This is reminiscent of the saddle-point analysis that we discussed in sec. \ref{sec_free_semiclassics} for the impurity in free theory.

In the double-scaling limit  $s\to\infty$ with fixed $g_\text{YM}^2s$  we can compute the first integral in eq.~\eqref{eq_WL_gLoc} by expanding around the saddle-point obtained via extremizing the exponent:
\begin{equation}\label{eq_expansion}
a_\text{saddle}=\frac{g^2_\text{YM} (2 s+1)}{16 \pi }\,.
\end{equation} 
The partition function in the denominator is obtained by expanding $Z(a)$ around $a=0$. Accounting for the measure, the general structure of the result is:
\begin{equation}
\log g=\frac{1}{g_\text{YM}^2}f_{-1}\left(g^2_\text{YM} s\right)-\log g_\text{YM}^2+
f_{0}\left(g^2_\text{YM} s\right)+O\left(g_\text{YM}^2\right)\,,
\end{equation}
where the $-\log g_\text{YM}^2$ term arises since the numerator and the denominator in eq.~\eqref{eq_WL_gLoc} are expanded around different saddle-points.  We are interested in the regime $g_\text{YM}^2 s\gg (4\pi)^2$, where $a_\text{saddle}\gg 1$ as anticipated.

For $\mathcal{N}=4$ SYM we find the following result:
\begin{align}\label{eq_f1N4}
f_{-1}\left(g^2_\text{YM} s\right)\vert_{\mathcal{N}=4} &= \frac{(g_\text{YM}^2 s)^2}{4 }\,,\\ \label{eq_f0N4}
f_0\left(g^2_\text{YM} s\right)\vert_{\mathcal{N}=4} &= 2 \log \left(g_\text{YM}^2 s\right)+e^{-g^2_\text{YM} s/2}+O\left(e^{-g^2_\text{YM} s}\right)\,.
\end{align}
Interestingly, we see that the $g$-function of the Wilson line grows exponentially with $s$.  We will see later that the structure of $1/s$ corrections can be nicely understood in terms of the derivative expansion for the theory in the Coulomb branch, where the coefficient of the term $\log \left(g_\text{YM}^2 s\right)$ is associated with the Wess-Zumino term.

Similarly, in $\mathcal{N}=2$ SQCD with $N_f=4$ we find:
\begin{align}\label{eq_f1N2}
f_{-1}\left(g^2_\text{YM} s\right)\vert_{\mathcal{N}=2} &= \frac{(g_\text{YM}^2 s)^2}{4 }\,,\\ \label{eq_f0N2}
f_0\left(g^2_\text{YM} s\right)\vert_{\mathcal{N}=2} &= -\frac{g_\text{YM}^4 s^2 \log 2}{8 \pi ^2} +3\log (g_\text{YM}^2 s)+O\left((g_\text{YM}^2 s)^0\right)\,.
\end{align}
The leading order results~\eqref{eq_f1N4} and~\eqref{eq_f1N2} in the double-scaling limit are the same for both theories. For $\mathcal{N}=2$ SCQD with $N_f=4$ however the \emph{one-loop} contribution \eqref{eq_f0N2} also corrects the $\sim s^2$ term, differently from eq.~\eqref{eq_f0N4}. Notice that the coefficient of the $\log \left(g_\text{YM}^2 s\right)$ term in $f_0$ is different between the two examples. The expansion in eq.~\eqref{eq_f0N2} also contains exponentially suppressed terms,  but we did not display them explicitly.

We now comment on the results in the large $s$ limit with fixed coupling $g_\text{YM}^2$. In this limit it is useful to use the representation~\eqref{eq_WL_gLoc1} of the integral. To perform the calculation we retain only the weights $q=-s$ and $q=s$ in the sum, since all the other give an exponentially small contribution in this limit. In $\mathcal{N}=4$ SYM the integrals can be performed exactly and one finds:
\begin{equation}\label{eq_WL_g_locSN4}
\log g\vert_{\mathcal{N}=4}\stackrel{s\rightarrow\infty}{=}
\frac{g^2_\text{YM} s^2}{4}+\log \left(\frac{g^2_\text{YM} s^2}{2}+1\right)+\log 2+O\left(e^{-\frac12 g^2_\text{YM}s}\right)\,.
\end{equation}

To perform the same calculation in this limit in the $\mathcal{N}=2$ with $N_f=4$ theory in principle we should retain the full instanton contribution in $Z(a)$. However, most of this can be avoided by the following observation: 
We can organize the full integrand besides the Wilson loop term in~\eqref{eq_WL_gLoc1} according to the genus expansion as 
\es{genusExp}{
e^{F_{-1} a^2R^2+F_0 \log(aR)+F_1 {1\over a^2R^2}+\ldots }\,.
} The large $Ra$ expansion is tantamount to the standard expansion at small $\epsilon_1=\epsilon_2$, which is analogous to the genus expansion of topological string theory, see for instance \cite{Klemm:2002pa}.
If we now insert the highest weight contribution from the Wilson line, $e^{4\pi sRa}$, then the saddle-point occurs for $aR\sim s$ and hence the genus expansion becomes the $1/s$ expansion.

The instantons are needed to determine the $F_i$. But for large s we only need $F_{-1}$ and $F_0$.  The latter however does not receive instanton corrections due its relation with the conformal anomalies.  Hence all the instantons would do in the large $s$ limit is to dictate the exact expression for the coefficient of the $ s^2$  term. Denoting the latter by $g_\text{CB}$,
we therefore find in the large $s$ limit with fixed $g_\text{YM}$:
\begin{equation}\label{eq_WL_g_locSN2}
\displaystyle \log g=\frac{g_\text{CB}^2 s^2}{4}+3\log s+O\left(s^0\right)\,.
\end{equation}
In general $g^2_\text{CB}=g^2_\text{CB}(g_\text{YM}^2)$ is a complicated function that receives contributions from instantons, but it is easy to determine it to all orders in perturbation theory: $g_\text{CB}^2= \frac{g_\text{YM}^2}{1+\frac{g_\text{YM}^2}{2\pi^2}\log 2}$. 
Therefore, the expansion of~\eqref{eq_WL_g_locSN2} for small $g_\text{YM}$ agrees with the double-scaling limit results~\eqref{eq_f1N2} and~\eqref{eq_f0N2}.

Let us now focus on the coefficient $h_D$ of the one-point function of the stress-tensor.  We define it according to the conventions of \cite{Lewkowycz:2013laa}:
\begin{equation}
\langle T_{00}(x)\rangle_{\mathds{R}^4} =\frac{h_D}{r^4}\,.
\end{equation}
In supersymmetric theories $h_D$ is related to the Bremsstrahlung function parametrizing the energy radiated by the line and it has been computed in various examples \cite{Correa:2012at,Fiol:2015spa,Fucito:2015ofa,Bianchi:2019dlw,Galvagno:2021qyq}.  In localization  this is given by the derivative with respect to the squashing parameter $b$ of the ellipsoid partition function $g_b$ \cite{Fiol:2015spa,Bianchi:2019dlw}:
\begin{equation}\label{eq_WL_gB}
h_D=\frac{1}{12\pi^2}\frac{\pd\log g_b}{\pd b}\Big\vert_{b=1}\,,
\end{equation}
for $b$ close to one $g_b$ is obtained from eq.~\eqref{eq_WL_gLoc1} by replacing $q\rightarrow q b$:
\begin{equation}\label{eq_WL_gBLoc}
g_b=\frac{\int_{\mathds{R}} da (2a^2)e^{-\frac{16\pi^2}{g_\text{YM}^2} R^2 a^2}Z(a R)\sum_{q=-s}^s e^{4\pi R q b a}}{\int_{\mathds{R}} da (2a^2)e^{-\frac{16\pi^2}{g_\text{YM}^2} R^2 a^2}Z(a R)}+O\left((1-b)^2\right)\,.
\end{equation}
We may evaluate this integral in the $s\to\infty$ limit with fixed $g_\text{YM}^2s$. Dropping exponentially small corrections, we retain only the terms with $q=-s $ and $q=s$ in eq.~\eqref{eq_WL_gBLoc}. One can therefore trade derivatives with respect to $b$ for derivatives with respect to $s$ in eq.~\eqref{eq_WL_gB} in this limit. We thus obtain the relation:
\begin{equation}\label{eq_hD_localization}
h_D\simeq  \frac{s}{12\pi^2}\frac{\pd \log g}{\pd s}\,.
\end{equation}
Exactly the same argument allows to conclude that eq.~\eqref{eq_hD_localization} holds up to exponentially small corrections in the limit $s\rightarrow \infty$ with $g_\text{YM}^2=\text{fixed}$.  The relation~\eqref{eq_hD_localization} thus holds for the Lagrangian $\mathcal{N}=2$ rank-1 SCFTs up to theory-dependent exponentially small corrections in the large $s$ limit.

\subsection{Coulomb branch interpretation}

We previously observed that for large $s$ the localization integrals are peaked around large values of $a\sim s$. Since $a$ represents the bottom component of the vector multiplet, we expect that we should be able to reproduce the previous results by studying the Wilson loop using the effective theory on the Coulomb branch. In this section we show explicitly that this is indeed case.

\paragraph{Coulomb branch action}
Let us first review the Coulomb branch EFT.
To leading order in derivatives the action consists of the free action for a single $\mathcal{N}=2$ vector multiplet plus, in some cases (as in $\mathcal{N}=4$ SYM), some free decoupled Hypermultiplets:
\begin{equation}\label{eq_WL_FreeSusy}
\mL/\sqrt{g}=\frac{1}{g_\text{CB}^2}\left(\pd_\mu\phi^\dagger\pd^\mu\phi
+\frac16 \mathcal{R}\phi^\dagger \phi+\frac{1}{4}F^2\right)+\text{fermions}+\text{Hypers}\,,
\end{equation}
where $\mathcal{R}$ is the Ricci scalar.  The free action of the vector multiplet in particular depends on a unique Wilson coefficient $g_\text{CB}$. 

Higher derivative corrections are suppressed by inverse powers of
$\phi^\dagger\phi$ and naively start at order $O\left(|\phi|^0\right)$ (see e.g. \cite{Argyres:2003tg} for examples).  
However, there is an exception to this expectation:  the (supersymmetric) Wess-Zumino term. Its presence is required to match the UV conformal anomaly in the EFT \cite{Schwimmer:2010za,Komargodski:2011vj} and scales as $O\left(\log |\phi|\right)$. Its supersymmetric form on the Coulomb branch of $\mathcal{N}=2$ theories can be found in \cite{deWit:1996kc,Henningson:1995eh,Dine:1997nq} (see also \cite{Hellerman:2017sur} for the explicit integration over superspace). Here we will only need the leading term, whose form is:
\begin{equation}\label{eq_WL_WZsusy}
\mL_\text{WZ}\supset -\Delta a \,E_4\log \left(\phi^\dagger \phi\right)
\,,
\end{equation}
where $E_4$ is the Euler invariant, normalized so that $\int_{S^4}E_4=2$.  The contribution from the $c$-anomaly is proportional to  $W^2$, the square of the Weyl tensor, but we will work only on conformally flat manifolds, for which $W^2=0$, so we do not discuss it further. The coefficient $\Delta a= a_\text{UV}-a_\text{IR}$ represents the difference in the conformal $a$-anomalies between the SCFT and the Coulomb branch contribution, in units such that an Abelian free vector multiplet contributes with $a_\text{VM}=\frac{5}{24}$ and a free Hypermultiplet with $a_\text{HM}=\frac{1}{24}$.   We will not need additional subleading EFT terms for what follows.

Finally, the $1/2$-BPS Wilson loop in the Coulomb branch is represented as follows \cite{Zarembo:2002an}:\footnote{More formally, the loop is represented through the sum of its components on the Coulomb branch \cite{Gaiotto:2010be}. Those with maximal electric charge (in the duality frame in which the magnetic one is zero) provide the leading contribution, while the other simple line contributions are exponentially suppressed in $s$; these are negligible in the EFT. Therefore in the appropriate duality frame the loop generically reduces to a sum of two electric lines as in eq.~\eqref{eq_WL_WLrepSUSY}. }
\begin{equation}\label{eq_WL_WLrepSUSY}
D_s^\text{BPS}\longrightarrow \exp\left[ is\int_{\mathcal{C}} d\tau\left(
\dot{x}^\mu A_\mu-i\frac{\phi+\phi^*}{\sqrt{2}} \right)\right]
+ \exp\left[ -is\int_{\mathcal{C}} d\tau\left(
\dot{x}^\mu A_\mu-i\frac{\phi+\phi^*}{\sqrt{2}} \right)\right]
,
\end{equation}
where we omit for simplicity the possibility that there are nontrivial integer multiplicities for some of the lines (adding multiplicities is straightforward, of course).  The relative sign in front of the scalar terms in~\eqref{eq_WL_WLrepSUSY} is fixed by supersymmetry.

\paragraph{Results}  We can now use equations~\eqref{eq_WL_FreeSusy} and~\eqref{eq_WL_WLrepSUSY} to reproduce the results of the previous subsection for the $g$-function and the stress tensor one-point function. To this aim, we expand the gauge field and the scalar around the saddle-point profile. In flat space, the scalar profile reads
\begin{equation}\label{eq_WL_saddle_scalar}
\phi=\pm {1\over \sqrt{2}}\frac{g_\text{CB}^2s}{4\pi^2}\int_{\mathcal{C}} d\sigma\frac{1}{\left[x-x(\sigma)\right]^2}\,,
\end{equation}
where the two signs refer to the two different simple components in eq.~\eqref{eq_WL_WLrepSUSY} and we left the line contour unspecified.  As before, a Weyl rescaling allows to obtain the profile on other manifolds of interest.  Notice that $\phi\sim s$, analogously to the localization result \eqref{eq_expansion}. We saw that, at least for protected observables, this implies that the derivative expansion on the Coulomb branch coincides with the $1/s^2$ expansion. 

We can use the classical solution as we did before to evaluate the expectation value of the circular loop (i.e. the $g$ function). It turns out that the scalar contributes only to a perimeter divergence without affecting the defect entropy. The leading order contribution to the defect entropy arises from the gauge field and is given by\footnote{
Let us review the $g$ function in Abelian gauge theory. The saddle-point solution in Feynman gauge reads:
\begin{equation}\label{eq_WL_saddle_general}
A^\nu_{\pm}=\pm i\frac{g_\text{CB}^2s}{4\pi^2}\int_{\mathcal{C}} d\sigma\frac{\dot{x}^\nu(\sigma)}{\left[x-x(\sigma)\right]^2}\,,
\end{equation}
where the $+$ ($-$) refers to the solution sourced by the first (second) simple line in eq.~\eqref{eq_WL_WLrepSUSY}.  The profile~\eqref{eq_WL_saddle_general} can be translated to $S^4$ or $AdS_2\times S^2$ by a Weyl rescaling. In particular, the solution~\eqref{eq_WL_saddle_general} can be used to compute the defect partition function by evaluating the DCFT action, e.g. in flat space or on $S^4$, with the defect placed on a great circle. Subtracting a perimeter divergent term,  the result for the universal part is quoted in the text, see for instance \cite{Kobayashi:2018lil}.}
\begin{equation}\label{eq_g_EFT}
\log g=\frac{s^2g_\text{CB}^2}{4}+O\left(\log s\right)\,.
\end{equation}

The leading correction arises from the Wess-Zumino term~\eqref{eq_WL_WZsusy} and it is of order $O\left(\log s\right)$. To evaluate it, it is convenient to consider the theory on the sphere.  The results is obtained by integrating eq.~\eqref{eq_WL_WZsusy} with $\phi$ evaluated on the saddle-point profile~\eqref{eq_WL_saddle_scalar}; in practice we only need that $\phi $ scales as $ g_\text{CB}^2 s$ since we neglect $O\left(s^0\right)$ contributions and the field appears as the argument of a logarithm. In conclusion, we find the following result for the $g$-function\footnote{The one-loop contribution to the partition function of the theory coincides with the partition function of a relativistic free theory; its UV divergent part is associated with the Weyl anomaly $a_\text{IR}$ \cite{Giombi:2014xxa} and its finite part contributes to~\eqref{eq_g_EFTsusy} at order $O\left(s^0\right)$.}
\begin{equation}\label{eq_g_EFTsusy}
\log g=\frac{g_\text{CB}^2 s^2}{4}+4\Delta a\log \left(g_\text{CB}^2s\right)+O\left(s^0\right)\,.
\end{equation}
A table with the values of $\Delta a$ for all rank-1 $\mathcal{N}=2$ theories can be found in \cite{Hellerman:2017sur}. 

We can compare \eqref{eq_g_EFTsusy} with the previous localization results. Using the known value $\Delta a=1/2$ for $\mathcal{N}=4$ \cite{Hellerman:2017sur}, we see that \eqref{eq_g_EFTsusy} agrees with the double-scaling limit result \eqref{eq_f1N4} and \eqref{eq_f0N4}, as well as with the result \eqref{eq_WL_g_locSN2} for fixed $g_{YM}^2$ and large $s$,  with the identification $g_\text{YM}=g_\text{CB}$ (as expected since the gauge coupling does not receive corrections in this case).\footnote{The exponentially small correction in eq.~\eqref{eq_f0N4} instead are interpreted as the worldline action associated with the propagation of a massive BPS particle - see e.g. \cite{Grassi:2019txd,Hellerman:2018xpi,Hellerman:2021duh,Hellerman:2021yqz} for discussions of similar contributions in a related context. } Similarly,  using that $\Delta a=3/4$ for $\mathcal{N}=2$ with $N_f=4$ \cite{Hellerman:2017sur}, we find that eq.~\eqref{eq_g_EFTsusy} agrees with the previous results ~\eqref{eq_f1N2}, \eqref{eq_f0N2}  and \eqref{eq_WL_g_locSN2}.

We now compute the coefficient $h_D$ of the one-point function.  At leading order in $s$,  this is just given by the sum of the results for a free gauge field and a free real scalar. To leading order in $s$ we find
\begin{equation}\label{eq_hD_EFTsusy}
h_D=h^\text{gauge}_D+h^\text{scalar}_D=\frac{s^2 g_\text{CB}^2}{32\pi^2}+\frac{s^2g_\text{CB}^2}{96\pi^2}=\frac{s^2 g_\text{CB}^2}{24\pi^2}\,.
\end{equation}
We can rewrite this formula in terms of a derivative of the partition function~\eqref{eq_g_EFTsusy}
\begin{equation}\label{eq_hD_EFTsusy2}
h_D\simeq \frac{s}{12\pi^2} \frac{\pd \log g}{\pd s}\,.
\end{equation}
This is in perfect agreement with eq.~\eqref{eq_hD_localization}.

\paragraph{Comments} We have seen that for protected observables the derivative expansion on the Coulomb branch coincides with the $1/s$ expansion.\footnote{This is quite similar to recent results for correlation functions of large $R$-charge operators in SCFTs \cite{Hellerman:2017sur,Hellerman:2017veg,Hellerman:2018xpi,Hellerman:2020sqj,Hellerman:2021yqz,Hellerman:2021duh}, which are obtained using EFT techniques analogous to the one described in this section. Relatedly, monopole operators in $3d$ gauge theories, which are roughly analogous to 't Hooft lines in $4d$ gauge theories (see e.g. \cite{Murthy:1989ps,Dyer:2013fja,Dyer:2015zha,Chester:2015wao}), were also argued to admit a universal EFT description at large charge \cite{Cuomo:2017vzg,delaFuente:2018qwv}.}
It is therefore natural to conjecture that the results \eqref{eq_g_EFT} and \eqref{eq_hD_EFTsusy2} hold for supersymmetric lines with large gauge charge in arbitrary non-Lagrangian rank-$1$ SCFTs. 

We have not discussed non-protected observables and we have not discussed Wilson lines in non-supersymmetric theories. A crucial point to understand is the effect of the large Coulomb electric field, which is expected to create an instability related to the Schwinger effect~\cite{pomeranchuk1945energy,Schwinger:1951nm}.\footnote{We thank O. Aharony,  S. Bolognesi and E. Rabinovici for discussions on this.  The Schwinger effect was studied in holographic CFTs too~\cite{Semenoff:2011ng,Bolognesi:2012gr}. } We plan to come back to this issue in the future.

\section*{Acknowledgements}

We thank O. Aharony, S. Bolognesi, S. Komatsu,  G. Korchemsky, P. Kravchuk, E. Lauria, Y. Oz, E. Rabinovici, L. Rastelli, S. Sachdev, A. Sever, S. Shao,  B. Van Rees, and S. Yankielowicz  for useful discussions.  
GC is supported by the Simons Foundation (Simons Collaboration on the Non-perturbative Bootstrap) grants 488647 and 397411.  ZK, MM and ARM are supported in part by the Simons Foundation grant 488657 (Simons Collaboration on the Non-Perturbative Bootstrap) and the BSF grant no. 2018204.  The work of ARM was also supported in part by the Zuckerman-CHE STEM Leadership Program.

\appendix

\section{Details of the diagrammatic calculations in free theory}\label{App_Details_Diagrammatics}

\subsection{One-loop contribution to the one-point function}\label{app:EasyInt}

The one-loop contribution~\eqref{phi2_diag2} to the one-point function $\langle\phi_a^2(\mathbf{x},0)\rangle$ in the presence of the defect is proportional to the following integral:
\begin{equation}
\begin{split}
\mathcal{I}_1&=\int\limits_{\tau_1>\tau_2>\tau_2>\tau_4} d[\tau] 
G\left(x-x(\tau_1)\right) G\left(x(\tau_2)-x(\tau_4)\right)G\left(x-x(\tau_3)\right) \\
&=
\frac{1}{\left[(2-\varepsilon)\Omega_{3-\varepsilon}\right]^3}
\int\limits_{\tau_1>\tau_2>\tau_2>\tau_4} d[\tau] 
\frac{1}{\left(\mathbf{x}^2+\tau_1^2\right)^{\frac{2-\varepsilon}{2}}\,\left(\mathbf{x}^2+\tau_3^2\right)^{\frac{2-\varepsilon}{2}}\,
\left|\tau_2-\tau_4\right|^{2-\varepsilon}}
\,.
\end{split}
\end{equation}
It is convenient to evaluate first the integrals over $\tau_2$ and $\tau_4$. This leads to
\begin{equation}\label{eq_app_easyInt2}
\mathcal{I}_1=\frac{1/2}{\varepsilon(1-\varepsilon)\left[(2-\varepsilon)\Omega_{3-\varepsilon}\right]^3}
\iint d\tau_1 d\tau_3\frac{|\tau_1-\tau_3|^{\varepsilon}}{\left(\mathbf{x}^2+\tau_1^2\right)^{\frac{2-\varepsilon}{2}}\,\left(\mathbf{x}^2+\tau_3^2\right)^{\frac{2-\varepsilon}{2}}}\,,
\end{equation}
where we used the symmetry of the integrand under exchanges of $\tau_1$ and $\tau_3$ to extend the integration over the full real axis for both variables. Notice that, while the prefactor of eq.~\eqref{eq_app_easyInt2} diverges for $\varepsilon\rightarrow 0$, the remaining integral is convergent. We may therefore evaluate $\mathcal{I}_1$ expanding the integrand to first order in $\varepsilon $. To this aim we rescale $(\tau_1,\tau_2)\rightarrow |\mathbf{x}|(\tau_1,\tau_2)$ and use the following results:
\begin{align}
&\int d\tau\frac{1}{1+\tau^2}=\pi\,,\qquad
\int d\tau\frac{\log\left(1+\tau^2\right)}{1+\tau^2}
= 2\pi\log 2\,,\\[1em]
&\iint d\tau_1 d\tau_2\frac{\log\left|\tau_1-\tau_2\right|}{\left(1+\tau_1^2\right)\left(1+\tau_2^2\right)}=\pi^2\log 2\,.
\end{align}
Eventually, we arrive at
\begin{equation}
\mathcal{I}_1=-\frac{1}{128 \left(\pi ^4 \mathbf{x}^2\right) \varepsilon }-\frac{\log \left(64 \pi ^3 \mathbf{x}^6\right)+3 \gamma_E +2}{256 \pi ^4 \mathbf{x}^2}+O\left(\varepsilon\right)\,,
\end{equation}
where $\gamma_E$ is the Euler constant.

\subsection{Two-loop contribution to \texorpdfstring{$g_{\gamma}$}{gGamma} for the bulk free theory}\label{app:NotSoHardInt}

The two-loop contribution to the $g$-function studied in sec.~\ref{subsec_free_g} is proportional to the following integral, see  eq.~\eqref{eq_diagram_int2} :
\begin{equation}
I_2^{(4-\varepsilon)}=
\int_0^{2\pi} d\phi_1\int^{\phi_1}_0 d\phi_2 
\int^{\phi_2}_0 d\phi_3 \int^{\phi_3}_0 d\phi_4 
\frac{1}{\left(16\sin^2\frac{\phi_{13}}{2} \sin^2\frac{\phi_{24}}{2}\right)^{\frac{2-\varepsilon}{2}}}\,.
\end{equation}
We can evaluate it following the strategy outlined in the appendix B.2 of \cite{Beccaria:2017rbe}.  This consists in expanding the denominator using the following Fourier representation
\begin{equation}\label{eq_app_Fourier}
\frac{1}{\left[4\sin^2\left(\frac{\phi}{2}\right)\right]^{\frac{2-\varepsilon}{2}}}=\frac{1}{2\pi}c_0(\varepsilon)+\frac{1}{\pi}\sum_{n=1}^\infty c_n(\varepsilon)\cos(n\phi)\,,
\end{equation}
where we defined for $n\in\mathds{N}$:
\begin{equation}\label{master_integral}
\begin{split}
c_n(\varepsilon)&=\int_0^{2\pi} d\phi\frac{\cos (n\phi)}{\left[4\sin^2\frac{\phi}{2}\right]^{\frac{2-\varepsilon}{2}}}=\frac{2 \cos (\pi  n) \Gamma (\varepsilon -1)}{\Gamma \left(\frac{\varepsilon }{2}-n\right) \Gamma \left(n+\frac{\varepsilon }{2}\right)}
=-|n|\pi+O\left(\varepsilon\right)\,.
\end{split}
\end{equation}
One then evaluates the integrals over the Fourier components using the following identity:
\begin{equation}
\int_0^{2\pi} d\phi_1\int^{\phi_1}_0 d\phi_2 
\int^{\phi_2}_0 d\phi_3 \int^{\phi_3}_0 d\phi_4 \cos(n\phi_{13})\cos(m\phi_{24})=\begin{cases}
\displaystyle 0  & \text{if }m\neq n>0\,,\\[0.6em]
\displaystyle \frac{\pi^2}{m} &\text{if } m=n>0\,,\\[0.6em]
\displaystyle -\frac{2\pi^2}{m^2} &\text{if } n=0\,,m>0\,,\\[0.6em]
\displaystyle \frac{2\pi^4}{3} &\text{if } n=m=0\,.
\end{cases}
\end{equation}
(The rest of the cases follow from the symmetry $n\leftrightarrow m$.)
As a result we find
\begin{equation}\label{eq_app_int2_intermediate}
I_2^{(4-\varepsilon)}=\frac{\pi^2}{6}\left[c_0(\varepsilon)\right]^2
-2c_0(\varepsilon)\sum_{n=1}^{\infty} \frac{c_n(\varepsilon)}{n^2}
+\sum_{n=1}^{\infty}\left[c_n\left(\varepsilon\right)\right]^2\,.
\end{equation}
Notice that $c_0(\varepsilon)=O\left(\varepsilon\right)$ due to eq.~\eqref{master_integral}. Therefore the first term in eq.~\eqref{eq_app_int2_intermediate} is of order $O(\varepsilon^2)$ and we will neglect it in what follows. One might similarly conclude that the second term is of order $O(\varepsilon)$, but this would be incorrect. This is because $c_0(\varepsilon)=O(\varepsilon)$ is multiplied by a sum which is logarithmically divergent for $\varepsilon=0$,  as it can be noticed using $c_n(0)/n^2=-\pi/ n$. The sum therefore results in a $1/\varepsilon$ pole in dimensional regularization, which compensates the simple zero of $c_0(\varepsilon)$ and leads to a finite result. 

In conclusion, we need to evaluate the two infinite sums in eq.~\eqref{eq_app_int2_intermediate}. This is easily achieved using the substitution \cite{Beccaria:2017rbe}
\begin{equation}\label{eq_app_int2_sub}
c_n(\varepsilon)\rightarrow -\pi n^{1-\varepsilon}\,,
\end{equation}
which is exact up to $O(\varepsilon)$ corrections. One may further check from the asymptotic expansion of $c_n(\varepsilon)$ for $n\rightarrow \infty$ that the terms neglected in eq.~\eqref{eq_app_int2_sub} do not lead to logarithmically divergent sums in eq.~\eqref{eq_app_int2_intermediate}, and therefore remain $O(\varepsilon)$ suppressed with respect to the leading order also after the summation.  Using eq.~\eqref{eq_app_int2_sub} both infinite sums are convergent when $\varepsilon$ is analytically continued to a sufficiently large value and can be evaluated using
\begin{equation}
\sum_{n=1}^{\infty} n^{\alpha}=\zeta(-\alpha)\,.
\end{equation}
Expanding the final result in $\varepsilon$ we finally arrive at eq.~\eqref{eq_diagram_int2}.

\section{Details of the semiclassical calculations in free theory}\label{App_Details_Semiclassics}

\subsection{The \texorpdfstring{$1/s$}{1/s} corrections to the one-point function of \texorpdfstring{$\phi_a^2$}{phiA2} close to four dimensions}\label{app:1Ovs_phi2}

In this appendix we evaluate the correction in eq.~\eqref{phi2_intermediate} to the leading order result~\eqref{eq_phi2_LO} for the one-point function $\langle \phi_a^2(\mathbf{x},0)\rangle$ in $d=4-\varepsilon$ with $\varepsilon\ll 1$. This amounts at evaluating the following integral in dimensional regularization:
\begin{equation}\label{eq_app_I3a}
\mathcal{I}_3=\lim_{\eta\rightarrow 0^+}\int \frac{d\omega}{2\pi}G_\chi(\omega)\left[\left|h_{\mathbf{x}}(\omega)\right|^2
-e^{i\omega \eta}\left|h_{\mathbf{x}}(0)\right|^2\right]\,,
\end{equation}
where $G_\chi (\omega)$ is defined in eq.~\eqref{eq_psi_prop} and $h_{\mathbf{x}}(\omega)$ is in eq.~\eqref{eq_hxOmega}. 

It is convenient to further rewrite the integral as:
\begin{equation}\label{eq_app_I3a2}
\mathcal{I}_3=\lim_{\eta\rightarrow 0^+}\int_0^{\infty} \frac{d\omega}{2\pi}\left\{\left[G_\chi(\omega)+G_{\chi}(-\omega)\right]
\left|h_{\mathbf{x}}(\omega)\right|^2-
\left[G_\chi(\omega)e^{i\omega\eta}+G_{\chi}(-\omega)e^{-i\omega\eta}\right]\left|h_{\mathbf{x}}(0)\right|^2\right\}.
\end{equation}
To evaluate the integral in eq.~\eqref{eq_app_I3a} we will expand the propagator as a series in $\alpha_0$ and commute the sum with the integral.  For this procedure to work, we must perform the expansion in both terms inside the first parenthesis of eq.~\eqref{eq_app_I3a2}. Indeed, only in this way do we obtain a series whose individual terms can be integrated without encountering an IR singularity. Using the identity
\begin{equation}
\begin{split}
G_\chi(\omega)e^{i\omega\eta}+G_{\chi}(-\omega)e^{-i\omega\eta}=&-2\frac{\sin(\eta\omega)}{\omega}\sum_{n=0}^\infty (-1)^n\alpha_0^{2n}\left(\frac{c^{(4-\varepsilon)}(\omega)}{\omega}\right)^{2n}\\
&+2\frac{\cos(\eta\omega)}{\omega}\sum_{n=0}^\infty (-1)^{n+1}\alpha_0^{2n+1}\left(\frac{c^{(4-\varepsilon)}(\omega)}{\omega}\right)^{2n+1}\,,
\end{split}
\end{equation}
we recast the integral as:
\begin{equation}\label{eq_app_I3b}
\begin{split}
\mathcal{I}_3=\lim_{\eta\rightarrow 0^+}\int_0^{\infty} \frac{d\omega}{2\pi}&\left\{\left[\left|h_{\mathbf{x}}(\omega)\right|^2-\cos(\eta\omega)
\left|h_{\mathbf{x}}(0)\right|^2\right]
\frac{2}{\omega}\sum_{n=0}^\infty (-1)^{n+1}\alpha_0^{2n+1}\left(\frac{c^{(4-\varepsilon)}(\omega)}{\omega}\right)^{2n+1}\right.\\
& \quad\left.
+2\left|h_{\mathbf{x}}(0)\right|^2\frac{\sin(\eta\omega)}{\omega}\sum_{n=0}^\infty (-1)^n\alpha_0^{2n}\left(\frac{c^{(4-\varepsilon)}(\omega)}{\omega}\right)^{2n}
\right\}\,.
\end{split}
\end{equation}
One might naively conclude that the term proportional to $\sin(\eta\omega)$ in the second line can be set to zero, by commuting the limit with the integral. We shall momentarily see that this is not the case. 

We can now commute the sum and the integral in eq.~\eqref{eq_app_I3b}. Notice that $c^{(4-\varepsilon)}(\omega)=\omega^{1-\varepsilon}c^{(4-\varepsilon)}(1)$ from eq.~\eqref{eq_psi_c}. Therefore, we can evaluate the integrals in the first line in eq.~\eqref{eq_app_I3b} using the following identity in dimensional regularization:\footnote{Notice that the limit $\eta\rightarrow 0^+$ is taken within dimensional regularization, hence before  the limit $\varepsilon\rightarrow 0$.}
\begin{multline}
\lim_{\eta\rightarrow 0^+}\int_0^{\infty} \frac{d\omega}{2\pi}
\omega ^{-(2 n+1)\varepsilon-1}
\left[\left|h_{\mathbf{x}}(\omega)\right|^2-\cos(\eta\omega)\left|h_{\mathbf{x}}(0)\right|^2\right]\\
=\frac{-\pi }{2  (2 n+1) \mathbf{x}^2}\left[\frac{1}{\varepsilon}+
(2 n+3) \log \left(2|\mathbf{x}|\right)+(2n+1) \gamma_E   +O\left(\varepsilon\right)\right]\,.
\end{multline}
The integrals in the second line are instead evaluated using:
\begin{equation}\label{eq_app_I3_split}
\lim_{\eta\rightarrow 0^+}\int_0^{\infty} \frac{d\omega}{2\pi} \frac{\sin(\eta\omega)}{\omega^{1+2n\varepsilon}}=\begin{cases}
\displaystyle \frac{1}{4} &\text{for }n=0\,,\\
\displaystyle 0 & \text{for }n\geq 1\,.
\end{cases}
\end{equation}
As anticipated,  the limit and the integral do not commute for $n=0$. Physically, this is because the propagator at $\alpha_0=0$ is discontinuous,  $\langle\chi(\tau)\bar{\chi}(0)\rangle_{\alpha_0=0}=\frac{1}{2}\text{sgn}(\tau)+\text{const}$ (the constant term drops from all physical observables).

Performing the series over $n$ we finally arrive at:
\begin{equation}\label{eq_app_I3c1}
\begin{split}
\mathcal{I}_3=&
-\frac{1}{\varepsilon}\frac{2\pi \arctan(\pi  \alpha_0)}{ \mathbf{x}^2}
-\frac{2\pi \arctan(\pi  \alpha_0) \log (4\mathbf{x}^2)}{\mathbf{x}^2}
-\frac{\pi^2\alpha_0\left[\log (4\mathbf{x}^2)+2\right]}{(1+\pi^2\alpha_0^2)\mathbf{x}^2}
+\frac{\pi^2}{2\mathbf{x}^2}+O\left(\varepsilon\right)\,,
\end{split}
\end{equation}
where the last term in eq.~\eqref{eq_app_I3c1} arises because of the point-splitting procedure from eq.~\eqref{eq_app_I3_split}.
Restoring the prefactor in eq.~\eqref{phi2_intermediate}, we arrive at the result~\eqref{eq_free_bulk_NLO}.

\subsection{Calculation of the \texorpdfstring{$\tilde{f}_0$}{f0} function}
\label{app_details_of_calc_f0}

In this appendix, we provide some technical details associated with the calculation of the $\tilde{f}_0$ function~\eqref{eq_f0_final} close to four and three dimensions.  This follows from the one-loop fluctuation determinant around the saddle-point.

In terms of the fluctuations~\eqref{eq_fluct} the quadratic action~\eqref{eq_free_defect_action_nonlocal} on the defect reads:
\begin{equation}
S^{(2)}\simeq\int d\phi\bar{\chi}\dot{\chi}-\frac{\tilde{\alpha}_0}{2R}
\int d\phi \int d\phi'
\frac{\left(\bar{\chi}\chi'+\bar{\chi}'\chi-\bar{\chi}\chi-\bar{\chi}'\chi'\right)}{\left(4\sin^2\frac{\phi-\phi'}{2}\right)^{\frac{d-2}{2}}}\,,
\end{equation}
where we defined the dimensionless combination $\tilde{\alpha}_0=\alpha_0R^{4-d}$. It is useful to decompose the fields into Fourier modes on the circle
\begin{equation}
\chi(\phi)=\sum_n \frac{e^{-i n\phi}}{\sqrt{2\pi}}\chi_n\,,\qquad
\bar{\chi}(\phi)=\sum_n \frac{e^{i n\phi}}{\sqrt{2\pi}}\bar{\chi}_n\,.
\end{equation}
The action then reads:
\begin{equation}\label{eq_Free_quadratic}
S_\text{eff}^{(2)}=\sum_n\bar{\chi}_n\left[-i n-
\tilde{\alpha}_0\left(c_n(\varepsilon)-e^{i n\eta}c_0(\varepsilon)\right)\right]\chi_n\,,
\end{equation}
where $c_n(\varepsilon)$ is defined in eq.~\eqref{master_integral} and $\eta$ is a positive infinitesimal parameter which follows from the point-splitting regularization in eq.~\eqref{eq_def_z_ordering}.
Notice the action~\eqref{eq_Free_quadratic} is independent of the zero modes, as expected since these are associated with the action of the symmetry group. 

We now perform the Gaussian integration over the fields $\bar{\chi}_n$ and $\chi_n$ in eq.~\eqref{eq_Free_quadratic}. Normalizing the result by the partition function $g_0$ of a decoupled defect, we find:
\begin{equation}\label{eq_fluct_det}
\begin{split}
\tilde{f}_0(\gamma_0^2 s,R,d) &=-\lim_{\eta\rightarrow 0^+}\sum_{n\neq 0}\log\left[-i n-\tilde{\alpha}_0\left(c_n(\varepsilon)-e^{i n\eta}c_0(\varepsilon)\right)\right]+\sum_{n\neq 0}\log\left(-i n\right)\\
&=-
\sum_{n=1}^{\infty}\log\left[1+\tilde{\alpha}_0^2
\left(\frac{c_n(\varepsilon)-c_0(\varepsilon)}{n}\right)^2\right]+
\lim_{\eta\rightarrow 0^+}\sum_{n=1}^\infty\frac{2\tilde{\alpha}_0 c_0(\varepsilon)n\sin(n\eta)}{n^2+\tilde{\alpha}_0^2
\left(c_n(\varepsilon)-c_0(\varepsilon)\right)^2}
\,,
\end{split}
\end{equation}
where we already neglected terms that mainfestly vanish in the limit $\eta\rightarrow 0$.  The result~\eqref{eq_fluct_det} holds for any value of $d=4-\varepsilon$. Our task is thus to evaluate the sums in eq.~\eqref{eq_fluct_det} in dimensional regularization.

Let us first consider the limit $\varepsilon\rightarrow 0$. We start with the second term  in the second line of eq.~\eqref{eq_fluct_det}: it is a convergent sum times $c_0(\varepsilon)=O\left(\varepsilon\right)$, hence it vanishes in the  $\varepsilon\rightarrow 0$ limit.
 One might naively conclude that $c_0(\varepsilon)$ may also be neglected also in the first term of eq.~\eqref{eq_fluct_det}. However, similarly to the discussion in appendix~\ref{app:NotSoHardInt}, this is not the case. Indeed it multiplies a $1/\varepsilon$ term arising from a logarithmically divergent sum. 
In light of this comment, we expand eq.~\eqref{eq_fluct_det} in $c_0(\varepsilon)$ and get
\begin{equation}
\tilde{f}_0(\gamma_0^2 s,R,4)=-\lim_{\varepsilon\rightarrow 0}
\sum_{n=1}^{\infty}\left\{\log\left[1+\tilde{\alpha}_0^2\left(\frac{c_n(\varepsilon)}{n}\right)^2\right]-c_0(\varepsilon)\frac{2 c_n(\varepsilon)  \tilde{\alpha}_0^2}{\left[c_n(\varepsilon)\right]^2 \tilde{\alpha}_0^2+n^2}+O\left(\frac{(c_0(\varepsilon))^2}{n^2}\right)\right\},
\end{equation}
where we retained the term linear in $c_0(\varepsilon)$, while all the higher order terms in $c_0(\varepsilon)$ are multiplied by sums which are finite in dimensional regularization and thus can be safely neglected in four dimensions. We finally evaluate the sums by expanding in $\alpha_0$ and replacing (just as in~\eqref{eq_app_int2_sub})
\begin{equation}
c_n(\varepsilon)\rightarrow-\pi  n^{1-\varepsilon} \,,
\end{equation}
which is exact up to $O(\varepsilon)$ terms. Performing the summation over $n$ in dimensional regularization and then expanding for  $\varepsilon\to 0 $, we arrive at:
\begin{equation}
\begin{split}
\tilde{f}_0(\gamma_0^2 S,R,4)&=-\sum_{k=1}^\infty\frac{(-1)^k \pi ^{2 k} \alpha_0^{2 k}}{2 k}+\sum_{k=0}^\infty\frac{(-1)^k \pi ^{2 k+2} \alpha_0^{2k+2}}{2 k+1}\\
&=
\frac{1}{2} \log \left(1+\pi ^2 \alpha_0^2\right)+\pi  \alpha_0 \arctan(\pi  \alpha_0)\,.
\end{split}
\end{equation}

Let us now consider the case of $d<4$.  
As for the correlation function $\langle\phi_a^2(\mathbf{x},0)\rangle$ in sec.~\ref{subsec_phi2_and_beta}, we focus on the IR limit $\tilde{\alpha}_0\rightarrow \infty$. 
The second term in eq.~\eqref{eq_fluct_det} can be evaluated by expanding the summand in $\tilde{\alpha}_0$. Only the first term of the expansion contributes in the limit $\eta\rightarrow 0^+$ and thus we find
\begin{equation}\label{eq_app_g_partial1}
\lim_{\eta\rightarrow 0^+}\sum_{n=1}^\infty\frac{2\tilde{\alpha}_0 c_0(\varepsilon)n\sin(n\eta)}{n^2+\tilde{\alpha}_0^2
\left(c_n(\varepsilon)-c_0(\varepsilon)\right)^2}=2\tilde{\alpha}_0 c_0(\varepsilon) \lim_{\eta\rightarrow 0^+}\sum_{n=1}^\infty\frac{\sin(n\eta)}{n}=\pi\tilde{\alpha}_0 I_1^{(d)}\,,
\end{equation}
where we used $c_0(\varepsilon)=I_1^{(d)}$ in the last line, where $I_1^{(d)}$ is given in eq.~\eqref{eq_diagram_int1}.  Eq.~\eqref{eq_app_g_partial1} is the only contribution linear in $\tilde{\alpha}_0$ to $\log g$. Adding eq.~\eqref{eq_app_g_partial1} to the leading order result~\eqref{eq_ftminus1}, one finds $\log g_{\gamma}/g_0=\pi\alpha_0(s+1)R^{4-d} I_1^{(d)} $ to linear order in $\alpha_0 R^{4-d}$, in agreement with the diagrammatic result~\eqref{eq_gFreeDiagBare}.

We now turn to the evaluation of the first term in eq.~\eqref{eq_fluct_det}. The sum converges for $d<7/2$, and can be evaluated in dimensional regularization for general values of $d$.  Each individual term in the sum scales as $\log(\tilde{\alpha}_0^2)$ and is therefore subleading with respect to eq.~\eqref{eq_app_g_partial1} in the IR. Large IR contributions can thus only arise from the large $n$ tail of the sum. We therefore expand the argument of the summand using
\begin{equation}\label{eq_app_g_expansion}
c_n(\varepsilon)-c_0(\varepsilon)
\stackrel{n\rightarrow\infty}{\simeq} 
\begin{cases}
\displaystyle-\frac{\pi  \sec \left(\frac{\pi  d}{2}\right)}{\Gamma (d-2)}n^{d-3} -c_0(\varepsilon)+O\left(n^{d-5}\right) 
& \text{for }d>3\,, \\[1em]
\displaystyle
-2 (\log n+\gamma_E +\log 4)+O\left(n^{-2}\right)
 &\text{for } d=3\,.
\end{cases}
\end{equation}
We conclude that most relevant contribution of the sum arises from the region where $n^{4-d}\sim\tilde{\alpha}_0$ for $3<d<4$ and from the one in which $n\sim\tilde{\alpha}_0\log\tilde{\alpha}_0$ for $d=3$.

To obtain a honest asymptotic expansion of the result we should proceed as in appendix D of \cite{Cuomo:2021cnb}, separating the first sum in eq.~\eqref{eq_fluct_det} into two pieces with a cutoff $\Lambda= a\tilde{\alpha}_0^{\frac{1}{4-d}}$ with $a\ll 1$. The sum over \emph{small} $n$'s must then be evaluated analytically by expanding the summand for large $\tilde{\alpha}_0$, while the sum of large $n$ can be approximated to arbitrary precision via the Euler-Maclaurin formula. In practice, if we are not interested in the $O(\tilde{\alpha}_0^{0})$ terms (not counting logarithms) we can simply replace the sum with an integral:\footnote{More precisely, this is true because the resulting integral is convergent for $n\rightarrow 0$.}
\begin{equation}\label{eq_app_g_integral}
-\sum_{n=1}^{\infty}\log\left[1+\tilde{\alpha}_0^2
\left(\frac{c_n(\varepsilon)-c_0(\varepsilon)}{n}\right)^2\right]
=-\int_0^{\infty} dn \log\left[1+\tilde{\alpha}_0^2
\left(\frac{c_n(\varepsilon)-c_0(\varepsilon)}{n}\right)^2\right]+
O\left(\tilde{\alpha}_0^0\right)\,,
\end{equation}
where we can replace $c_n(\varepsilon)-c_0(\varepsilon)$ with the expansion~\eqref{eq_app_g_expansion}. The evaluation of eq.~\eqref{eq_app_g_integral} is conveniently performed separately for $d>3$ and $d=3$.

To compute the integral~\eqref{eq_app_g_integral} in $d>3$, we rescale $n\rightarrow\tilde{\alpha}_0^{\frac{1}{4-d}} n$ and expand the integrand for large $\tilde{\alpha}_0$. We obtain
\begin{equation}\label{eq_app_g_integralD}
\begin{split}
-\int_0^{\infty} dn \log\left[1+\tilde{\alpha}_0^2
\left(\frac{c_n(\varepsilon)-c_0(\varepsilon)}{n}\right)^2\right]=&
-\pi\tilde{\alpha}_0^{\frac{1}{4-d}}  \csc \left(\frac{\pi }{8-2 d}\right) \left[\frac{\pi   \left| \sec \left(\frac{d \pi }{2}\right)\right| }{\Gamma (d-2)}\right]^{\frac{1}{4-d}}\\
&+\pi\tilde{\alpha}_0\frac{2 \pi \Gamma (3-d)}{(4-d) \Gamma \left(2-\frac{d}{2}\right)^2}+O\left(\tilde{\alpha}_0^{1-\frac{d-3}{4-d}}\right)\,.
\end{split}
\end{equation}
The first term on the right hand side of eq.~\eqref{eq_app_g_integralD} is proportional to $\tilde{\alpha}_0^{\frac{1}{4-d}} = R\alpha_0^{\frac{1}{4-d}}$.  Thus this is a pure cosmological constant term and it can be renormalized away. The second term is the leading physical contribution and it is linear in $\tilde{\alpha}_0$ as the leading order~\eqref{eq_ftminus1}.  Adding eq.~\eqref{eq_app_g_integralD} to eq.~\eqref{eq_app_g_partial1} we obtain the result~\eqref{eq_f0_fixedD} in the main text.

In $d=3$ the expansion is more subtle due to the logarithm in eq.~\eqref{eq_app_g_expansion}. In this case it is convenient to rescale $n\rightarrow 2\tilde{\alpha}_0\log\tilde{\alpha}_0\, n$, so that the integral reads:
\begin{equation}
-2\tilde{\alpha}_0\log\tilde{\alpha}_0\int_0^{\infty} dn \log\left[1+
\frac{1}{n^2}\left(1+\frac{\log\left(\log\tilde{\alpha}_0\right)+\log n+\gamma_E+\log 8}{\log\tilde{\alpha}_0}\right)^2\right]\,.
\end{equation}
We may now expand the integrand for large $ \log\tilde{\alpha}_0$ (notice $ \log\tilde{\alpha}_0\gg\log\left(\log\tilde{\alpha}_0\right)$ for $\tilde{\alpha}_0\gg 1$) and evaluate the resulting integrals. We obtain:
\begin{equation}\label{eq_app_g_integral3}
\begin{split}
-\int_0^{\infty} dn \log\left[1+\tilde{\alpha}_0^2
\left(\frac{c_n(\varepsilon)-c_0(\varepsilon)}{n}\right)^2\right]
\stackrel{d=3}{=}&
-2\pi\tilde{\alpha}_0\log\tilde{\alpha}_0 -2\pi\tilde{\alpha}_0\left[
\log\left(\log\tilde{\alpha}_0\right)+\log 8+\gamma_E\right]\\
&+
O\left(\tilde{\alpha}_0\frac{\log^2\left(\log\tilde{\alpha}_0\right)}{\log\tilde{\alpha}_0}\right).
\end{split}
\end{equation}
Adding eq.~\eqref{eq_app_g_integral3} with the limit $d\rightarrow 3$ of eq.~\eqref{eq_app_g_partial1} and neglecting the terms linear in $\tilde{\alpha}_0=\alpha_0 R$, which represent a cosmological constant contribution, we obtain the result~\eqref{eq_f0_fixedD_3d} in the main text. Notice that the term proportional to $\tilde{\alpha}_0\log R$ cancels between eqs.~\eqref{eq_app_g_partial1} and~\eqref{eq_app_g_integral3}.

\section{Running from the classical profile for the defect coupling in the \texorpdfstring{$O(3)$}{O(3)} model}\label{App_ClassicalBeta}

In this appendix we obtain the four-dimensional beta function for the defect coupling $\gamma$ in eq.~\eqref{thirdForm} to leading order in the triple-scaling limit~\eqref{TripleScaling}. As we explain below,  remarkably, the beta function in this limit can be extracted from the solution of the classical saddle-point equations.

\subsection{Beta function in the physical renormalization scheme}\label{App_ClassicalBeta1}

In this and the next section (and only in these two sections) it will prove useful to work in a \emph{physical} regularization scheme,  rather than within dimensional regularization. In particular, we will analyze the theory~\eqref{thirdForm} directly in four spacetime dimensions with a cutoff scale $M$ (whose precise definition will be given below). Thus, the couplings in the action~\eqref{thirdForm} are to be interpreted directly as the physical couplings at the scale $M$.

The saddle-point equations demand $\bz\frac{\sigma^a}{2}z=n^a$ with $n^a$ a unit vector independent of the defect position, while varying~\eqref{thirdForm} we get the bulk equation of motion:
\begin{equation}\label{eq_saddle_eq}
-\pd^2\phi_a+\frac{1}{3! }\,\phi_a(\phi_b)^2=-4\pi y \le( \bz\frac{\sigma^a}{2}z\ri)\,\delta^{d-1}(x_\bot)\,.
\end{equation}
The source imposes the following boundary condition:
\begin{equation}\label{eq_bc_sing}
\phi_a\overset{r\rightarrow 0}{\longrightarrow}-c_d\frac{y \le( \bz\frac{\sigma^a}{2}z\ri)}{ r^{d-3}}\,,
\end{equation}
where $r$ denotes the distance from the defect and $c_d=\Gamma\le((d-3)/2\ri)/\pi^{(d-3)/2}$, which was chosen such that $c_4=1$. This implies that $\phi_a\propto \bz\sigma^a z$.  As in the free theory, there is really a family of saddle-points, related by the global action of the zero-mode in the path integral.

To leading order in the semiclassical limit~\eqref{TripleScaling} the saddle-point is given by the solution of the equation~\eqref{eq_saddle_eq} in four spacetime dimensions. However, it turns out that in $d=4$
the eq.~\eqref{eq_saddle_eq} does not have a solution that is compatible with the boundary condition~\eqref{eq_bc_sing}, since the cubic term unavoidably leads to logarithmic corrections to a power law profile, making the profile more singular than required at $r\rightarrow 0$. The resolution of this conundrum is to introduce a running coupling. This is how the beta function 
$\beta_0^{(4d)}(y)$ will emerge from classical physics.

Technically, the absence of solutions to the problem~\eqref{eq_saddle_eq} in four dimensions is associated with the necessity of regularizing the source term on the right-hand side. In principle, this can be done by finding the solution in $d$ spacetime dimensions. This would lead to a singular result in the limit $d\rightarrow 4$.  By reabsorbing the singularities in the definition of the defect coupling as in eq.~\eqref{eq_alpha_ct}, we could then extract its beta function.\footnote{Notice that in four dimensions and to leading order in the triple-scaling limit~\eqref{TripleScaling} we can neglect the beta function of the bulk coupling $\lambda$~\eqref{eq_LambdaBeta}.} This strategy however suffers of some drawbacks. First, it requires finding an analytical solution to the nonlinear problem~\eqref{eq_saddle_eq} for arbitrary values of $d$. Second, dimensional regularization hides the physical origin of the \emph{classical} running. For these reasons, in the following we will introduce a more physically transparent approach, which allows studying the problem directly in four spacetime dimensions and that lends itself to a straightforward numerical implementation, bypassing the complication of finding an analytical solution to a nonlinear boundary value problem. Nonetheless, as a proof of concept,  in app.~\ref{app:MS} we will show how to find the beta function from the equation~\eqref{eq_saddle_eq} in dimensional regularization as a series expansion for small values of $y$.

To obtain a mathematically consistent formulation we require  that the boundary conditions~\eqref{eq_bc_sing} are satisfied at a certain distance $r=M^{-1}>0$ from the defect rather than for $r\rightarrow 0$:
\begin{equation}\label{eq_bc_nonSing}
\left.\phi_a=-\frac{y \le( \bz\frac{\sigma^a}{2}z\ri)}{ r}\right\vert_{r=M^{-1}}\,.
\end{equation}
Physically, the length scale $M^{-1}$ can be interpreted as the \emph{thickness} of the line defect. The boundary condition~\eqref{eq_bc_nonSing} introduces a mass scale and thus breaks the scale invariance of the action~\eqref{thirdForm}; by requiring that observables be independent of the scale $M$ we can thus obtain the beta function of the defect coupling.

 To do so, let us define the function $\chi$ by
\es{chiDef}{
\phi_a\equiv-\frac{\le( \bz\frac{\sigma^a}{2}z\ri)}{ r} \chi(\log(rM))\,,
}
where henceforth we will use a new coordinate $u\equiv \log(rM)$. At $r=1/M$ we have $u=0$, and from comparing~\eqref{eq_bc_sing} and~\eqref{chiDef}, we find that $\chi(0)$ has the interpretation of the running coupling $y$ at the scale $M$. 

$ \chi(u)$ solves the translationally invariant equation:
\begin{equation}\label{chieq}
 0=\chi''(u)-\chi'(u)-\frac16\chi^3(u)\,,
\end{equation}
where by $'$ we refer to derivatives with respect to $u$, such that $\chi'(u)\equiv \partial_u\chi(u)$.
To solve this equation we need to choose boundary conditions. One is given by the interpretation of $\chi(u)$ as the running coupling; we also impose that $\chi(\infty)=0$, which is consistent with the defect being infrared free in four dimensions. In summary, we have 
\begin{equation}\label{BC}
\chi_y(0)=y\,, \qquad \chi_y(\infty)=0\,,
\end{equation}
where the subscript labels the boundary conditions.

Next, we use the fact that the modulus squared of the bulk scalar field coincides with the leading order value of the $\phi_a^2$ one-point function and it is thus a measurable physical quantity given by
\es{BulkScalar}{
{\phi_a^2\ov\lambda}={{\chi^2}(u)\ov \lam\, r^2}\,,
}
where we divided by $\lambda$ to compensate for the field redefinition that we implemented in~\eqref{thirdForm}. Since $\chi_y(u)$ is proportional to a physical quantity, it satisfies the Callan-Symanzik equation:
\begin{equation}\label{CSeq}
\left[\frac{\partial}{ \partial \log M}+\beta^{(4d)}_y(y) \frac{\partial}{ \partial y}\right]\chi_y\left(\log(rM)\right)=0\,.
\end{equation}
From eq.~\eqref{CSeq} we find the following result for the beta function of the coupling $y$:
\begin{equation}\label{CSeqBeta}
\beta^{(4d)}_y(y)=-\frac{\chi_y'(u)}{  \partial_y\chi_y(u)}\,.
\end{equation}
As promised, eq.~\eqref{CSeqBeta} relates the beta function of the defect coupling with the classical solution to the saddle-point equations. Notice that eq.~\eqref{CSeqBeta} does not depend on $u$, and  we can exploit this $u$-independence to write a simpler formula.
We use the translation invariance of~\eqref{chieq} to write 
\begin{equation}\label{chi1}
\chi_y(u)=\chi_1(u+u_0(y))\quad \implies \quad y=\chi_1(u_0(y))\,,
\end{equation}
which defines the function $u_0(y)$ as the inverse function of $\chi_1$, in terms of which
\begin{equation}\label{betaySimp}
\begin{split}
\beta^{(4d)}_y(y)&=-\frac{\chi_y'(u)}{  \partial_y\chi_y(u)}\\
&=-\frac{\chi_1'(u+u_0(y))}{  \partial_y\chi_1(u+u_0(y))}=-\frac{1}{ u_0'(y)}\,,
\end{split}
\end{equation}
or alternatively equal to $-\chi_1'(u_0(y))$.
In practice, we can obtain $\beta_y(y)$ from the parametric plot $\{\chi_1(u_0),-\chi_1'(u_0)\}$, which is easy to make once we are in possession of $\chi_1(u)$; see fig.~\ref{fig:betay}. This function does not have a nontrivial zero, it grows monotonically and diverges as $y\to\infty$.

Let us comment on one detail of the numerics. The way we set up the problem in eq.~\eqref{betaySimp} is not ideal for numerics, since we have to shoot to find the ideal value of $\chi_1'(0)$ that gives a decaying function for $u\to\infty$. It is easier to choose a very small fixed initial value  $\de$ for $u=0$, use the asymptotic solution~\eqref{chiAsymp} that we obtain below
 in the vicinity of  $u=0$, and numerically integrate forwards and backwards to obtain $\chi_\de(u)$. (Or for $y<\de$ simply use the perturbative beta function that can be read off from the asymptotic solution, see~\eqref{betaBehaviors}.) From the derivation of eq.~\eqref{betaySimp} it should be clear that for any $\de$ we have the formula
 \es{betaySimp2}{
 \beta^{(4d)}_y(y)=-\chi_\de'(\chi_\de^{-1}(y))\,.
 }
 Then the parametric plot $\{\chi_\de(u_0),-\chi_\de'(u_0)\}$ will still give us the graph of $\beta_y(y)$, see fig.~\ref{fig:betay}.

One may wonder if the salient features of the beta function can be understood analytically, without needing a numerical solution. Next, we show that this is indeed possible. 

\subsection{Analytic results on the semiclassical beta function}\label{app:classbeta}

An elementary argument implies the absence of zero of the beta function. This is because such a zero would be unavoidably associated with a scale invariant solution to the equation~\eqref{eq_saddle_eq} in four dimensions, which however does not exist for $\lambda\neq 0$ as we discussed at length. Technically this is reflected in the fact that the beta function~\eqref{CSeqBeta} is proportional to $\chi_y'(u)$, while eq.~\eqref{eq_saddle_eq} does not admit a constant nonzero solutions. 

Notice that this observation alone does not rule out the possibility that the function $\beta^{(4d)}_y(y)$ tends to zero for $y\rightarrow \infty$, which would imply a strongly coupled fixed point for any $\ep>0$. However, even this scenario is ruled out both by numerics and by the analytic argument presented below.

We continue with the analysis of $\chi(u)$ for large $u$. We are interested in a decaying positive solution to the equation of motion~\eqref{chieq}. Performing the asymptotic analysis with this input yields:
\es{chiAsymp}{
\chi(u)={\sqrt3\ov \sqrt {u-u_*}}\le(1+{\log (u-u_*)\ov u-u_*}+\dots\ri)\,,
}
where $u_*$ is a free parameter that can be used to match the solution at finite $u$.\footnote{We could have further expanded~\eqref{chiAsymp} at large $u$, but that would have obscured the meaning of $u_*$.}
Asymptotically, we have a positive, monotonically decreasing function.\footnote{This result is actually a theorem, see \cite{bellman2013stability} Chap.~7.5 Theorem 4.} Going towards smaller $u$, we wonder if the function can have a maximum. At this maximum at $u=u_m$, we would have $\chi''(u_m)<0,\, \chi'(u_m)=0,\, \chi(u_m)>0$, which contradicts the equation of motion~\eqref{chieq}. We conclude that the function cannot have a maximum.

$\chi(u)$ thus continues to grow as we decrease $u$; it can either asymptote to a constant at $u\to-\infty$ or diverge (either at a finite $u=u_s$ or $u\to-\infty$). A simple asymptotic analysis rules out the possibility of a finite limit as $u\to-\infty$. Inspired by Chap.~7 of the book \cite{bellman2013stability}, we make an attempt at understanding the large $\chi$ behavior of the equation  of motion~\eqref{chieq}. We introduce the notation $p(\chi)\equiv \chi'(\chi)$ and rewrite eq.~\eqref{chieq} as:
\es{HardyRewrite}{
0=p {d p \ov d\chi}-p-{\chi^3\ov 6}\,.
}
It is simple to guess that the large $\phi$ behavior of $p$ is 
\es{pAsym}{
p(\chi)=-{\chi^2\ov 2 \sqrt 3}+\dots\,.
}
This result is rigorously established by Hardy's theorem, see  Chap.~5 of \cite{bellman2013stability}. We can then use the relation $p(\chi(u))=d\chi/du$ to write
\es{uprel}{
u(\chi)&=u(\chi_0)+\int_{\chi_0}^\chi {d\tilde\chi\ov p(\tilde\chi)}\,
}
and using the asymptotics~\eqref{pAsym} derive that the position of the divergence of $\chi$, $u_s$ is finite:
\es{uprel2}{
u_s&=u(\chi_0)+\int_{\chi_0}^\infty {d\tilde\chi\ov p(\tilde\chi)}<\infty\,.
}
For completeness, we have determined the near singularity behavior of $\chi(u)$:
\begin{equation}
\begin{aligned}\label{chiSing}
\chi(u)={2\sqrt{3}\ov u-u_s}&\le[1+{u-u_s\ov 6}-{(u-u_s)^2\ov 36}+{(u-u_s)^3\ov 54}
\right.\\
&\left.+\le({2\ov 135}\,\log(u-u_s)+a \ri)(u-u_s)^4+\dots\ri]\,,
\end{aligned}
\end{equation}
where the two undetermined parameters are $u_s$ and $a$: these can be used to match to the solution of interest.

Thus we have established that $\chi_1(u)$ (or any $\chi_\de(u)$) is a monotonically decreasing function on $(u_s,\infty)$ and it goes from $\infty$ to $0$. Then the function $u_0(y)$ defined in eq.~\eqref{chi1} is a function on $(0,\infty)$ that interpolates monotonically between $\infty$ and $u_s$. In eq.~\eqref{betaySimp} we derived that $\beta^{(4d)}_y(y)=-{1/u_0'(y)}$ and it follows that it is a function on $[0,\infty)$ monotonically increasing from $0$ to $\infty$.

The asymptotic formulas in eqs.~\eqref{chiAsymp} and~\eqref{chiSing} can be used to obtain the asymptotic behaviors of the beta function. We simply plug them in into eq.~\eqref{betaySimp2}, and evaluate in the case of the large $u$ asymptotics from eq.~\eqref{chiAsymp} for large $u$, and in the case for the near singularity behavior from eq.~\eqref{chiSing} small $(u-u_s)$ to obtain:
\es{betaBehaviors}{
\beta^{(4d)}_y(y)=\begin{cases}
{y^3\ov 6}-{y^5\ov 12}+{y^7\ov 9}+\dots \qquad &y\ll1\,,\\
{y^2\ov 2\sqrt{3}}-{y\ov 3}+{2\ov 3\sqrt{3}}+\dots \qquad &y\gg1\,.
\end{cases}
}
Note that in both cases the dependence on the shift of $u$, denoted by $u_*$ and $u_s$, is guaranteed to drop out, since the formula~\eqref{betaySimp2} works for any $\chi(u)$.
\footnote{On the other hand, the constant $a$ from~\eqref{chiSing} shows up in the large $y$ expansion of the beta function at $O(1/y^2)$. Its value can only be determined numerically. (At the same order we also have a $\log(y)$ correction to the power series in $1/y$.) }

\begin{figure}[t]
\centering
\includegraphics[scale=0.57]{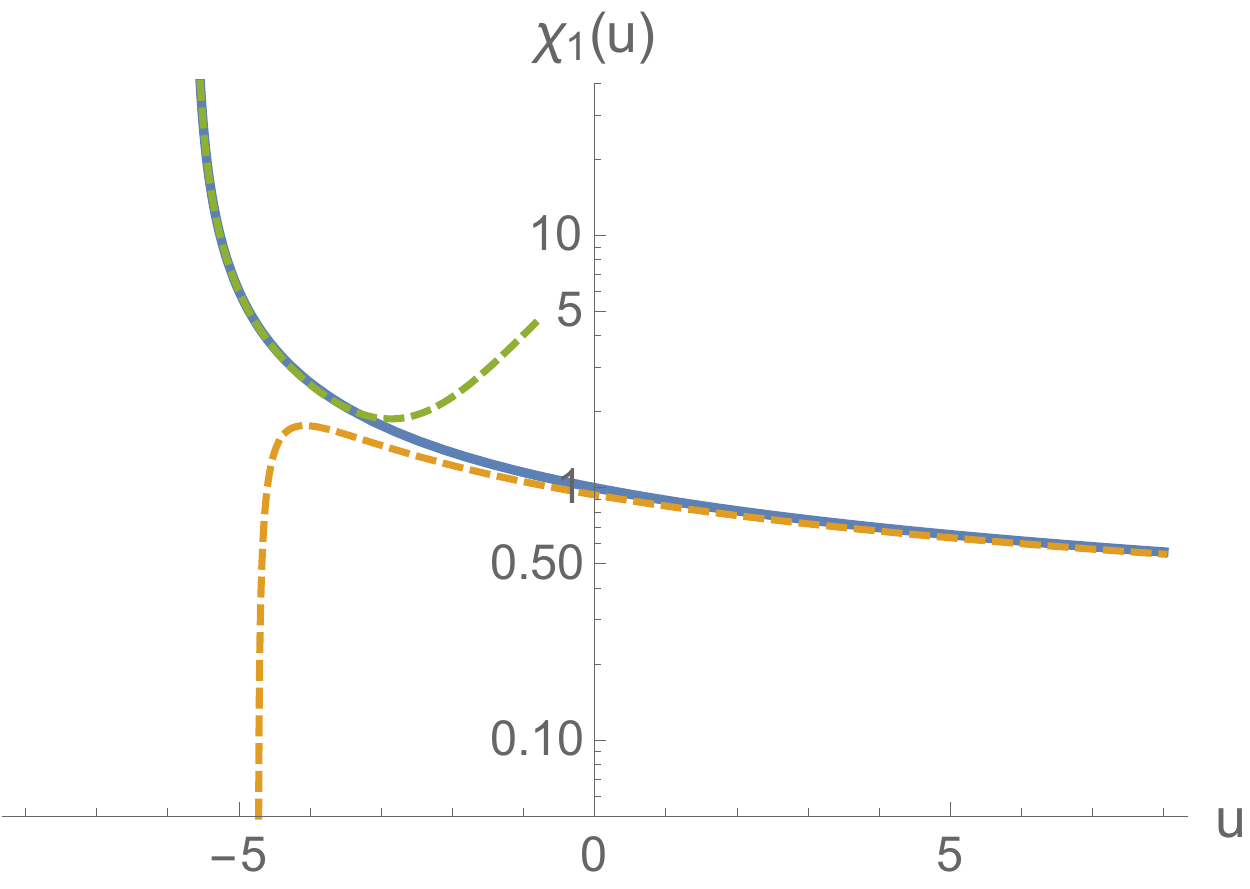}\hspace{0.5cm}
\includegraphics[scale=0.6]{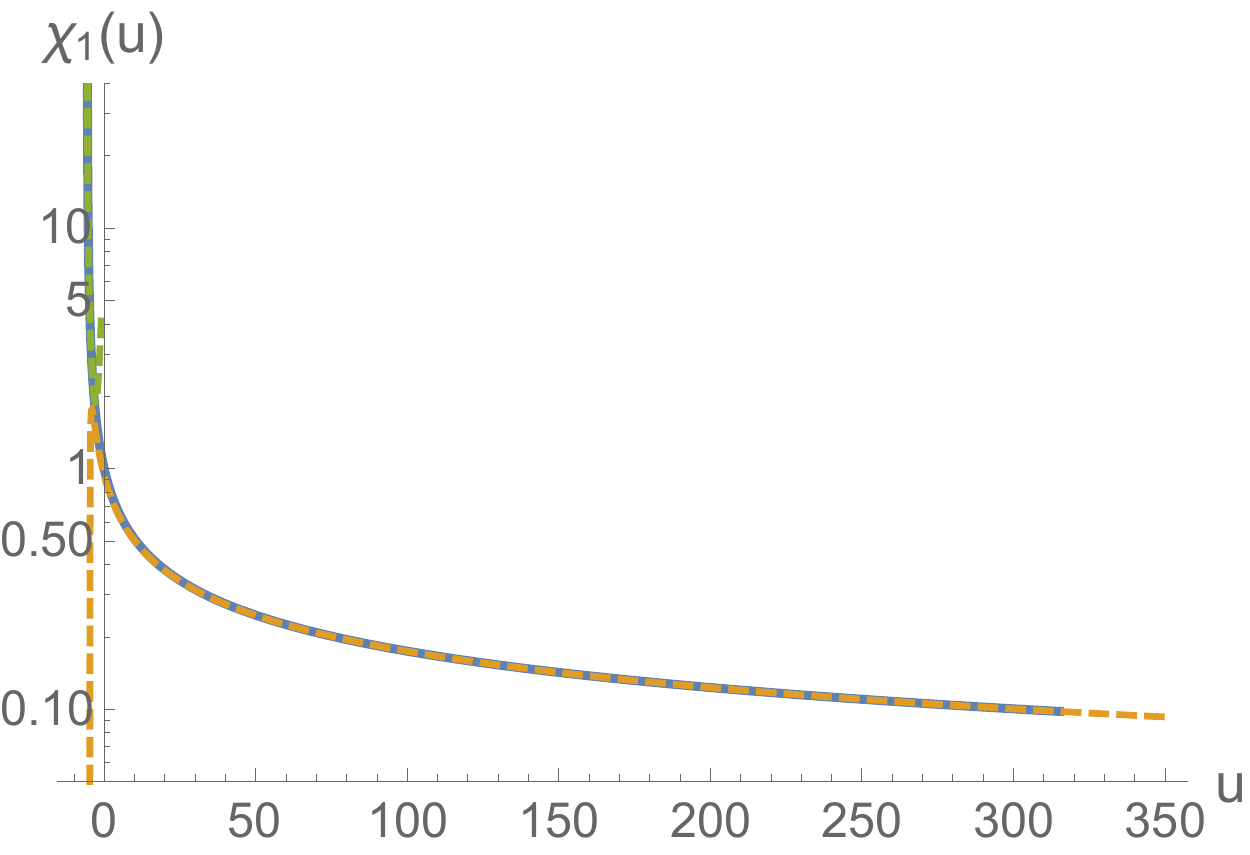}
\caption{Plot of $\chi_1(u)$ together with its asymptotic behaviors. On the left we are showing a zoomed in version of the plot on the right. The solid line is the numerical solution for  $\chi_1(u)$ and the green and orange dashed lines are the near singularity and large $u$ asymptotics respectively. Note that except for a small window of $u$ these asymptotics describe the full curve very accurately. The fitted values of the matching parameters for the asymptotics are $u_*=-5.2$ (from eq.~\eqref{chiAsymp}) and $u_s=-5.63$  (from eq.~\eqref{chiSing}; also note we did not obtain a reliable value for $a$). }\label{fig:chi1u}
\end{figure}

Finally in fig.~\ref{fig:chi1u} we plot the numerical solution for $\chi_1(u)$ together with the asymptotics from eqs.~\eqref{chiAsymp} and~\eqref{chiSing}.\footnote{We obtained this function by shifting $\chi_\de(u)$ appropriately in $u$, as discussed around eq.~\eqref{chi1}.}
Recall that we used this numerical solution to obtain the beta function in fig.~\ref{fig:betay}.

Finally, we also present a series solution (in the amplitude $y$) of the equation of motion~\eqref{chieq}.
The linearized equation has an exponentially blowing up solution $\chi\sim e^u$. We tune its amplitude to zero to satisfy the boundary condition~\eqref{BC}, and get  the small $y$ solution
\es{chiseries}{
\chi_y(u)=y-{u\ov 6}\,y^3+{u(u+2)\ov 24}\,y^5-{u(5u^2+24 u+48)\ov 432}\,y^7+O(y^9)\,,
}
which if plugged into eq.~\eqref{CSeqBeta} reproduces the beta function we got in eq.~\eqref{betaBehaviors}:
\es{chiseriesBeta}{
\beta^{(4d)}_y(y)={y^3\ov 6}-{y^5\ov 12}+{y^7\ov 9}+O(y^9)\,.
}
This had to be the case, as the series solution can be matched to the asymptotic solution from eq.~\eqref{chiAsymp}.\footnote{To match the two expressions, we have to set
\es{Adet}{
u_*=-{3\ov y^2}\le(1+{y^2\ov 2}\, \log\le({3\ov y^2}\ri)+\dots\ri)\,.
} }
To compare~\eqref{chiseriesBeta} with the diagrammatic result $\beta_0^{(4d)}={1\over 3}y^2+...$, remember that $\gamma^2\beta^{(4d)}_0$ described the flow of $\gamma^2$, which can then be translated to the flow of $y$ since we are strictly in four dimensions and the flow of $\lambda$ can be ignored.

\subsection{Beta function in the interacting \texorpdfstring{$O(3)$}{O(3)} model in dimensional regularization}\label{app:MS}

The dimensional regularization method we present here will follow the conventional machinery. However, it hides the physical meaning of the running coupling as a saddle-point field profile, which was abundantly clear in the previous computation presented in apps.~\ref{App_ClassicalBeta1} and~\ref{app:classbeta}.

Our starting point is to solve eq.~\eqref{eq_saddle_eq} in fractional dimension with the boundary condition~\eqref{eq_bc_sing}.  The equation is not scale invariant for $d<4$, therefore we were not able to find an exact solution. 
Nonetheless one can find a solution perturbatively in the bare double scaling parameter $y_0$:
\es{dSol}{
\phi_a(r)&=-c_d\frac{y_0\le( \bz\frac{\sigma^a}{2}z\ri)}{ r^{d-3}}\le[1+a_1 U^2+a_2 U^4+\dots\ri]\,,\\
U&\equiv y_0 r^{4-d}\,,
}
where we have determined the first 20 $a_i$ coefficients, e.g.
\begin{equation}
a_1=\frac{c_d^2}{12 (d-4) (3 d-11)}\,,\quad
a_2=\frac{c_d^4}{96 (d-4)^2 (3 d-11) (5 d-19)}\,.
\end{equation}
We notice the pattern that as $d\rightarrow 4$ these coefficients blow up as $a_k\sim 1/\varepsilon^{k}$. As a result, we have that 
\es{akU2k}{
a_k U^{2k}\sim \le[y_0^2\le({1\ov \ep}+\log r+O(\ep)\ri)\ri]^k\,.
}

We renormalize these divergences by requiring that $\phi_a^2/\lambda_0$, which is a physical expectation value, is finite. The only way to consistently do so is to shift the coupling $\ga_0$. We do so in the following way: 
\begin{equation}\label{gammaRenorm}
\ga_0^2= \frac{M^{\varepsilon}}{c_{4-\varepsilon}^2}\left[\ga_\text{DR}^2+\frac{\delta\ga^2(y_\text{DR})}{\varepsilon}+
\frac{\delta_2\ga^2(y_\text{DR})}{\varepsilon^2}+\ldots\right]\,,
\end{equation}
which is equivalent to
\begin{equation}\label{gammaRenorm2}
y_0=\frac{M^{\varepsilon}}{c_{4-\varepsilon}}\left[y_\text{DR}+\frac{\delta y(y_\text{DR})}{\varepsilon}+
\frac{\delta_2 y(y_\text{DR})}{\varepsilon^2}+\ldots\right]\,,
\end{equation}
where the DR subscript stands for dimensional regularization, and we conventionally divided by the factor $c^2_{4-\varepsilon}=1+O\left(\varepsilon\right)$ for convenience; therefore the physical coupling defined by eqs.~\eqref{gammaRenorm},~\eqref{gammaRenorm2} does not coincide with the one in the usual minimal subtraction scheme. To the order we are working we can neglect the running of the coupling $\lam$, the counterterms for $\ga_\text{DR}$ and $y_\text{DR}$ are identical and they are functions of the renormalized triple-scaled coupling $y_\text{DR}$. 

The beta functions of the defect coupling is conveniently written in terms of $y$ and is obtained from
\begin{equation}\label{eq_beta_double}
\beta_{y_\text{DR}}=-{\ep}y_\text{DR}+\beta^{(4d)}_{y_\text{DR}}\,,\qquad
\beta^{(4d)}_{y_\text{DR}}={y_\text{DR}}\,\frac{d\,\delta y(y_\text{DR})}{d\,y_\text{DR}}-\delta y(y_\text{DR})\,.
\end{equation}
Our result reads:
\begin{equation}\label{beta4dMS}
\begin{split}
\beta^{(4d)}_{y_\text{DR}}=&{y_\text{DR}^3\ov 6}-{y_\text{DR}^5\ov 12}+{11y_\text{DR}^7\ov 144}-\frac{59 y_\text{DR}^9}{648}+
\frac{2609 y_\text{DR}^{11}}{20736}-\frac{14869 y_\text{DR}^{13}}{77760}
+\frac{2318219 y_\text{DR}^{15}}{7464960}
\\&
-\frac{1729831 y_\text{DR}^{17}}{3265920}
+\frac{14116674883 y_\text{DR}^{19}}{15049359360}
-\frac{241476805 y_\text{DR}^{21}}{141087744}
+O\left(y_\text{DR}^{23}\right)
\,,
\end{split}
\end{equation}
where we reported several orders to illustrate the simplicity of the procedure compared to standard loop calculations.
The first two terms of eq.~\eqref{beta4dMS} agree with the result for $\beta^{(4d)}_{y}$ given in eq.~\eqref{betaBehaviors}, which was obtained in a different renormalization scheme as discussed in app.~\ref{App_ClassicalBeta1}. Indeed, to the order we are working $y$ is the only running coupling in four dimensions, and it is a well known fact that for such a setup the first two coefficients of the beta function are scheme-independent \cite{Weinberg:1996kr}.  This does not apply to the higher order terms, and indeed eq.~\eqref{beta4dMS} differs from the eq.~\eqref{betaBehaviors} in the $O(y_\text{DR}^7)$ term (and beyond).  As a consistency check, in the following we independently determine the translation between $y$ and $y_\text{DR}$ and show that the beta functions computed from the classical profile and from minimal subtraction match precisely once this is take into account.

The renormalization procedure described above leads to the following bulk scalar profile in 4 dimensions:
\es{4dProfile}{
{\phi_a\ov \sqrt{\lambda}}=-\frac{\le( \bz\frac{\sigma^a}{2}z\ri)}{ \sqrt{\lam} r}\,  y_\text{DR}\le[1
-\frac{2 u+3}{12 }\, y_\text{DR}^2+
\frac{4 u^2+20 u+31}{96}\,  y_\text{DR}^4+\ldots\ri]\,.
}
where to simplify formulas we use the coordinate $u=\log(rM)$.
Recalling the definition $\chi(u=0)=y$ in the scheme used in app.~\ref{App_ClassicalBeta1}, we get the relation:
\es{SchemeRelation}{
y=y_\text{DR}\le[1
-\frac{1}{4 }\, y_\text{DR}^2+
\frac{31}{96}\,  y_\text{DR}^4+\ldots\ri]\,.
}
Inverting this relation, it is simple to verify that the beta functions from eqs.~\eqref{betaBehaviors} and~\eqref{beta4dMS} match on the nose.

\bibliography{Biblio}
	\bibliographystyle{JHEP.bst}

\end{document}